\documentclass[3p,authoryear,preprint]{elsarticle}

\usepackage[T1]{fontenc} %

\usepackage[utf8]{inputenc} %

\usepackage{graphicx} %

\def\graphpath{./}
\graphicspath{\graphpath}

\usepackage{enumitem} %
\usepackage{subfig} %
\usepackage{amsmath,amssymb,amsthm} %
\usepackage{varioref} %
\usepackage{mathtools}
\usepackage[subnum]{cases}

\usepackage{mathrsfs}

\def\psecdv#1#2{{\frac {\partial^2 #1} {\partial #2^2} }}  %
\def\ddv#1#2{{\frac{\mathrm d #1} {\mathrm d #2} }}  %
\def\dsecdv#1#2{{\frac{\mathrm d^2 #1} {\mathrm  d #2^2} }}  %

\def\ie{\textit{i.e.}\ }  %
\def\eg{\textit{e.g.} } %

\def\ol#1{\overline{#1}}
\def\wt#1{\widetilde{#1}}

\def\exp{\text{exp}}
\def\cos{\text{cos}}
\def\sin{\text{sin}}

\def\Re{\text{Re}} %

\usepackage{bm}

\usepackage[colorlinks,allcolors=red]{hyperref}

\usepackage[overload]{empheq} %

\usepackage{ stmaryrd } %

\newcommand{\majrevisioncolor}{black}
\newcommand{\majrev}[1] {{\color{\majrevisioncolor}#1}}

\newcommand{\improvecolor}{black}
\newcommand{\improve}[1] {{\color{\improvecolor}#1}}

\usepackage{import}

\usepackage{supertabular}
\usepackage{siunitx}

\bibpunct{\color{cyan}[}{\color{cyan}]}{,}{n}{}{;}

\newcommand{\uu}[1]{\underline{#1}}

\makeatletter
\def\ps@pprintTitle{%
 \let\@oddhead\@empty
 \let\@evenhead\@empty
 \def\@oddfoot{}%
 \let\@evenfoot\@oddfoot}
\makeatother
\makeatletter
\providecommand{\doi}[1]{%
  \begingroup
    \let\bibinfo\@secondoftwo
    \urlstyle{rm}%
    \href{http://dx.doi.org/#1}{%
      doi:\discretionary{}{}{}%
      \nolinkurl{#1}%
    }%
  \endgroup
}
\makeatother

\begin{document}

\begin{frontmatter}

\date{}

\title{On the use of circulant matrices for the stability analysis of recent weakly compressible SPH methods}

\author[aff1]{G.~Chaussonnet}
\ead{geoffroy.chaussonnet@kit.edu}
\author[aff1]{R.~Koch}
\author[aff1]{H.-J.~Bauer}
\address[aff1]{Karlsruher Institut f{\"u}r Technolgie - Institut f{\"u}r Thermische Str{\"o}mungsmaschinen, Karlsruhe, Germany}

\begin{abstract}
In this study, a linear stability analysis is performed for different Weakly Compressible Smooth Particle Hydrodynamics (WCSPH) methods on a 1D periodic domain
\improve{describing an incompressible base flow}.
The perturbation equation can be vectorized and written as an ordinary differential equation where the coefficients are circulant matrices.
The diagonalization of the system is equivalent to apply a spatial discrete Fourier transform. This leads to stability conditions expressed by the discrete Fourier transform of the first and second derivatives of the kernel. Although spurious modes are highlighted, no tensile nor pairing instabilities are found in the present study,
\improve{suggesting that the perturbations of the stresses are always damped if the base flow is incompressible.}
The perturbations equation is solved in the Laplace domain, allowing to derive an analytical solution of the transient state.
Also, it is demonstrated analytically that a positive background pressure combined with the uncorrected gradient operator leads to a reordering of the particle lattice. It is also shown that above a critical value, the background pressure leads to instabilities.
Finally, the dispersion curves for inviscid and viscous flows are plotted for different WCSPH methods and compared to the continuum solution.
\improve{It is observed that a background pressure equal to $\rho c^2$ gives the best fidelity to predict the propagation of a sound wave. When viscosity effects are taken into account, the damping of pressure fluctuations show the best agreement with the continuum for $p_{back} \sim \rho c^2/2$.}
\end{abstract}


\end{frontmatter}

\section{Introduction}

The Smoothed Particles Hydrodynamics (SPH) method was initiated by \citet{gingold1977smoothed} and \citet{lucy77} to study star formation. It is a second-order \citep{monaghan92} Lagrangian method resolving and storing physical quantities (mass, momentum, energy) on spatial discretization points, so-called particles, which move with the local fluid velocity. 
The interaction between several particles is taken into account with a weighting function which promotes the influence of closer particles. This function is called the \emph{kernel}.
SPH was later extended to fluid mechanics \citep{monaghan1994simulating}, based on a weakly compressible approach that is commonly used nowadays.
In the context of multiphase flow simulation, the main advantage of SPH over traditional grid-based methods is the intrinsic capturing of the phase interface by the natural rearrangement of the particles. In contrast to Eulerian methods, no interface capturing algorithms nor local mesh refinement is required.
Hence, the SPH method is broadly used for simulating free surface flow configurations where the liquid motion is driven by means of gravity, inertia or pressure \citep{liu10}. However, despite its advantages, SPH suffers from peculiar weaknesses.
First, mostly in shear driven flows, an instability can lead to voids in the particle lattice, which in turns, lead to numerical divergence. 
Second, due to the motion of the particles with the fluid velocity, a disorder of the particle arrangement may be encountered, that continuously changes during the simulation and compromises the accuracy of the spatial operators. Therefore, a correction algorithm has to be applied at every time step, which significantly increases the computational costs. On the other hand, it was observed by \citet{Colagrossi20121641} that the inaccuracy induced by the particle disorder counteracts the onset of the instability mentioned above. This phenomenon will be demonstrated analytically in the present study.\\
Several authors investigated the stability of the SPH method.
\citet{fulk1994numerical} extensively studied the stability in a \majrev{discrete} 1D configuration on a infinite domain for different sets of SPH equations. He emphasized the onset of the instability at the highest wavenumber $\pi/\Delta x$ ($\Delta x$ being the inter-particle distance), when the summation of the second derivative of the kernel ($W''$) over the neighbors and the local pressure ($P$) have different signs. The author proposed many strategies to overcome this instability, such as using (i) concave up/down kernels, (ii) a background pressure or (iii) a particle motion correction. It is worth to note that the two latter tweaks are still among the most popular.
At the same time, \citet{swegle1995smoothed} conducted a von-Neumann linear stability analysis (LSA) of viscous \majrev{discrete} 1D SPH with an equation of state (EoS) similar to the Tait's EoS. They derived their analysis by supposing only two neighbors, and they found the same condition as \citet{fulk1994numerical} for the onset of the instability at highest wavenumber. They finally extended the results to an arbitrary number of neighbors and found the same condition on $\sum W''$ to determine the stability.
\citet{balsara1995neumann} conducted a von-Neumann LSA of 1D compressible SPH where different time integration schemes and different inter-particle spacings were investigated. The author derived the dispersion relation for medium to large wavelength, contrary to \citet{fulk1994numerical} and \citet{swegle1995smoothed} who focused on the largest wavenumber.
 The author advised to keep the ratio of smoothing length to inter-particle distance between 1.0 and 1.4 to minimize unrealistic acoustic wave dispersion. However, no Taylor series expansion of the kernel was taken into account, and no tensile instability was found.
\citet{morris1996study} applied LSA to \majrev{discrete in-space} 1D smooth particles magneto-hydrodynamics (SPMHD) on an infinite domain. The author considered a weakly compressible approach and both a viscous and inviscid flow. For the first time, a non-zero background pressure was taken into account, and different approximations of the pressure gradient were investigated. The author found that the same instability as \citet{fulk1994numerical} for the highest wavenumber, but he attributed its origin to a negative background pressure and not depending on the product $(\sum W'') \times P$. Hence, he implied that the sum of $W''$ has always the same sign. He found this instability independent of the viscosity.
In a similar study, \citet{rasio2000particle} derived the numerical sound speed depending on several parameters, among which are the number of neighbors and the first and second derivative of the kernel.
\citet{belytschko2000unified} proposed a unified framework to study the stability of meshless particle methods including SPH. They found new conditions of instability based on the relation between the wavenumber and the number of neighbors, and attributed it to the tensile instability found by \citet{swegle1995smoothed}. Like \citet{morris1996study}, they highlighted a spurious mode at $\pi/\Delta x$ that they assigned to a rank deficiency of the stiffness matrix in the linearized equations.
\citet{borve2004two} applied a LSA to 2D SPMHD without background pressure in the EoS. They found too that the ratio of smoothing length to inter-particle distance should be between 1.0 and 1.4.
More recently, \citet{dehnen2012improving} conducted a LSA in 3D of WCSPH and showed the superiority of the Wendland kernel. They highlighted the difference between tensile and pairing instability and explored the link between negative values of the kernel Fourier transform and the pairing instability. Due to their multidimensional analysis, they were able to separate the longitudinal and transverse modes. Furthermore, they stressed the importance of the density estimator in SPH method, especially with regards to the pairing instability.
\majrev{\citet{violeau2014maximum} performed a von-Neumann LSA on multidimensional viscous weakly-compressible SPH, with a continuous spatial scheme. Unlike the authors previously mentioned, they considered a discretized time variable, which highlighted the importance of the time integration scheme on the stability.}
\\\\
As pointed out by \citet{dehnen2012improving}, the estimation of the density is an important issue that influences the stability of SPH method. Classically, there are two different methods to estimate the density of the fluid in WCSPH. First, the density is algebraically determined by the ratio of the particle mass and its effective volume, based on the position of its neighbors \citep{monaghan05}. This method is referred to as \emph{sum-SPH} subsequently.
Second, the density is estimated from the SPH formulation of the continuity equation \citep{monaghan05}. This method will be referred to as \emph{div-SPH}.
Recently, a method called $\delta$-SPH was presented by \citet{antuono2010free}. It was successfully applied in the field of free surface flows \citep{marrone2013accurate}. It is based on div-SPH with additional dissipation terms in each of the Navier-Stokes equations, which causes the modification of the estimation of density.
Finally, a method based on an artificial transport velocity formulation was developed by \citet{adami2013transport}. It is referred to as \emph{TV-SPH} in the following. Since this method is based on sum-SPH with artificially convected particles, its density estimator is different from sum-SPH.\\
The objectives of the present study are
\improve{(i) to present a new method for the linear stability analysis of homogeneous media with the use of circulant matrices, and (ii) to apply this method}
to investigate the stability and the dispersion curves of the four methods for estimating the density. \majrev{Note that the time variable will be considered continuous here, and the investigation of the integration scheme will be done in a subsequent study.} \\
\improve{The traditional approach to study the stability and the dispersion of a numerical scheme is to consider perturbations as plane progressive waves (\eg $\delta \phi =\epsilon \, \text{e}^{i(\omega t - k x)}$). The perturbations are injected into the constitutive equations and the dispersion relation is obtained by the equation $f(\omega,k)=0$. This method is referred to as the normal mode analysis.
The new method presented here makes no assumption on the shape of the perturbations. Hence, it allows non-oscillating transient solutions for the perturbations. Furthermore, as presented in the course of this paper, it is possible to derive the \majrev{analytical} temporal solution for each particle.
The method is applied in 1D for the sake of simplicity, and still allows some findings. There is no theoretical limitations to apply this method in greater number of dimensions.
}\\
\improve{To give a synthetic view of the method, the steps are described as follows.}
The equations of perturbations are cast into a matrix form of an ordinary differential equation in time. Its coefficients are circulant matrices, whose eigenvalues and eigenvectors are easily obtained using the circulant matrix theory. The matrix form of the perturbations equations is then diagonalized, which corresponds to \improve{express the normal modes of the system and allow to decouple the motion of each particle}. These equations are solved in the Laplace domain, which features the advantage over the Fourier transform to capture the transient state, in addition to the stability and the dispersion relation.
Then, after simplification of the transfer function, the solution of the perturbation equation is transformed back into the temporal geometrical form. 
\improve{This temporal form is later used to validate the whole derivation against the results of the numerical resolution of the starting equation.}\\\\
The plan of this paper is as follows. In Section~\ref{sec_WCSPH}, the method described above is applied to the classical sum-SPH approach, and every steps are explained thoroughly.
Then the same method is applied to the div-SPH, $\delta$-SPH and TV-SPH density estimator in Sections~\ref{sec_divSPH}, \ref{sec_deltaSPH} and \ref{sec_TVSPH}, respectively. The dispersion curve of all methods, both for inviscid and viscous flows are discussed in Section~\ref{sec_dispersion}. An illustration of the spurious mode is presented in Section~\ref{sec_illus_spurious}.
\improve{The expression of the perturbations equation with circulant matrices is illustrated in~\ref{appendix_circulant_matrix}, and the validation of the derivation is provided in~\ref{appendix_validation} where the analytical temporal solutions of the perturbations are compared to the resolution by a numerical scheme.}

\section{Application of LSA on sum-SPH \label{sec_WCSPH} density estimator}

\subsection{Preliminary definitions}

\begin{figure}[!htb]
	\centering
	\def \svgwidth {0.45\textwidth}
	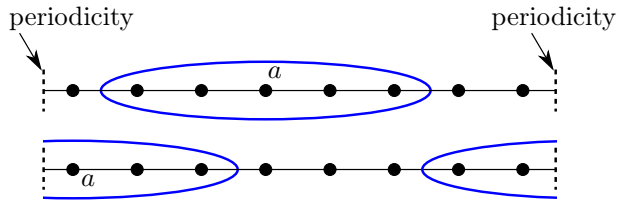
	\caption{Top: 1D domain. Bottom: illustration of split sphere of influence}
	\label{fig_1D_domain_01}
\end{figure}

The 1D domain consists of $N$ particles equidistantly placed along a segment \improve{of length $L$}, with an inter-particle distance of $\Delta x$. This is representative of the equilibrium state (Fig.~\ref{fig_1D_domain_01}).
In order to avoid any boundary effects, the domain is periodic. 
The number of neighbors on one side of the particle is labelled $M$. Hence, a complete sphere of influence consists of $2 M + 1$ particles.
The smoothing length $h$ is chosen so that its multiple exactly covers the neighbors included in the sphere of influence as in \citep{quinlan2006truncation}. Therefore, $h$, $\Delta x$ and $M$ are related by:
\begin{equation}
k h = (2 M + 1) \, \Delta x = D
\label{eq_euler1}
\end{equation}
where $k$ is an integer and $D = k h$ is the diameter of the sphere of influence.
Due to the periodicity, the sphere of influence of particles located near the boundaries may be split
into two parts located at either boundary of the domain,
as illustrated on Fig.~\ref{fig_1D_domain_01} (\textit{bottom}).
\\
In this study, two different kernels were investigated. The so-called quintic kernel \citep{morris97} is defined in 1D by:
\begin{equation}
W(r,h) = \frac{1}{120 h}
\begin{cases}
	(3-r)^5 - 6 \, (2-r)^5 + 15 \, (1-r)^5
	& \quad \text{for }  0 \le r \le 1 \\
	(3-r)^5 - 6 \, (2-r)^5
	& \quad \text{for }  1 < r \le 2 \\
	(3-r)^5
	& \quad \text{for }  2 < r \le 3
\end{cases}
\end{equation}
where $r=| x_a - x_b | / h$ is the normalized distance between particle $a$ and $b$. The second investigated kernel belongs to the family of Wendland kernels \citep{wendland95}. It is expressed in 1D as:
\begin{equation}
\label{eq_kernel_wendland}
W(r,h) = \frac{1}{2 h} \, (1-r)^5 \, (1 + 5r + 8 r^2)
\end{equation}
\majrev{which is a seventh-order polynomial. 
Note that even though the fifth-order polynomial $(1-r)^4 \, (1+4 \,r)$ is often applied in the SPH community, its use here is inappropriate because its degree depends on the integer part of $d/2$ where $d$ is the number of dimension, leading to a $C^2$-function of degree 4 in 1D \citep{wendland95}.
As we want to investigate kernels of at least degree 5, we dismiss the aforementioned polynomials.
In the following, the kernel defined by Eq.~\ref{eq_kernel_wendland} is simply referred to as the \emph{Wendland kernel}.}
In this case, the particle distance is normalized by the radius of the sphere of influence $r=| x_a - x_b | / 3h$.
\\\\
In the rest of this paper, the dependency of the kernel on $h$ will be dropped in the notation: $W(r) \equiv W(r,h)$ for the sake of simplicity. The radius of the sphere of influence is set constant to $3 \, h$, which leads to discrete values of $\Delta x$ (Eq.~\ref{eq_euler1}):
\begin{equation}
\Delta x = \frac{6 \, h}{2 M + 1} = \frac{D}{2 M + 1}
\label{eq_non_dimensional_delta_x}
\end{equation}
\majrev{
Furthermore, \citet{dehnen2012improving} showed that a more universal measure of the kernel characteristic length was given by its standard deviation $\tilde{h}$:
\begin{equation}
\tilde{h} ^2 = \int{ r^2 \, W(r) \, \mathrm dr}
\label{eq_std_kernel_h}
\end{equation}
This length is equal to $h/\sqrt{2}\approx 0.707 \, h$ and $h\sqrt{3/5}\approx 0.775 \, h$ for the quintic and the Wendland kernel, respectively. It will be used later as the reference length scale to non-dimensionalize the equations, except with $\Delta x$ when it is used to quantify the number of neighbors (Eq.~\ref{eq_non_dimensional_delta_x}) or the kernel compacity.
In this case, the diameter of the sphere of influence $D$ will be used.
}\\
Finally, the letter $a$ refers to the particle under consideration whereas $b$ refers to its neighbors. Hence $\sum_b$ is a summation over the neighbors of particle $a$. The difference between a quantity at particle $a$ and at particle $b$ is indexed $ab$. For instance, $x_a-x_b$ is abbreviated by $x_{ab}$.
For the kernel, $W(x_a-x_b)$ is abbreviated by $W_{ab}$.
\improve{Tensor of $n$\textsuperscript{th} order are $n$-time underlined, \eg \uu{V} and \uu{\uu{M}} denote a vector and a matrix, respectively.}

\subsection{System of equations}

In the absence of gravity, the 1D WCSPH approach with the algebraic density formulation (sum-SPH) is defined by the following system of equation:
\begin{subequations}
\label{eq_WCSPH01}
\begin{align}[left = \empheqlbrace\,]
\rho_a = & \ m_a \sum\limits_b {W_{ab}}  \\
\dsecdv {x_a} t = & \  - \frac{1}{\rho_a} \sum\limits_{b} V_b (p_b + p_a) W'_{ab} + 2 \nu  \, \sum\limits_b {V_b \frac{u_a - u_b}{r_{ab}} \, W'_{ab}} \label{eq_WCSPH01b} \\
p_a = & \  \frac{\rho_0 \, c^2 } {\gamma} \left[ \left( \frac{\rho_a}{\rho_0} \right)^{\gamma} - 1 \right]  + p_{back} \label{eq_state_WCSPH}
\end{align}
\end{subequations}
In the momentum conservation (Eq.~\ref{eq_WCSPH01b}), the term $W'_{ab}$ is the derivative $\partial W_{ab}/ \partial x_a$ which corresponds to the kernel gradient $\nabla W_{ab}$ in 1D. The pressure gradient is expressed by the so-called $G_+$ operator where the pressure $p_a$ and $p_b$ are added, ensuring conservation of linear momentum.
The second term on the RHS of Eq.~\ref{eq_WCSPH01b} represents the viscous effects. It exhibits the kinematic viscosity $\nu$ multiplied by the Laplacian of the velocity in the SPH formalism \citep{cleary99}. 
Another frequent expression of the Laplacian was proposed by \citet{morris97}, but in 1D, its expression reduces to the same as the one of \citet{cleary99}.
Equation~\ref{eq_state_WCSPH} is the Tait equation of state, valid for weakly compressible flows \citep{batchelor00}. The term $\gamma$ is the polytropic ratio, usually set to 7.

\subsection{Linearization}

\begin{figure}[!htb]
	\centering
	\def \svgwidth {0.45\textwidth}
	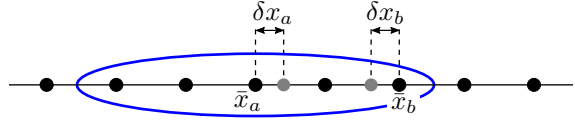
	\caption{Linearization of the kernel}
	\label{fig_1D_domain_02}
\end{figure}

The linearization of the physical variables (density, position,
\improve{velocity}
and pressure) is written:
\begin{equation}
\label{eq_physical_linear}
\rho = \ol{\rho} +  \delta \rho
\text{,} \quad 
x = \ol{x}  +  \delta x
\text{,} \quad 
\improve{u =} \improve{\ol{u}  +  \delta u}
\text{,} \quad 
p = \ol{p} +  \delta p 
\end{equation}
where overlined quantities are constant and $\delta X$ is a small perturbation of the quantity $X$.
\improve{In the rest of the paper, the mean velocity $\ol{u}$ is set to a constant, so that the base flow is incompressible.}
The particle volume $V$ is defined as $V = m / \rho$ and linearized to the first order using Eq.~\ref{eq_physical_linear}:
\begin{equation}
	V = {m}/{\rho} \approx  \ol{V} \left(1 - \frac{\delta \rho}{\ol{\rho}} \right) 
	\label{eq_volume_linear}
\end{equation}
The value of the kernel is linearized by a Taylor expansion of first order as depicted in Fig.~\ref{fig_1D_domain_02}:
\begin{subequations}
\label{eq_kernel_linear_heavy}
\begin{align}%
W(x_a - x_b) \approx  \ & W(\ol{x_a} - \ol{x_b}) + (\delta x_a - \delta x_b) W'(\ol{x_a} - \ol{x_b}) \\
W'(x_a - x_b) \approx \ &  W'(\ol{x_a} - \ol{x_b}) + (\delta x_a - \delta x_b) W''(\ol{x_a} - \ol{x_b})
\end{align}
\end{subequations}
\\
For the sake of clarity in the following, Eqs.~\ref{eq_kernel_linear_heavy} are written as:
\begin{equation}
W_{ab} \approx W_{\overline{ab}} + \delta x_{ab} W'_{\overline{ab}}
\quad \text{and} \quad
W'_{ab} \approx W'_{\overline{ab}} + \delta x_{ab} W''_{\overline{ab}}
\label{eq_kernel_linear}
\end{equation}
\noindent Inserting Eqs.~\ref{eq_physical_linear}, \ref{eq_volume_linear} and \ref{eq_kernel_linear} into the Eqs.~\ref{eq_WCSPH01}, the linearized equations of perturbations around an equilibrium state is expressed as:
\begin{subequations}
\label{eq_euler_perturbation_01_for_dummies}
\begin{align}[left = \empheqlbrace\,]
\delta \rho_a = & \  m \,  \sum\limits_b { (\delta x_a - \delta x_b) \,   W'_{\overline{ab}}}  \label{eq_sumSPH_mass_for_dummies} \\
\dsecdv {\delta x_a} t = & \   \ol{V} \, \sum\limits_{b}
\left[2 \,  \frac{\ol{p}}{\ol{\rho}^2} \, (\delta \rho_a + \delta \rho_b) \, W'_{\overline{ab}}
- 2 \, \frac{\ol{p}}{\ol{\rho}}\, \delta x_{ab}  \, W''_{\overline{ab}} 
- \frac{\delta p_a + \delta p_b}{\ol{\rho}} \, W'_{\overline{ab}}  
+ 2 \nu \ddv {}{t} (\delta x_{ab})  \frac{\, W'_{\overline{ab}}}{x_{ab}} \right]  \label{eq_sumSPH_momentum_for_dummies}  \\
\delta p_a = & \ c^2 \, \delta \rho_a  \label{eq_sumSPH_EoS_for_dummies}
\end{align}
\end{subequations}
\majrev{
By combining Eqs.~\ref{eq_sumSPH_momentum_for_dummies} and \ref{eq_sumSPH_EoS_for_dummies}, one can express the acceleration of the perturbed particles as:
\begin{equation}
\label{eq_euler_perturbation_02_for_dummies}
\dsecdv {\delta x_a} t =  -m \, \sum\limits_{b}
\frac{\delta \rho_a +\delta \rho_b}{\ol{\rho}^2} \left[  c^2 -  \frac{2\ol{p}}{\ol{\rho}} \right] \, W'_{\overline{ab}}
+ \frac{2\ol{p}}{\ol{\rho}^2} \, \delta x_{ab}  \, W''_{\overline{ab}} 
- 2 \frac{\nu}{\ol{\rho}} \ddv {}{t} (\delta x_{ab})  \frac{\, W'_{\overline{ab}}}{x_{ab}}
\end{equation}
This equation is well-known in the literature, and was already derived, for instance by \citet{monaghan05}. However, we show in the following that Eqs.~\ref{eq_euler_perturbation_01_for_dummies} and \ref{eq_euler_perturbation_02_for_dummies} can be further simplified. 
Indeed, on an ordered particle lattice, the pairwise summation of the kernel gradient for two particles $b$ equidistant from particle $a$, is equal to zero. This is because the kernel gradient is an odd function. Therefore, each time that the kernel gradient is the only variable depending on $b$ under the summation, it is can be simplified. For instance, Eq.~\ref{eq_sumSPH_mass_for_dummies} can be decomposed as:
\begin{equation}
\delta \rho_a = m \sum_b (\delta x_a - \delta x_b) W'_{\overline{ab}}
= m \, \delta x_a \, \underbrace{\sum_b W'_{\overline{ab}}}_{\substack{=0 \\ \text{on an} \\ \text{ordered lattice}}} - m \sum_b \delta x_b W'_{\overline{ab}}
\end{equation}
So that:
\begin{equation}
\delta \rho_a = - m \sum_b \delta x_b W'_{\overline{ab}}
\end{equation}
Therefore, all perturbations at particle $a$ vanish when they are multiplied by the kernel gradient, and Eqs.~\ref{eq_euler_perturbation_01_for_dummies} can be simplified to:
\begin{subequations}
\label{eq_euler_perturbation_01}
\begin{align}[left = \empheqlbrace\,]
\delta \rho_a = & \   - m \,  \sum\limits_b { \delta x_b \,   W'_{\overline{ab}}}  \label{eq_sumSPH_mass} \\
\dsecdv {\delta x_a} t = & \   \ol{V} \, \sum\limits_{b}
\left[2 \,  \frac{\ol{p}}{\ol{\rho}^2} \, \delta \rho_b \, W'_{\overline{ab}}
- 2 \, \frac{\ol{p}}{\ol{\rho}}\, \delta x_{ab}  \, W''_{\overline{ab}} 
- \frac{\delta p_b}{\ol{\rho}} \, W'_{\overline{ab}}  
+ 2 \nu \ddv {}{t} (\delta x_{ab})  \frac{\, W'_{\overline{ab}}}{x_{ab}} \right]  \label{eq_sumSPH_momentum}  \\
\delta p_a = & \ c^2 \, \delta \rho_a 
\end{align}
\end{subequations}
}
The motion of particle $a$ (Eq.~\ref{eq_sumSPH_momentum}) depends on the perturbations of its neighbors via four different terms on the right-hand side.
The first term depends on the neighbor density due to the mass conservation (Eq.~\ref{eq_volume_linear}) and represents the cross effects of compressibility and constant pressure.
The second term depends on the particle distance from their equilibrium position and is proportional to the equilibrium pressure. It has no physical meanings and stems from the inaccuracy of the SPH gradient operator on disordered particle lattices.
The third term results from the pressure perturbations and the last term comprises the viscous effects.

\subsection{Converting the perturbation equation into a circulant matrix system}

In this section the transformation of the perturbation equations (Eqs.~\ref{eq_euler_perturbation_01}) to a matrix system is explained.
All perturbed variables $\delta x$, $\delta \rho$, $\delta p$ of Eq.~\ref{eq_euler_perturbation_01} are written down for all particles \ie $\delta x_0$, $\delta x_1$, ..., $\delta \rho_0$, $\delta \rho_1$, ..., $\delta p_0$, $\delta p_1$, ... and then gathered into perturbation vectors:
\begin{equation}
\begin{cases}
	\uu{X} = (\delta x_0, \delta x_1, \delta x_2, ... , \delta x_{N-1})^T \\
	\uu{R} = (\delta \rho_0, \delta \rho_1, \delta \rho_2, ... , \delta \rho_{N-1})^T \\
	\uu{P} = (\delta p_0, \delta p_1, \delta p_2, ... , \delta p_{N-1})^T
\end{cases}
\end{equation}
As illustrated in \ref{appendix_circulant_matrix}, Eqs.~\ref{eq_euler_perturbation_01} can be expressed as a matrix system:
\begin{subequations}
\label{eq_euler_perturbation_02}
\begin{align}[left = \empheqlbrace\,]
\uu{R} = & \   m \,  \uu{\uu{A}}' \, \uu{X}  \\
\dsecdv {} t ({\uu{X}}) = & \  -2 \frac{\ol{p} \, \ol{V}}{\ol{\rho}^2} \, \uu{\uu{A}}' \,  \uu{R}
+ 2 \, \frac{\ol{p} \, \ol{V}}{\ol{\rho}} \, \uu{\uu{A}}'' \,  \uu{X}
+ \frac{\ol{V}}{\ol{\rho} } \, \uu{\uu{A}}' \,  \uu{P}
+ 2 \ol{V} \, \nu \, \uu{\uu{A}}''' \ddv{}{t}  {\uu{X}} 
 \label{momentum_linearizerd_WCSPH} \\
\uu{P} = & \ c^2 \, \uu{R}
\end{align}
\end{subequations}
where the matrices $\uu{\uu{A}}'$, $\uu{\uu{A}}''$ and $\uu{\uu{A}}'''$ enjoy interesting properties. First, $\uu{\uu{A}}'$ is antisymmetric while $\uu{\uu{A}}''$ and $\uu{\uu{A}}'''$ are symmetric. Second, and most important for the rest of the study, all three matrices are \emph{circulant matrices} \citep{SWB-011280778}. Therefore, they can be expressed by their first row vector, $\uu{C}'$, $\uu{C}''$ and $\uu{C}'''$ respectively:
\begin{equation}
\uu{C}' = 
\begin{cases}
	c_j' = -c_{n-j}' = W'(j \Delta x) \quad \text{ for } \quad j \in [1,M] \\
	c_j' = 0 \quad \text{otherwise}\\
\end{cases}
\end{equation}

\begin{equation}
\uu{C}'' = 
\begin{cases}
	c_0'' =
 - 2 \, \sum\limits_{j=1}^{M} W''(j \Delta x) \\
	c_j'' = c_{n-j}'' = W''(j  \Delta x) \quad \text{ for } \quad j \in [1,M] \\
	c_j'' = 0 \quad \text{otherwise} \\
\end{cases}
\end{equation}
\begin{equation}
\label{eq_C_third_expression}
\uu{C}''' = 
\begin{cases}
	c_0''' = 
2 \, \sum\limits_{j=1}^{M} {W'(j  \Delta x)} / {(j \Delta x)} \\
	c_j''' = c_{n-j}''' = - W'(j \Delta x) / {(j \Delta x)}  \quad \text{ for } \quad j \in [1,M] \\
	c_j''' = 0 \quad \text{otherwise} \\
\end{cases}
\end{equation}
Note that in the expressions of $\uu{C}'$, $\uu{C}''$ and $\uu{C}'''$, the properties $W'(-j \Delta x) = - W'(j \Delta x)$ and $ W''(-j \Delta x) =  W''(j \Delta x)$ were used. 
In Eq.~\ref{momentum_linearizerd_WCSPH}, the pressure and density terms are multiplied by $\uu{\uu{A}}'$ while the viscous term is multiplied by $\uu{\uu{A}}'''$. Therefore, this approach can be regarded as an expression of the linearized SPH operators in a matrix form. The linearized gradient is expressed by $\uu{\uu{A}}'$, and the linearized Laplacian is expressed by $\uu{\uu{A}}'''$.
The matrix $\uu{\uu{A}}''$ comes from the linearization of the gradient of a constant term. It should be zero if the SPH gradient operator was first order consistent.
\majrev{Note that these matrices are similar to other terms in the literature. For instance, $\uu{\uu{A}}'^2$ and $\uu{\uu{A}}'''$ are similar to $F_1$ and $F_2$ by \citet{violeau2014maximum}, and to $G_1$ and $G_2$ (Section 4.1.2) by \citet{fulk1994numerical}.}
\\\\
Expressing the pressure fluctuation with the help of the EoS (Eq.~\ref{eq_state_WCSPH}) leads to $\ol{p} = p_{back}$. Therefore, the constant part of the pressure is the background pressure. Note that originally, the Tait EoS does not include any background pressure term.
The system~\ref{eq_euler_perturbation_02} is further simplified with $\ol{V} =  \Delta x$ and $m = \ol{\rho} \, \Delta x$. Introducing the non-dimensional background pressure as $p_{bg} =  {p_{back}} / {\ol{\rho} \, c^2}$ and substituting the perturbed pressure and density into Eq.~\ref{momentum_linearizerd_WCSPH} finally leads to:
\begin{equation}
\dsecdv {} t ({\uu{X}}) = \Delta x \, c^2 \left[ \Delta x \,\uu{\uu{A}}'^2 (1 - 2 p_{bg})
+  2  p_{bg}   \, \uu{\uu{A}}'' 
 \right] \uu{X}
 + 2 \, \nu \,  \Delta x \, \uu{\uu{A}}'''  \ddv {} {{t}} ({\uu{X}})  
\label{eq_euler_perturb_matrix2}
\end{equation}
Equation~\ref{eq_euler_perturb_matrix2} can be written in a non-dimensional form using:
\begin{equation}
\begin{cases}
\uu{X}^* = {\uu{X}}/{\tilde{h}}
\text{ ,}\quad
t^* = {ct}/{\tilde{h}}
\text{ ,}\quad
\Delta x^* = {\Delta x}/{\tilde{h}} 
\text{ ,}\quad   \\
\uu{\uu{A}}'^{*} = \tilde{h}^2 \, \uu{\uu{A}}'
\text{ ,} \quad
\uu{\uu{A}}''^* = \tilde{h}^3 \, \uu{\uu{A}}'' 
\text{ ,} \quad
\uu{\uu{A}}'''^* = \tilde{h}^3 \, \uu{\uu{A}}''' 
\text{ ,} \quad
\nu^* = \nu/(\tilde{h} c) 
\end{cases}
\end{equation}
It leads to:
\begin{equation}
\frac{1}{\Delta x} \dsecdv {} {{t}} ({\uu{X}})
= \left[ \Delta x \,\uu{\uu{A}}'^{2} (1 - 2 p_{bg})
+  2 p_{bg} \, \uu{\uu{A}}''
\right] \, \uu{X}
+ 2 \, \nu \,  \uu{\uu{A}}'''  \ddv {} {{t}} ({\uu{X}})  
\label{eq_NS_perturb_matrix1}
\end{equation}
where the variables are written in their non-dimensional form without the asterisk ($^*$) for the sake of simplicity. In the following, all variables are written in the non-dimensional form unless it is explicitly mentioned.
\majrev{When it is used to illustrate the number of neighbors, the non-dimensional inter particle distance will be written $\tilde{\Delta} x = \Delta x / D$.}
Equation~\ref{eq_NS_perturb_matrix1} is a matrix ordinary differential equation where the unknown is the vector ${\uu{X}}$.

\subsection{Properties of matrices $\uu{\uu{A}}'$, $\uu{\uu{A}}''$ and $\uu{\uu{A}}'''$}

\begin{figure}[h]
	\centering
	\includegraphics[width=0.49\textwidth,keepaspectratio]{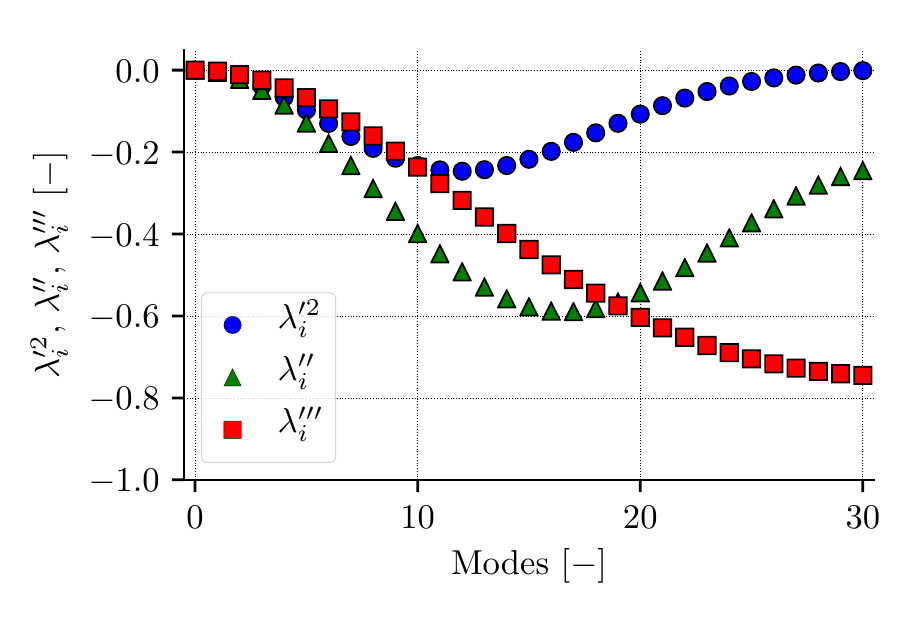}
	\includegraphics[width=0.49\textwidth,keepaspectratio]{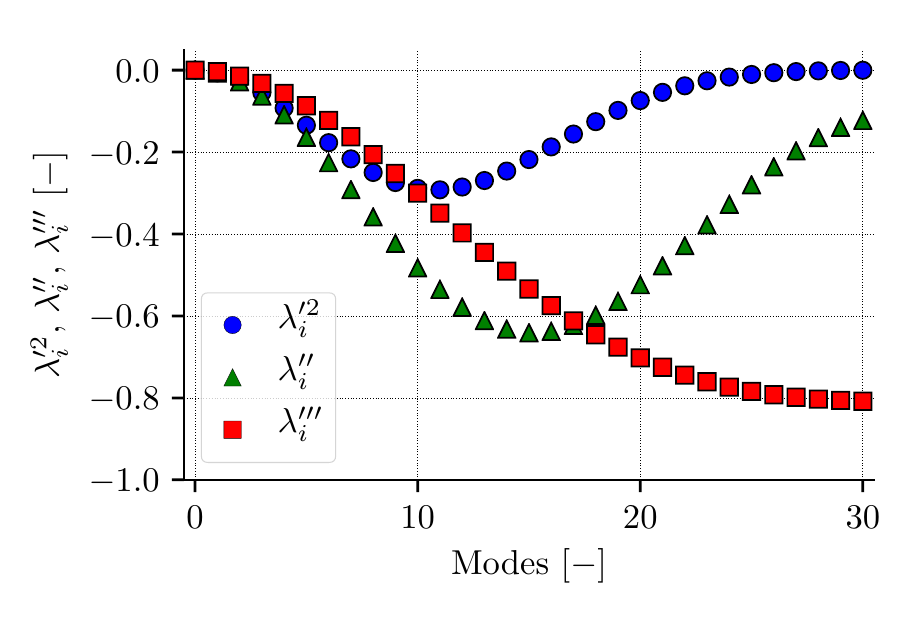}
	\caption{Values of $(\lambda_i'^2,\lambda_i'',,\lambda_i''')$ for $N$=63 and $\tilde{\Delta} x$=0.14. Left: quintic kernel. Right: Wendland kernel.}
	\label{fig_modes_lambdas_illus}
\end{figure}

Since matrices $\uu{\uu{A}}'$, $\uu{\uu{A}}''$ and $\uu{\uu{A}}''$ are circulant matrices, they are diagonalizable in the same basis \citep{SWB-011280778}. Their eigenvalues are labeled $\lambda_j'$, $\lambda_j''$ and $\lambda_j'''$, respectively:
\begin{equation}
\lambda_j' = 2 \, \hat{i} \, \sum\limits_{k=1}^{M} c_k' \, \sin \left(  \frac{2 \pi}{N} k j  \right)
\label{eq_def_lambda_prime}
\end{equation}
\begin{equation}
\lambda_j'' = c_0'' + 2 \sum\limits_{k=1}^{M} c_k'' \, \cos \left(  \frac{2 \pi}{N} k j  \right)
\label{eq_def_lambda_second}
\end{equation}
\begin{equation}
\lambda_j''' = c_0''' + 2 \sum\limits_{k=1}^{M} c_k''' \, \cos \left(  \frac{2 \pi}{N} k j  \right)
\label{eq_def_lambda_third}
\end{equation}
where $\hat{i}$ is the imaginary number. These eigenvalues depend only on (i) the kernel, (ii) the total number of particles $N$, and (iii) the number of neighbours $M$.
The matrix $\uu{\uu{A}}'^2$ is also a circulant matrix.
Since the eigenvalues of $\uu{\uu{A}}'$ are pure imaginary, the eigenvalues of $\uu{\uu{A}}'^2$ are negative.
In addition, eigenvalues are symmetric in the sense:
\begin{equation}
\lambda_j'^2 = \lambda_{N-j}'^2
\text{ ,} \quad 
\lambda_j'' = \lambda_{N-j}''
\text{ ,} \quad
\lambda_j''' = \lambda_{N-j}'''
\quad \text{for}  \quad j \in [1,N_{mid}]
\label{eq_def_lambda_symetry}
\end{equation}
where 
\begin{equation}
N_{mid} = 
\begin{cases}
(N-1)/ 2 & \quad \text{for }N \text{ odd} \\
N/2 & \quad \text{for }N \text{ even} \\
\end{cases}
\end{equation}
Furthermore, the eigenvalues exhibit particular values for the mode $j$=0 with $\lambda_0'$=$\lambda_0''$=$\lambda_0'''$=0.
At the largest mode $j=N/2$, $\lambda_{N/2}'$ is equal to zero, meaning that matrix $\uu{\uu{A}}'$ has a rank deficiency that may lead to spurious modes \citep{belytschko2000unified}. As highlighted by \citet{fulk1994numerical}, the eigenvalue of $\uu{\uu{A}}''$ for the largest mode is different from zero, \ie $\lambda_{N/2}'' \ne 0$. Furthermore, it is found in the present study that the eigenvalue of $\uu{\uu{A}}'''$ at the largest mode is also non-zero, $\lambda_{N/2}''' \ne 0$.
Finally, the eigenvalues of a circulant matrix are equal to the coefficients of the discrete Fourier transform of the its first row \citep{SWB-011280778}. Hence, $\lambda_j'$, $\lambda_j''$ and $\lambda_j'''$ are the coefficients of the discrete Fourier transform of $W'(x)$, $W''(x)$ and $W'(x)/x$, respectively.
\majrev{\citet{violeau2018spectral} approximated $\lambda_i'''$ for a large number of particles whereas it is expressed exactly in the present study, independently of the number of particles.}
\\\\
Figure~\ref{fig_modes_lambdas_illus} shows an illustration of the eigenvalues, for N=63, M=3 (\ie $\tilde{\Delta} x$=0.14) and different kernels.
It is observed that all eigenvalues are always negative. At largest modes, $\lambda_i'^2$ goes to zero whereas the other values remain significantly negative.

\subsection{Projection in the modal space}
\begin{figure}[h]
	\centering
	\includegraphics[width=0.49\textwidth,keepaspectratio]{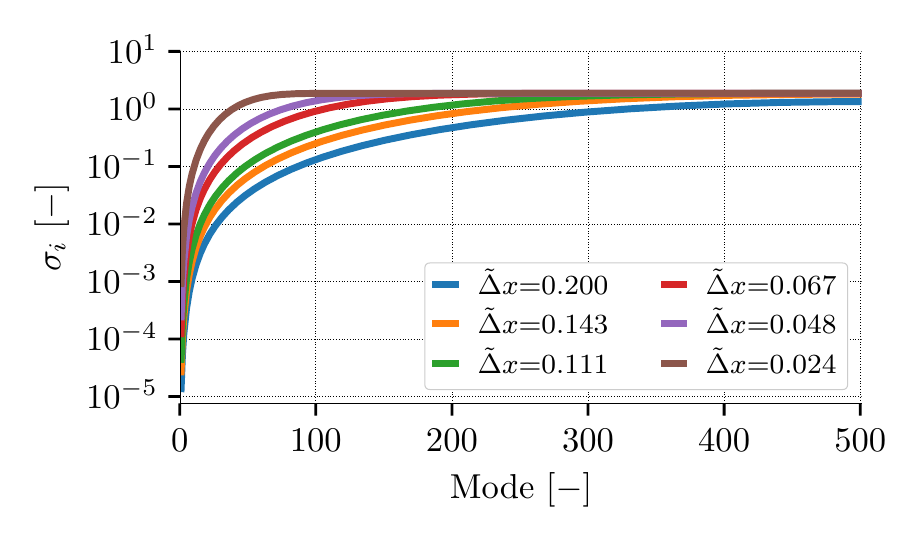}
	\caption{Values of $\sigma_i$ for $N$=1000, $\nu$=1, $p_{bg}=1$, quintic kernel and different $\tilde{\Delta} x$.}
	\label{fig_sigma_i}
\end{figure}

\begin{figure}[h]
	\centering
	\includegraphics[width=0.49\textwidth,keepaspectratio]{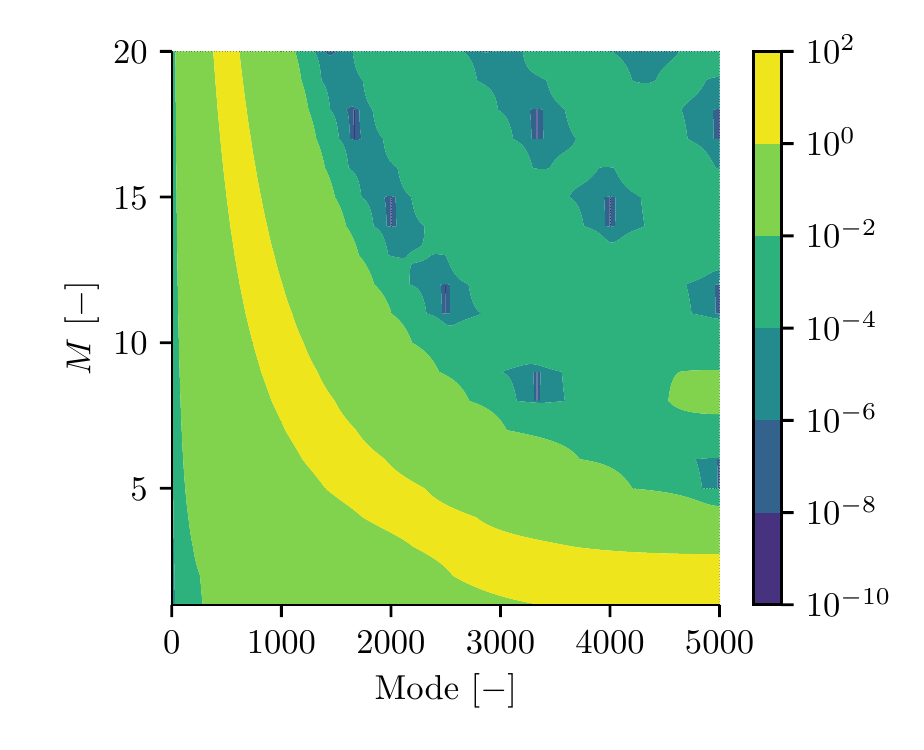}
	\includegraphics[width=0.49\textwidth,keepaspectratio]{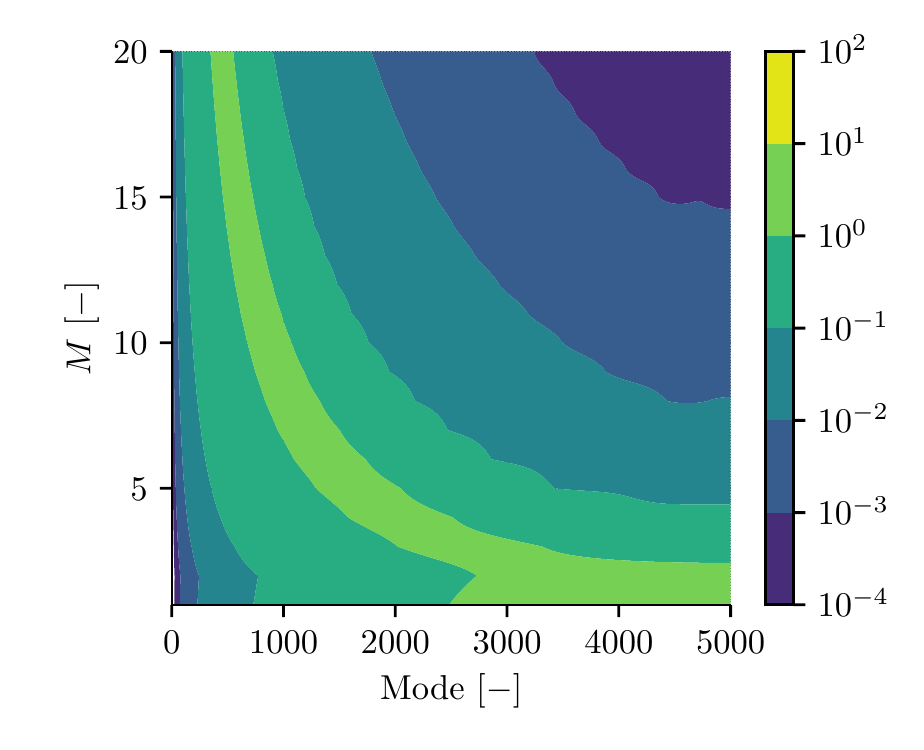}
	\caption{Color map of $\phi_i^2$ for $N$=10000, $p_{bg}=1$. Left: quintic kernel. Right: Wendland kernel.}
	\label{fig_phi_i2_map_M_vs_modes}
\end{figure}

Since $\uu{\uu{A}}'^2$, $\uu{\uu{A}}''$ and $\uu{\uu{A}}'''$ are circulant matrices, they can be diagonalized on the same basis. 
By expressing $\uu{\uu{A}}'^2$, $\uu{\uu{A}}''$  and $\uu{\uu{A}}'''$ as diagonal matrices:
\begin{equation}
\uu{\uu{A}}'^2  = \uu{\uu{P}}^{-1}  \, \uu{\uu{\Delta}}'^2  \, \uu{\uu{P}} 
\quad \text{and} \quad
\uu{\uu{A}}''  = \uu{\uu{P}}^{-1}  \, \uu{\uu{\Delta}}''  \, \uu{\uu{P}}
\quad \text{and} \quad
\uu{\uu{A}}'''  = \uu{\uu{P}}^{-1}  \, \uu{\uu{\Delta}}'''  \, \uu{\uu{P}}
\label{eq_diago_Asecond}
\end{equation}
and defining $\uu{Y} = \uu{\uu{P}} \, \uu{X}$, Eq.~\ref{eq_NS_perturb_matrix1} is expressed:
\begin{equation}
\frac{1}{\Delta x} \dsecdv {} t ({\uu{Y}}) = \left[\Delta x \, \uu{\uu{\Delta}}'^2
+ 2 p_{bg} \, (\uu{\uu{\Delta}}''   - \Delta x \, \uu{\uu{\Delta}}'^2) 
\right]\, \uu{Y} +  2 \, \nu \,  \uu{\uu{\Delta}}'''  \ddv {} {{t}} ({\uu{Y}})
\label{eq_euler_perturb_matrix4}
\end{equation}
As $\uu{\uu{\Delta}}'^2$, $\uu{\uu{\Delta}}''$ and $\uu{\uu{\Delta}}'''$ are diagonal matrices composed of the eigenvalues, Eq.~\ref{eq_euler_perturb_matrix4} can be reduced to $N_{mid}+1$ non-coupled ordinary differential equations:
\begin{equation}
\frac{1}{\Delta x} \dsecdv {} t ({y_i}) - 2 \, \nu \, \lambda_i'''  \ddv {} {{t}} (y_i)
- \left[\Delta x \, \lambda_i'^2
+ 2 p_{bg} (\lambda_i''  - \Delta x \, \lambda_i'^2) 
\right]\, y_i = 0
\label{eq_euler_perturb_matrix4b}
\end{equation}
with $i$ being the mode number and $i \in \llbracket 0,N_{mid} \rrbracket$.
Hence, Eq.~\ref{eq_euler_perturb_matrix4b} represents the 
perturbations expressed in the \emph{modal space}.
Furthermore, the change-of-basis rules from the matrix $\uu{\uu{A}}'$ to its eigenvalue is the classical definition of a discrete Fourrier transform \citep{SWB-011280778}. This means that expressing the equation of motion in the modal space is equivalent to apply a discrete spatial Fourier transform. Thus, studying the stability of SPH with $\lambda_j'$, $\lambda_j''$ and $\lambda_j'''$ is equivalent to study the values of the discrete Fourier transform of the different expressions involving the kernel derivatives.
This result is similar to those of (i) \citet{fulk1994numerical} and \citet{swegle1995smoothed} because it involves derivatives of the kernel, and similar to those of \citet{dehnen2012improving} because it involves the Fourier transform of the kernel, here in the discrete form. \improve{Equation~\ref{eq_euler_perturb_matrix4b} also shows the influence of the modes of $W'(x)/x$ represented by $ \lambda_i'''$.}
Equation~\ref{eq_euler_perturb_matrix4b} is formally rewritten as:
\begin{equation}
\dsecdv {} t ({y_i}) + \sigma_i  \ddv {} {{t}} (y_i) + \phi_i^2 \, y_i = 0
\label{eq_euler_perturb_matrix5}
\end{equation}
with
\begin{equation}
\sigma_i =  - 2 \, \nu \,  \Delta x \, \lambda_i'''
\quad \text{and} \quad
\phi_i^2 = 
- \Delta x^2 \, \lambda_i'^2
- 2 \Delta x \, p_{bg} (\lambda_i''  - \Delta x \, \lambda_i'^2) 
\label{eq_euler_perturb_matrix5b}
\end{equation}
In the following, the two terms appearing in the expression of $\phi_i^2$ are labeled $\psi_{1,i}$ and $\psi_{2,i}$, and they are given by:
\begin{equation}
\psi_{1,i}  = (\Delta x \, \lambda_i')^2  (1 - 2 p_{bg}) 
\quad \text{and} \quad
\psi_{2,i} = 2 p_{bg} \Delta x \, \lambda_i'' 
\label{eq_euler_perturb_matrix5c}
\end{equation}
Note that these terms resemble to those labeled $D_1$ and $D_2$ (Section 4.1.2) by \citet{fulk1994numerical}. However, $D_1$ and $D_2$ were defined with an infinite summation so they cannot be numerically determined. In the present study, $\psi_{1,i}$ and $\psi_{2,i}$ are exactly determined by a summation from 1 to $M$.\\
Equation~\ref{eq_euler_perturb_matrix5} is an ordinary differential equation of second order with a damping term $\sigma_i$ and an undamped angular frequency $\phi_i$.
Some values of the damping factor $\sigma_i$ are plotted in Fig.~\ref{fig_sigma_i} for $N$=1000 and $\nu$=1, with different $\Delta x$. The values are always positive, ensuring a damping for any selection of the parameters. It was observed (but not presented here) that $\sigma_i$ is very little dependent on the type of kernel.
It can be shown that the asymptote of the damping term in Fig.~\ref{fig_sigma_i} for high modes is given by:
\begin{equation}
\lim_{\substack{N \to \infty \\ M \to \infty}} \sigma_{N/2} = \int_{\Omega} \frac{W'(x)}{x} \, \mathrm dx
\end{equation}
\\In contrast, $\phi_i^2$ shows a more complex behavior with regards to $p_{bg}$, $\Delta x$ and the type of kernel. A map of $\phi_i^2$ is shown in Fig.~\ref{fig_phi_i2_map_M_vs_modes},  for a larger number of particles ($N$=10000), $p_{bg}=1$, and different number of neighbors $M$. First, $\phi_i^2$ is always $\ge$0, implying that $\phi_i$ is real. Hence, any onset of linear instability will be cancelled. However, for the quintic kernel, when $M\ge$9 (\ie below $\Delta x\le$0.32), some modes lower than $N_{mid}$ lead to $\phi_i^2$=0, generating a spurious mode which may trigger marginal instabilities at certain conditions. Such particular values are not observed with the Wendland kernel, suggesting at better stability.

\subsubsection{Effect of the background pressure on the particle disorder}

\begin{figure}[h]
	\centering
	\includegraphics[width=0.49\textwidth,keepaspectratio]{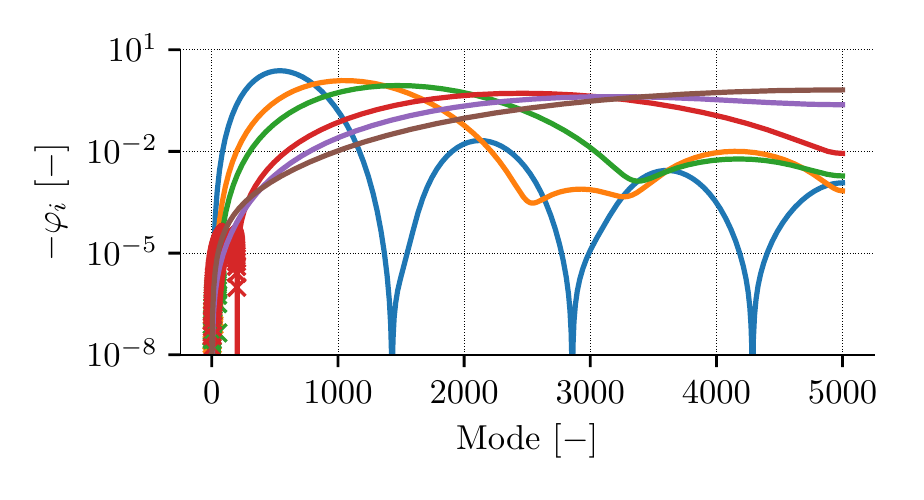}
	\includegraphics[width=0.49\textwidth,keepaspectratio]{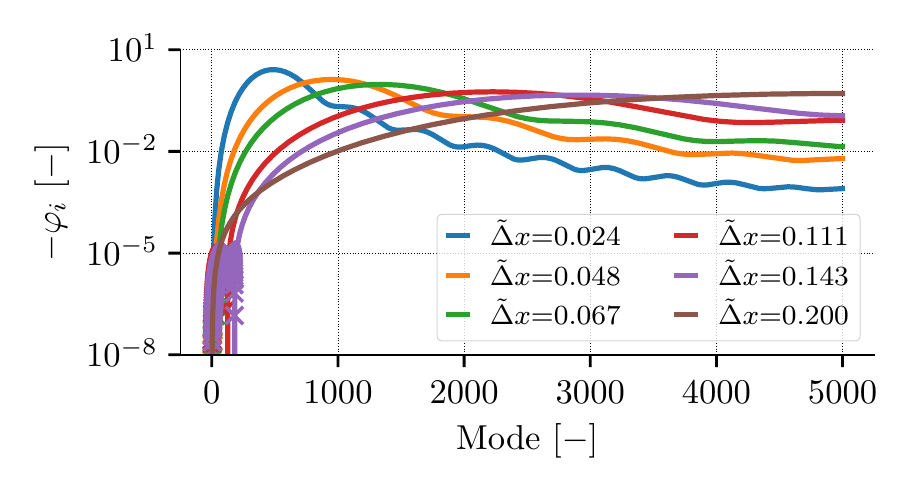}
	\caption{Values of $-\varphi_i$ for $N$=10000 with quintic (left) and Wendland (right) kernel.}
	\label{fig_Bi}
\end{figure}

This subsection details the effect of background pressure on one single particle and demonstrates the reordering effect of background pressure. Considering only the contribution of the background pressure to the acceleration of the perturbation (Eq.~\ref{eq_euler_perturb_matrix4b}), one obtains:
\begin{equation}
\dsecdv {y_i} t = 2 p_{bg} (\lambda_i''  - \Delta x \, \lambda_i'^2) \Delta x \, y_i
\label{eq_pback_details_01}
\end{equation}
Equation~\ref{eq_pback_details_01} states that the contribution of the background pressure to the acceleration of any perturbation is of the sign of the term between bracket $\varphi_i= \lambda_i''  - \Delta x \, \lambda_i'^2$. This term depends on $\Delta x$ and on the type of kernel. Figure~\ref{fig_Bi} shows its dependence on $\Delta x$ for the quintic and Wendland kernel. The crosses in Fig.~\ref{fig_Bi} correspond to negative values that could not be shown in a log-log plot. Hence, the $y$ position of crosses corresponds to $\varphi_i$ instead of $-\varphi_i$. They show the presence of positive values for lower modes, both for the quintic and the Wendland kernels. \majrev{These positive values are discussed in the following subsection.}
Apart from the points indicated by the crosses, $\varphi_i$ is always negative, meaning that the acceleration due to the background pressure is acting against the perturbation, which demonstrates the reordering effect of the background pressure. 
This results was observed by \citet{Colagrossi20121641},  \citet{marrone2013accurate}, and \citet{litvinov2015towards}, but to the authors knowledge, there was no formal proof of the feature. 
According to Eq.~\ref{eq_pback_details_01}, the time scales associated with the reordering effect of background pressure are:
$\tau_{bg} = 1 / \sqrt{2 p_{bg} \Delta x (\lambda_i''  - \Delta x \, \lambda_i'^2)}$.

\majrev{
\subsubsection{Critical value of the background pressure \label{sssec_critical_pback}}
The term $\varphi_i$ is positive (crosses in Fig.~\ref{fig_Bi}) for lower modes, which corresponds to large wavelength as it will be shown in Section~\ref{ssec_link_mode_length}. This means that the background pressure can increase the disorder for low frequency perturbations and particular values of $\tilde{\Delta} x$ (or M, equivalently). As shown in Fig.~\ref{fig_phi_i2_map_M_vs_modes}, for usual values of the background pressure, no instability is found ($\phi_i^2>0$), but for a background pressure above a critical value, the term proportional to $p_{bg}$ is dominant in $\phi_i^2$ (Eq.~\ref{eq_euler_perturb_matrix5b}), and $\phi_i^2$ becomes negative. Searching the critical $p_{bg}$ such as $\phi_i^2<0$ and $\varphi_i>0$ leads to:
\begin{equation}
p_{bg} > \frac{1}{2} \, \frac{\Delta x \, \lambda_i'^2}{\Delta x \, \lambda_i'^2 - \lambda_i'' }
\label{eq_phii2_negative}
\end{equation}
The critical values of $p_{bg}$ are plotted in Fig.~\ref{fig_pback_lim} for $N$=10000 and different $\tilde{\Delta} x$. Only the lower modes are subject to the instability, and therefore admit a critical background pressure. It is interesting to note that the Wendland kernel is also impacted by this instability. Contrary to the instabilities that may be triggered by the spurious mode (Fig.~\ref{fig_phi_i2_map_M_vs_modes}), which appears for a large number of neighbors, this instability occurs also at low $M$. The results of this subsection and Eq.~\ref{eq_phii2_negative} are validated with numerical simulations in \ref{appendix_pback_lim}.

\begin{figure}[h]
	\centering
	\includegraphics[width=0.49\textwidth,keepaspectratio]{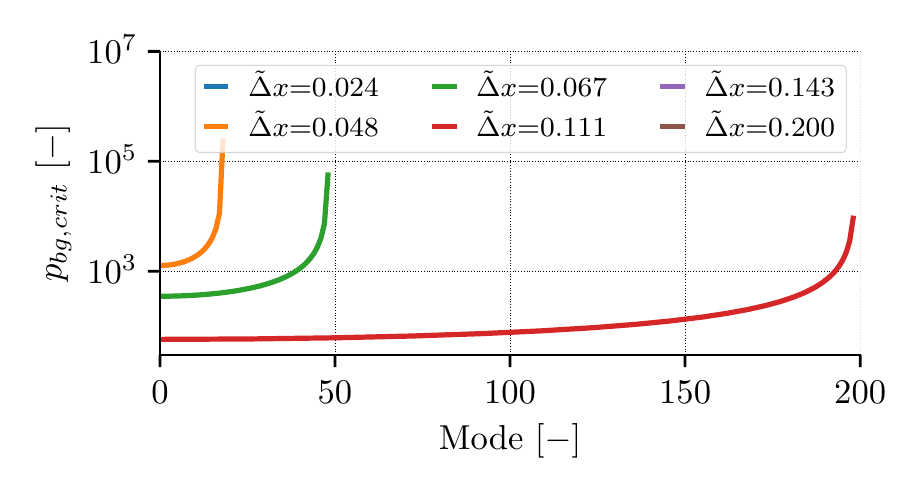}
	\includegraphics[width=0.49\textwidth,keepaspectratio]{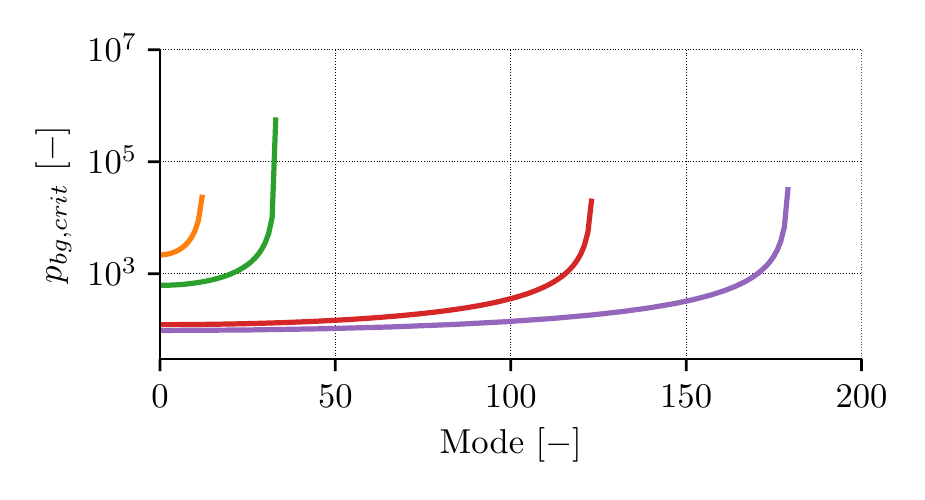}
	\caption{Critical values of $p_{bg}$ for $N$=10000 with quintic (left) and Wendland (right) kernel.}
	\label{fig_pback_lim}
\end{figure}
}

\subsection{Solutions of the perturbation equation}

This subsection aims to solve the equation of perturbation for the general case. It is organized as follows: the equations are transformed into the Laplace domain, where the transient and steady state as well as the stability can be investigated. Another advantage of the Laplace domain is to express ordinary differential equations by polynomials, which simplifies the analysis. Indeed, other WCSPH methods are ruled by a system of differential equations that cannot be directly integrated. After a simplification of the rational function, the solution are converted back into a temporal form of the modal space. Finally, the perturbations are written in the original geometric space.

\subsubsection{Projection into the Laplace domain}

Equation~\ref{eq_euler_perturb_matrix5} is transformed into a Laplace formulation. The rules of Laplace transformations are:
\begin{equation}
\label{eq_Laplace_rules}
\mathscr{L}(y_i) = Y_i
\  \text{,} \quad 
\mathscr{L} \left(\ddv {y_i} {t} \right) = p \, Y_i -  y_{i,0}
\quad \text{and} \quad 
\mathscr{L} \left(\dsecdv {y_i} {t} \right) = p^2 \, Y_i -  p \, y_{i,0} - y_{i,0}'
\end{equation}
where $y_{i,0}$ and $y_{i,0}'$ are the initial perturbation of position and velocity, respectively. The term $p$ is the Laplace variable and must not be mistaken for the pressure term. In the Laplace domain, the perturbation of positions (Eq.~\ref{eq_euler_perturb_matrix5}) is given by:
\begin{equation}
Y_i = \frac{p +  \sigma_i}{p^2 + \sigma_i \, p + \phi_i^2} \, y_{i,0} + \frac{1}{p^2 + \sigma_i \, p + \phi_i^2} \, y_{i,0}'
\label{eq_WCSPH_Laplace01}
\end{equation}
Hence, $Y_i$ depends on two polynomials in $p$ proportional to the initial perturbations. Their poles determine the stability of the system. The discriminant of the denominator is:
\begin{equation}
\Delta_i= \sigma_i^2 - 4 \, \phi_i^2
\label{eq_Delta_i}
\end{equation}
and the poles yield:
\begin{equation}
\pi_i^{\pm} = \frac{- \sigma_i }{2} \left(   1 \pm \sqrt{1 - 4(\phi_i / \sigma_i)^2 }  \right)
\label{eq_WCSPH_poles}
\end{equation}
The real part of $\pi_i^{\pm}$ is always of the sign of $- \sigma_i$, (\ie the sign of $\lambda_i'''$) because the square root in Eq.~\ref{eq_WCSPH_poles}  is either imaginary or lower than one. This is always true when the background pressure is positive. This implies that all modes will be damped in time, so that no instability is present, even when $\Delta_i$ becomes positive.
However, even though the real part of $\pi_i^{\pm}$ remains negative, it can be very close to zero, leading to an insufficient damping of the perturbation and a spurious mode may arise. 

\subsubsection{Temporal solution in the modal space}

In order to transform the solution back into the temporal modal space, the denominator of the polynomials in Eq.~\ref{eq_WCSPH_Laplace01} are written as:
\begin{equation}
p^2 + \sigma_i \, p + \phi_i^2 = \left( p + \frac{\sigma_i}{2} \right)^2 + \omega_i^2
\quad \text{with} \quad
\omega_i^2 = - \frac{\Delta_i}{4} = \phi_i^2 - \frac{\sigma_i^2}{ 4 }
\label{eq_euler_perturb_matrix6}
\end{equation}
so that Eq.~\ref{eq_WCSPH_Laplace01} can be expressed as:
\begin{equation}
Y_i =
\left[ \frac{p +  \sigma_i/2}{(p + \sigma_i / 2)^2 + \omega_i^2} + \frac{\sigma_i}{2\, \omega_i} \, \frac{\omega_i}{(p + \sigma_i / 2)^2 + \omega_i^2}    \right] \,   y_{i,0}
+ \left[ \frac{1}{\omega_i} \, \frac{\omega_i}{(p + \sigma_i / 2)^2 + \omega_i^2}    \right] \,   y_{i,0}'
\end{equation}
If $\Delta_i$ is negative, then $\omega_i^2$ is positive. Applying the inverse Laplace transform rules leads to:
\begin{subequations}
\label{eq_WCSPH_solution}
\begin{align}[left = \empheqlbrace\,]
y_i(t) = & \ f_{i}(t) \, y_{i,0} +  g_{i}(t) \, y_{i,0}'
\quad \text{with} \label{eq_WCSPH_solution_a} \\
f_{i}(t) = & \  \exp \left(  - \frac{\sigma_i}{2} t \right)
\left[ 
 \cos  \left( \omega_i t \right)  
+ \frac{\sigma_i}{2 \, \omega_i} \, \sin  \left(  \omega_i t \right) \right] \label{eq_f_i} \\
 g_{i}(t) = & \  \exp \left(  - \frac{\sigma_i}{2} t \right)
\left[ 
 \frac{1} {\omega_i} \,  \sin \left(  \omega_i t \right) \right] \label{eq_g_i}
\end{align}
\end{subequations}
When $\Delta_i>0$, the solution is obtained by replacing the sine and cosine functions by their hyperbolic expressions. %
When $\Delta_i$=0, which is equivalent to $\omega_i$=0, then $f_i(t)$ and $g_i(t)$ reduce to:
\begin{equation}
f_i(t) = \exp \left(  - \frac{\sigma_i}{2} t \right)  \, \left[  1 + \frac{\sigma_i}{2} \, t \right] 
\quad \text{and} \quad
g_i(t) = t \ \exp \left(  - \frac{\sigma_i}{2} t \right)
\label{eq_solution_yi_viscous_delta_zero}
\end{equation}
which correspond to a spurious mode. When $\nu>0$, this mode is damped.
However, in Eq.~\ref{eq_solution_yi_viscous_delta_zero}, $g_i(t)$ has a maximum at $t=2/\sigma_i$ which is equal to $ 2 / (\sigma_i e)$. This might cause instabilities in case of small damping ratio.
When $\nu$=0, the spurious mode is characterized by $f_i(t)$=1 and $g_i(t)$=$t$. Hence, any initial perturbation of velocity grows linearly with time.\\
Finally, the zero mode ($i=0$) is similar to a non-damped spurious mode:
\begin{equation}
\label{eq_spurious_inviscid}
f_0(t) = 1
\quad \text{and} \quad
g_0(t) = t
\end{equation}
Therefore, an initial perturbation of the velocity always leads to a constant velocity of all the particles. This is related to the conservation of linear momentum as illustrated in the next subsection.
It is interesting to note that the stability of the system depends also on which physical value is perturbed.

\subsubsection{Temporal solution in the geometrical space}

The matrix ${\uu{Y}}(t)$ in Eq.~\ref{eq_euler_perturb_matrix4} is a diagonal matrix whose diagonal elements are the right hand side of Eq.~\ref{eq_WCSPH_solution_a}. Multiplying ${\uu{Y}}(t)$ on the left by $ \uu{\uu{P}}^{-1}$ leads to the expression of  $\uu{X}$:
\begin{equation}
{\uu{X}}(t) = \uu{\uu{F}} (t) \  \uu{X}_0  + \uu{\uu{G}} (t) \  \uu{U}_0
\label{eq_sumSPH_solution_of_perturbations}
\end{equation}
where $\uu{X}_0$ and $\uu{U}_0$ are the vector of initial perturbations for the position and velocity, respecitvely. The terms $\uu{\uu{F}} (t)$ and $ \uu{\uu{G}}(t)$ are time-dependent matrices expressed by their general term:
\begin{subequations}
    \label{eq_WCSPH_F_and_G}
    \begin{align}
    \displaystyle
	F_{ij}(t) = & \ \frac{1}{N} \left[ f_0(t) + 2 \sum\limits_{k=1}^{N_{mid}}  f_k(t) \ \cos \left( \frac{2 \pi}{N} k (i-j)  \right) \right] \\
	G_{ij}(t) = & \ \frac{1}{N} \left[ g_0(t) + 2 \sum\limits_{k=1}^{N_{mid}} g_k(t) \ \cos \left( \frac{2 \pi}{N} k (i-j)  \right) \right]
    \end{align}
\end{subequations}
Equations~\ref{eq_WCSPH_F_and_G} highlights the coupling between the temporal solutions and the nodal oscillations. On the other hand, if only the position of the particle $i$ is perturbed, then its own response is given by $\delta x_i(t)=F_{ii}(t) \delta x_0$ which is a pure combination of time-oscillations independent of the oscillations of the neighbors.
\\Let us consider an initial velocity perturbation of $\delta u_0$ applied to the particle $0$ with $\nu > 0$. After a long time, due to the damping (Eq.~\ref{eq_g_i}), all $g_k(t)$ are zero except $g_0(t)=t$.
Therefore, the position of each particle follow the same law $\delta x_i = \delta u_0 t/N$ which means a constant momentum of $m \, \delta u_0/N$. After summing the velocity of all particles, one obtains $m \, \delta u_0$ which corresponds to the initial momentum of the system. Consequently the linear momentum of the system is conserved.
The validation of Eqs.~\ref{eq_WCSPH_solution} and \ref{eq_WCSPH_F_and_G} is given in \ref{appendix_validation}.

\subsection{Towards the dispersion relation\label{ssec_link_mode_length}}

\begin{figure}[h]
	\centering
	\includegraphics[width=0.5\columnwidth,keepaspectratio]{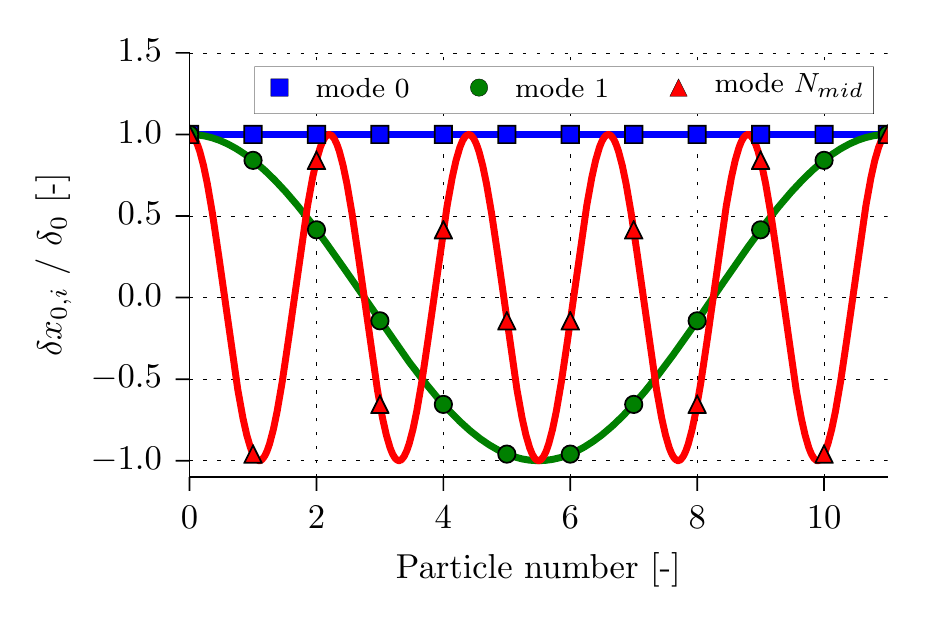}
	\caption{Illustration of different initial sinusoidal perturbations for $N$=11.}
	\label{fig_euler_spatial_mode_ini}
\end{figure}

As of now it was not possible to derive any dispersion equation because
no wavelength was present in the equation of motion.
However, it is possible to find a link between the time oscillations and the spatial modes by considering an initial spatial sinusoidal perturbation of mode $p$:
\begin{equation}
\delta x_{i,0} = \delta_0 \, \cos \left( \frac{2 \pi}{N} p \,  i  \right)
\end{equation}
where $\delta_0$ is the amplitude of the perturbation.
Because of the aliasing effect, the largest mode that can be resolved on the particle lattice is $N_{mid}$. Thus, p $\in \llbracket 0,N_{mid} \rrbracket$.
The wavelength associated to mode $p$ is $\lambda_p = L/p$ where $L=N \, \Delta x$ is the total length of the domain. Therefore, the initial sinusoidal perturbation can be written in terms of $\lambda_p$:
\begin{equation}
\delta x_{i,0} = \delta_0 \, \cos \left( 2 \pi \,  i  \frac{\Delta x}{\lambda_p} \right)
\end{equation}
and the smallest resolved wavelength is $\lambda_{min} = 2 \Delta x$ if $N$ is even.
Figure~\ref{fig_euler_spatial_mode_ini} shows an illustration of different initial sinusoidal perturbations.
\\Given these initial perturbations, the motion of particle $i$ is given by:
\begin{equation}
\delta x_i(t) = \sum\limits_{j=0}^{N-1} {F_{ij}(t) \delta x_{j,0}} 
\label{eq_euler_dispersion02}
\end{equation}
which simplifies, after some algebra, to:
\begin{equation}
\delta x_i(t) = f_p(t) \, \delta x_{i,0}
\label{eq_euler_dispersion03}
\end{equation}
where $f_p$ is given by Eq.~\ref{eq_f_i}.
This means that when a particle lattice is excited with by a spacial mode $p$, the particles will oscillate with the corresponding temporal mode $p$. Therefore, Eq.~\ref{eq_euler_dispersion03} establishes a link between spatial and temporal modes, \ie a dispersion relation which will be investigated in Section~\ref{sec_dispersion}.

\section{Application of LSA to div-SPH\label{sec_divSPH}}

In this section the motion of perturbations with div-SPH will be investigated, \ie when the density is estimated with the divergence of the velocity. The equations of motion are:

\begin{subequations}
\label{eq_divu01}
\begin{align}[left = \empheqlbrace\,]
\ddv {\rho_a} t = & \ \rho_a \sum\limits_b {V_b (u_a - u_b) \, W'_{ab}} \\
\dsecdv {x_a} t = & \  - \frac{1}{\rho_a} \sum\limits_{b} V_b (p_b + p_a) W'_{ab} + 2 \, \nu_a \, \sum\limits_b {V_b \frac{u_a - u_b}{r_{ab}} \, W'_{ab}} \\
p_a = & \  \frac{\rho_0 \, c^2 } {\gamma} \left[ \left( \frac{\rho_a}{\rho_0} \right)^{\gamma} - 1 \right]  + p_{back}
\end{align}
\end{subequations}
which leads, after linearization, non-dimensionalization and projection into the mode basis, to:
\begin{subequations}
\label{eq_divu02}
\begin{align}[left = \empheqlbrace\,]
\ddv {z_i} t = & \ \Delta x \, \lambda_i' \,  \ddv {y_i} {{t}} \\
\dsecdv {y_i} t = & \ \psi_{2,i} \,  y_i - \sigma_i \, \ddv {y_i} {{t}} + \frac{ \psi_{1,i}}{ \Delta x \, \lambda_i'} z_i
\end{align}
\end{subequations}
where $\sigma_i$, $\psi_{1,i}$ and $\psi_{2,i}$ are defined by Eqs.~\ref{eq_euler_perturb_matrix5b} and \ref{eq_euler_perturb_matrix5c}. The term $z_i$ is the general term of the density perturbations vector in the modal space $\uu{Z} = \uu{\uu{P}} \,  \uu{R}$.
The system is transformed into the Laplace domain: 
\begin{subequations}
\label{eq_divu03}
\begin{align}[left = \empheqlbrace\,]
pZ_{i} - z_{i,0}= & \ \Delta x \, \lambda_i' \,  (pY_{i} - y_{i,0})  \label{eq_laplace_density} \\
p^2 Y_i - p y_{i,0} - y'_{i,0}= & \  \psi_{2,i} \, Y_i - \sigma_i \, (pY_{i} - y_{i,0}) + \frac{ \psi_{1,i}}{ \Delta x \, \lambda_i'}  Z_i \label{eq_laplace_posi}
\end{align}
\end{subequations}
where $Y_i(p)$ and $Z_i(p)$ are the Laplace transforms of $y_i(t)$ and $z_i(t)$.
The initial conditions of position, velocity and density are denoted by $y_{i,0}$, $y'_{i,0}$ and $z_{i,0}$, respectively.
Substituting Eq.~\ref{eq_laplace_density} into Eq.~\ref{eq_laplace_posi} leads to:
\begin{subequations}
\begin{align}
Y_i = & \  P_1(p) \, y_{i,0} + P_2(p) \, y'_{i,0} + P_3(p) \, z_{i,0}
\quad \text{with}  \\
P_1(p) = & \  [p^2 + \sigma_i \, p - \psi_{1,i} ] / Q(p) \\ 
P_2(p) = & \  p / Q(p) \\ 
P_3(p) = & \   (\psi_{1,i} / \Delta x \, \lambda_i' ) / Q(p) \\
Q(p) = & \  p [p^2 + \sigma_i \, p + \phi_i^2] 
\end{align}
\end{subequations}
The poles of $P_1$, $P_2$, $P_3$ are the roots of $Q(p)$. They are given by: 
\begin{equation}
r_1 = 0   
\quad \text{and} \quad
r_{2,3} = \frac{- \sigma_i }{2} \left(   1 \pm \sqrt{1 - 4(\phi_i / \sigma_i)^2 }  \right)
\end{equation}
The roots $r_{2,3}$ are equal to $\pi_i^{\pm}$ as given by Eq.~\ref{eq_WCSPH_poles} with sum-SPH. Hence, their real part is either negative or zero, leading to a decreasing exponential or continuous component in the temporal domain, as for sum-SPH. In addition, the value $p=0$ is also a pole of $P_1$ and $P_3$, which means that the decomposition of $P_1$ and $P_3$ into rational functions will exhibit the term $1/p$. The inverse Laplace transform of $1/p$ is the Heaviside function, \ie a continuous component signal. In other words, an initial perturbation of position or density will remain unaffected.\\
Since the the roots of $Q$ are the same as $\pi_i^{\pm}$, it means that, apart from the continuous mode, the dispersion and the damping curves of div-SPH are the same as for sum-SPH. 
\majrev{Also, the div-SPH method is subject to the critical background pressure instability highlighted in Section~\ref{sssec_critical_pback}.}
Hence, the only particularity of using the velocity divergence to estimate the density is the non-damped perturbation of position and density.
\\\\
The rational functions $P_1$, $P_2$ and $P_3$ are simplified to apply the inverse Laplace transforms:
\begin{subequations}
\begin{align}[left = \empheqlbrace\,]
P_1 = & \ - \frac{1}{\phi_i^2} 
\left( \psi_{1,i} \, \frac{1}{p}
+ \psi_{2,i} \,   \frac{p+\sigma_i/2}{(p + \sigma_i / 2)^2 + \omega_i^2}
+ \frac{\psi_{2,i} \, \sigma_i}{2 \, \omega_i} \,   \frac{\omega_i}{(p + \sigma_i / 2)^2 + \omega_i^2}
\right) \\
P_2 = & \ \frac{1}{\omega_i} \,   \frac{\omega_i}{(p + \sigma_i / 2)^2 + \omega_i^2} \\
P_3 = & \ \frac{  - \psi_{1,i}}{\Delta x \, \lambda_i' \, \phi_i^2} \left(
 \frac{p+\sigma_i/2}{(p + \sigma_i / 2)^2 + \omega_i^2}
+ \frac{\sigma}{2 \, \omega_i} \,   \frac{\omega_i}{(p + \sigma_i / 2)^2 + \omega_i^2} 
- \frac{1}{p}
\right)
\end{align}
\end{subequations}
where $\omega_i$ is defined in Eq.~\ref{eq_euler_perturb_matrix6}. The rational functions $P_1$, $P_2$ and $P_3$ are then transformed into a temporal form to $f_i(t)$, $g_i(t)$, $h_i(t)$, respectively. When $\Delta_i$ (Eq.~\ref{eq_Delta_i}) is negative, the functions can be expressed as :
\begin{subequations}
\label{eq_divSPH_temporal_funcs}
\begin{align}[left = \empheqlbrace\,]
f_i(t) = & \ - \frac{1}{\phi_i^2} 
\left[ \psi_{1,i} 
+ \psi_{2,i} \, \exp \left( \frac{-\sigma_i}{2} t \right) \left( \cos (\omega_i t) + \frac{\sigma_i}{2 \, \omega_i} \,  \sin (\omega_i t) \right)  \right] \label{eq_divSPH_temporal_modal_position} \\
g_i(t) = & \ \frac{1}{\omega_i} \,   \exp \left( \frac{-\sigma_i}{2} t \right) \sin (\omega_i t)   \\
h_i(t) = & \ \frac{  - \psi_{1,i}}{\Delta x \, \lambda_i' \, \phi_i^2} \left[
\exp \left( \frac{-\sigma_i}{2} t \right) \left( \cos (\omega_i t) + \frac{\sigma_i}{2 \, \omega_i} \,  \sin (\omega_i t) 
 \right)  -1 \right] 
\end{align}
\end{subequations}
When $\Delta_i>0$, the sinusoidal functions are to be replaced by their hyperbolic variants. When $\Delta_i=0$, the cosine is replaced by $1$ and $ \sin (\omega_i t) / \omega_i$ is replaced by $t$. The zero mode are $f_0(t) = 1$, $g_0(t) = t$ and $h_0(t) = 0$. 
Note that $h_i(t)$ is a pure imaginary function due to the term $\lambda_i'$ in its prefactor.
The solution of the perturbation equation in the modal space yields:
\begin{equation}
\label{eq_divSPH_sum_temporal}
y_i(t) = f_i(t) \, y_{i,0} + g_i(t) \, y_{i,0}' + h_i(t) \, z_{i,0}
\end{equation}
Transforming $y_i(t)$ to $\uu{X}(t)$ is done using Eqs.~\ref{eq_WCSPH_F_and_G} for the position and velocity. For the density,  due to the property $\lambda_k' = - \lambda_{n-k}'$,
the inverse transformation of $h_i(t)$ is:
\begin{equation}
\label{eq_divSPH_to_geom_H}
H_{ij}(t) = \frac{1}{N} \left[ h_0(t) + 2 \, \hat{i} \sum\limits_{k=1}^{N_{mid}} h_k(t) \ \sin \left( \frac{2 \pi}{N} k (i-j) \right) \right]
\end{equation}
which is a real function given that $h_k(t)$ are purely imaginary. Equations.~\ref{eq_divSPH_temporal_funcs} and \ref{eq_divSPH_to_geom_H} are validated in \ref{appendix_validation}.

\section{Application of LSA to $\delta$-SPH \label{sec_deltaSPH}}

The same analysis is now applied to $\delta$-SPH. This scheme, originally proposed by \citet{antuono2010free}, is based on div-SPH, in which additional diffusion terms are added in the density and energy equation. The equations of motion are recalled in 1D:
\begin{subequations}
\label{eq_delta01}
\begin{align}[left = \empheqlbrace\,]
\ddv {\rho_a} t = & \ \rho_a \sum\limits_b {V_b (u_a - u_b) \, W'_{ab}} + \xi  h c \sum\limits_b {V_b \psi_{ab} \, W'_{ab}}  \label{eq_delta_SPH_01_density}\\
\rho_a \dsecdv {x_a} t = & \  - \sum\limits_{b} V_b (p_b + p_a) W'_{ab} + \alpha h c \ol{\rho}  \sum\limits_b {V_b \pi_{ab} \, W'_{ab}} \\
\rho_a \ddv {e_a} t = & \  p_a \sum\limits_{b}{V_b (u_a - u_b) \, W'_{ab}} -  \frac{\alpha h c \ol{\rho}}{2}  \sum\limits_b {V_b \pi_{ab} \, (u_a - u_b) W'_{ab}} + \chi h c \ol{\rho} \sum\limits_b {V_b \phi_{ab} \, W'_{ab}} \\
p_a = & \  c^2  \, (\rho_a - \ol{\rho}) \left[ 1 + \Gamma \left(  \frac{e_a}{\ol{e}} - 1  \right)  \right] + p_{back} \label{eq_delta_SPH_EoS}
\end{align}
\end{subequations}
where:
\begin{equation}
\psi_{ab} = 2 \frac{\rho_a - \rho_b}{x_{ab}}
\quad
\phi_{ab} =2 \frac{e_a - e_b}{x_{ab}}
\quad
\pi_{ab}  = \frac{u_a - u_b}{x_{ab}}
\label{eq_delta02}
\end{equation}
The system of equations~\ref{eq_delta01} shows, in order of appearance, the conservation of mass, momentum and energy, and the EoS.
The terms $\alpha$, $\xi$, $\chi$ and $\Gamma$ are the non-dimensional coefficients of (i) the viscosity (equal to $2\nu$), (ii) the mass diffusivity, (iii) the energy diffusivity and (iv) the internal energy response to compressibility, respectively. In their original paper, \citet{antuono2010free} omitted the background pressure in the EoS but later publications (\eg \citet{marrone2013accurate}) reintroduced it.
The intermediate quantities are linearized the first-order approximation as:
\begin{equation}
\label{eq_deltaSPH02b}
\psi_{ab} \approx 2 \frac{\delta \rho_a - \delta \rho_b}{x_{\ol{ab}}} 
\quad
\phi_{ab} \approx 2 \frac{\delta e_a - \delta e_b}{x_{\ol{ab}}} 
\quad
\psi_{ab} \approx \frac{\delta u_a - \delta u_b}{x_{\ol{ab}}} 
\end{equation}
As mentioned by \citet{antuono2010free}, the energy contribution inside the equation of state (Eq.~\ref{eq_delta_SPH_EoS}) is second order, so it does not appear in the density and motion equation after linearization.
Non-dimensionalization and projection into the mode basis lead to the following system:
\begin{subequations}
\label{eq_delta03}
\begin{align}[left = \empheqlbrace\,]
\ddv {z_i} t = & \ \Delta x \, \lambda_i' \,  \ddv {y_i} {{t}}  + 2 \Delta x \, \xi \, \lambda_i''' z_i \\
\dsecdv {y_i} t = & \ \psi_{2,i} \,  y_i - \sigma_i \, \ddv {y_i} {{t}} + \frac{ \psi_{1,i}}{ \Delta x \, \lambda_i'} z_i
\end{align}
\end{subequations}
where, in this case, the term $\sigma_i$ defined in Eq.~\ref{eq_euler_perturb_matrix5b} is now equal to $- \alpha \Delta x \, \lambda_i'''$.
It is worth to note that this system of equations \improve{differs from} the system of div-SPH (Eqs.~\ref{eq_divu02}) in that a diffusion term is added in the density equation. Translating this system into the Laplace domain and expressing the particle motion $Y_i$ leads to:
\begin{subequations}
\label{eq_deltaSPH_Laplace_Y}
\begin{align}
Y_i = & \  P_1(p) \, y_{i,0} + P_2(p) \, y'_{i,0} + P_3(p) \, z_{i,0}
\quad \text{with}  \\
P_1(p) = & \  [p^2 - (\alpha + 2\, \xi) \Delta x \,  \lambda_i''' \, p + 2 \alpha \xi (\Delta x \lambda_i''')^2 - \psi_{1,i} ] / Q(p) \\ 
P_2(p) = & \  (p - 2\, \xi \, \Delta x \,  \lambda_i''' ) / Q(p) \\ 
P_3(p) = & \  (\psi_{1,i} / \Delta x \, \lambda_i' ) / Q(p)
\end{align}
\end{subequations}
and 
\begin{equation}
\label{eq_Qp_deltaSPH}
Q(p) = p^3 - (\alpha + 2 \, \xi) \,  \Delta x \, \lambda_i''' \, p^2 + [2 \alpha \xi (\Delta x \lambda_i''')^2   + \phi_i^2  ] \, p + 4 \Delta x^2 \xi \, p_{bg}  \lambda_i'' \, \lambda_i''' 
\end{equation}

\begin{figure}[!htb]
	\centering
	\includegraphics[width=0.49\textwidth,keepaspectratio]{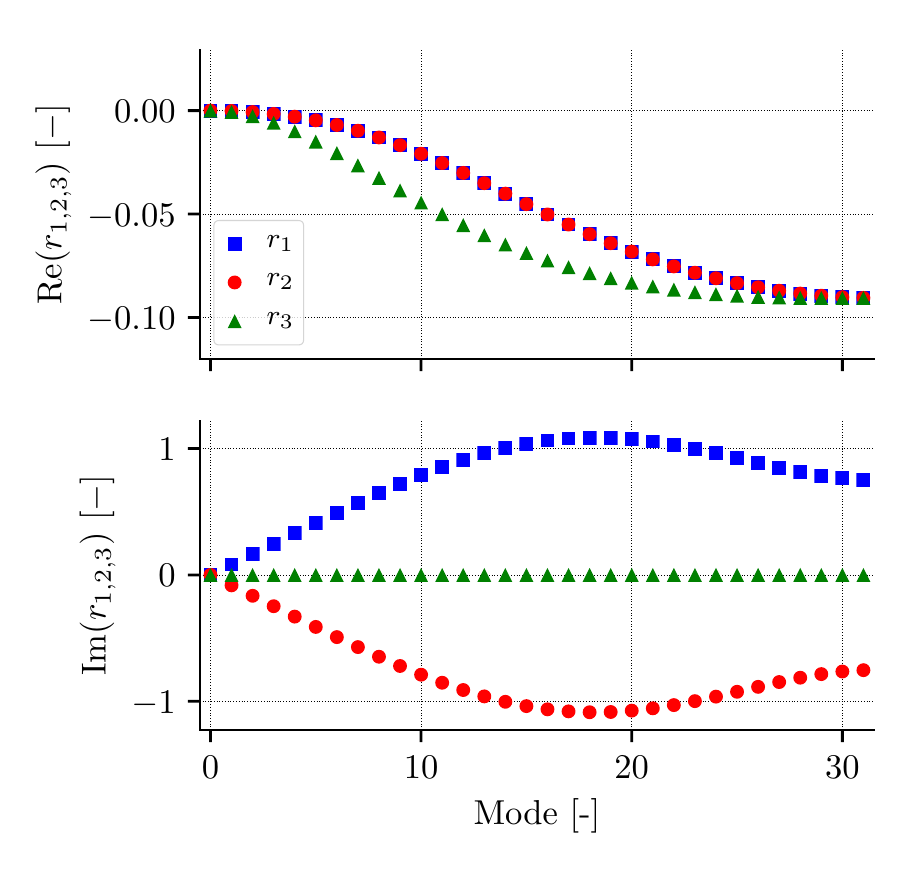}
	\includegraphics[width=0.49\textwidth,keepaspectratio]{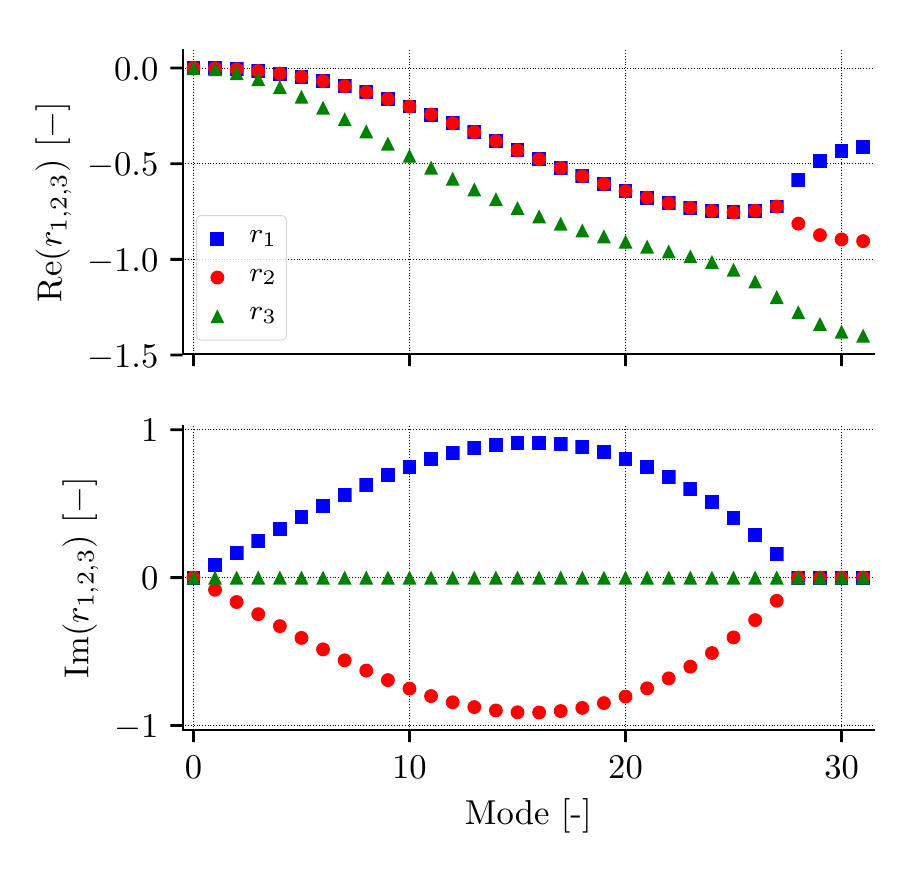}
	\caption{Real (top) and imaginary (bottom) parts of the roots of $Q$ for $\alpha$ = 0.2 (left) and 2 (right). Other parameters are $p_{bg}$=1 and $\tilde{\Delta} x$=0.143, M=3, N=63, $\xi=\alpha/4$ and the quintic kernel was used.}
	\label{fig_deltaSPH_modes_pbg1}
\end{figure}

\begin{figure}[!htb]
	\centering
	\includegraphics[width=0.49\textwidth,keepaspectratio]{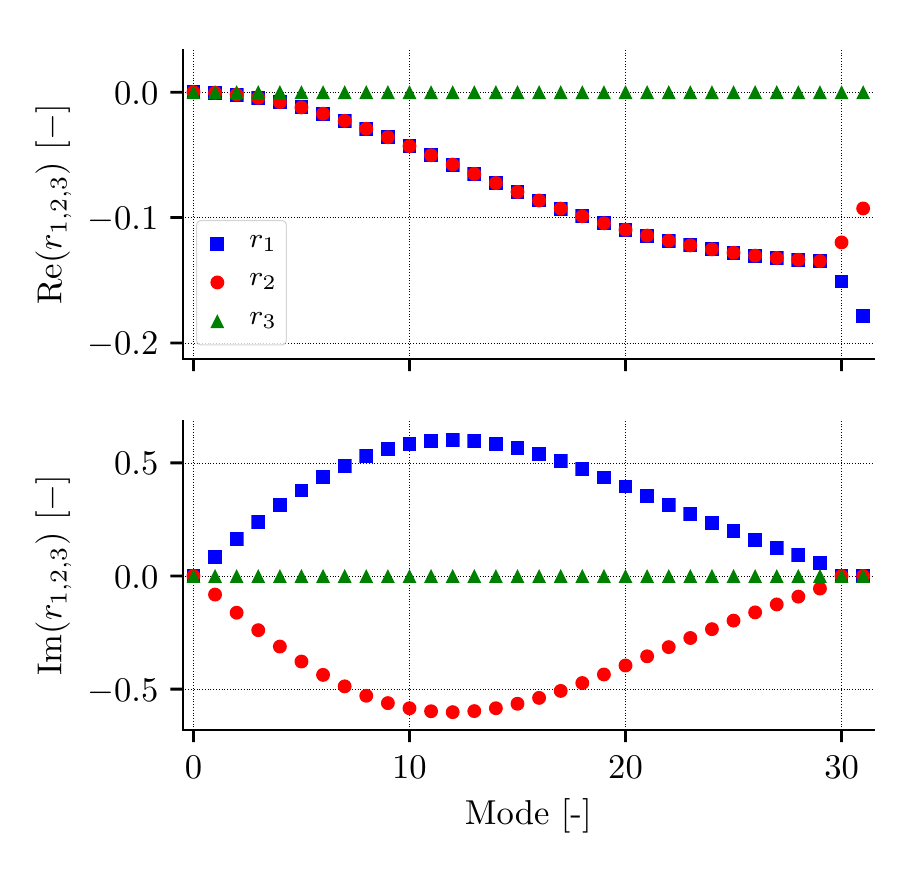}
	\includegraphics[width=0.49\textwidth,keepaspectratio]{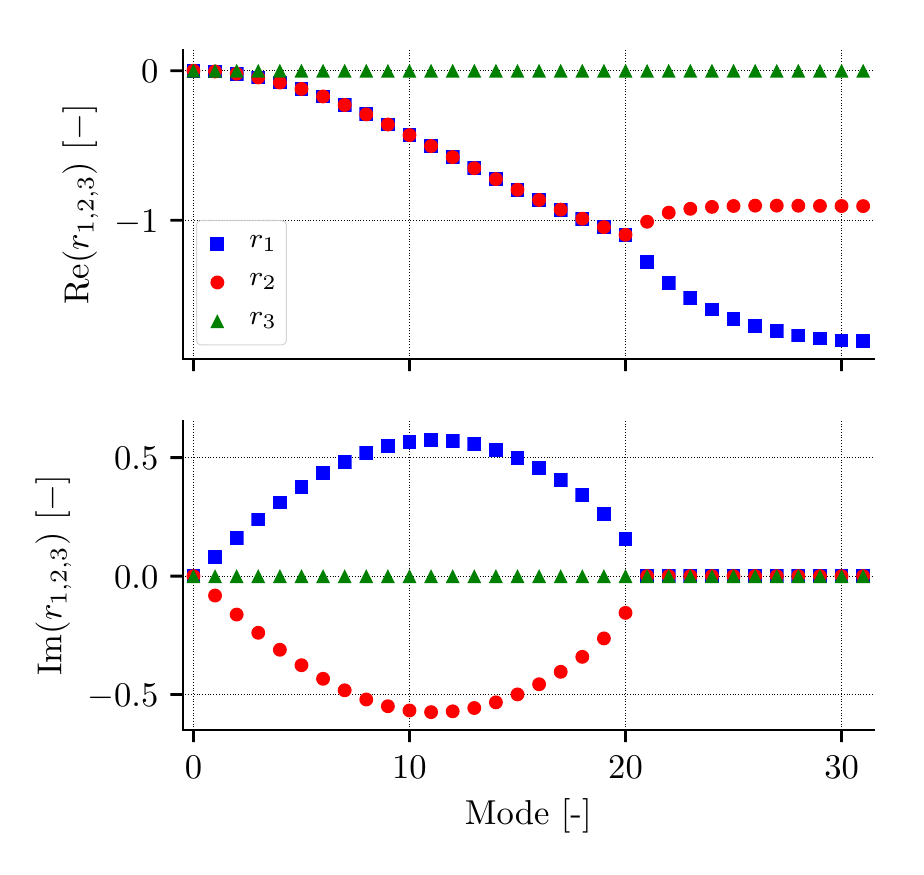}
	\caption{Real (top) and imaginary (bottom) parts of the roots $r_1$, $r_2$, $r_3$ of $Q$ for $\alpha$ = 0.2 (left) and 2 (right). Other parameters are $p_{bg}$=0 and $\tilde{\Delta} x$=0.143, M=3, N=63, $\xi=\alpha/4$ and the quintic kernel was used.}
	\label{fig_deltaSPH_modes_pbg0}
\end{figure}

In this case, the roots of $Q(p)$ cannot be analytically expressed in a compact form. However, it is worth to note that because of the constant term $4 \Delta x^2 \xi \, p_{bg}  \lambda_i'' \, \lambda_i''' $ , $p=0$ is not a root of $Q$, which means that no undamped continuous mode can be found, provided that both $\xi$ and $p_{bg}$ are non zero. Consequently, for $\xi > 0$ and $p_{bg} > 0$, any initial perturbation of any physical type is damped. This is a clear advantage of $\delta$-SPH over div-SPH when background pressure is added. In order to investigate the poles of $P_1$, $P_2$ and $P_3$, the equation $Q(p)=0$ is solved numerically and the roots are discussed below.
\\\\
Figure~\ref{fig_deltaSPH_modes_pbg1} shows the roots of $Q$ for different parameters. Their real part represents the stability while the imaginary part is the angular frequency of the oscillation.
When the viscosity increases (Fig.~\ref{fig_deltaSPH_modes_pbg1} right), a second regime is observed for higher modes. It is an aperiodic regime where one of the damping ratio decreases to zero ($r_1$ in Fig.~\ref{fig_deltaSPH_modes_pbg1} right), as also pointed out by \citet{antuono2012numerical}.
In this case, the continuous solution ($r_3$) is damped faster than the sinusoidal modes. This means that physical oscillations have a larger amplitude that the unphysical continuous perturbations in $\delta$-SPH.
It was also observed (but not illustrated here) that when $\xi > \alpha/2$, the real part of $r_1$ and $r_2$ becomes positive for the first modes, leading to the slow divergence of a perturbation of large wavelength. The value of $\xi = \alpha/2$ was found by \citet{antuono2010free} as an optimal value, but not a limiting one, probably due to the assumption of zero background pressure in their analysis.
\\\\
Figure~\ref{fig_deltaSPH_modes_pbg0} shows the roots of $Q$ for the same parameters as in Fig.~\ref{fig_deltaSPH_modes_pbg1} except that background pressure is set to zero.
It is observed that the angular frequency decreases to zero at higher modes, and that the aperiodic mode starts at lower modes. 
As mentioned earlier, the continuous mode is not damped ($r_3$=0), for any values of $\alpha$. Therefore to fully benefit of the advantage of $\delta$-SPH, it is demonstrated here that a strictly positive background pressure is mandatory.
However, for a large viscosity (Fig.~\ref{fig_deltaSPH_modes_pbg0} right), the damping of the aperiodic mode is larger than when the background pressure is positive. This means that high frequency perturbations are faster damped without background pressure, which could be an advantage.\\
In the present case where $p_{bg}$=0, the value of $\xi$ is not limited by $\alpha$, as mentioned by \citet{antuono2010free}, as it appears as a regular viscosity coefficient in the expression of $Q$ (Eq.~\ref{eq_Qp_deltaSPH}). 
In this context, \citet{antuono2012numerical} studied the stability of the $\delta$-SPH method when the background pressure is equal to zero. They applied a 1D linear stability analysis with a modified diffusive term in the density equation ($\psi_{ab}$ in Eq.~\ref{eq_delta_SPH_01_density}) which is also valid on an incomplete sphere of influence.
For a given $\alpha$, they provided a range of $\xi$ to avoid the aperiodic regime. However, they did not mention the undamped continuous mode.
Applying the same type of analysis with the diffusion operator presented in Eq.~\ref{eq_delta_SPH_01_density} leads to the condition on the diffusion constants $\xi$ and $\alpha$:
\begin{equation}
\label{eq_condition_delta_SPH_xi_alpha}
\frac{\alpha}{2} + \frac{\phi_i}{\Delta x \, \lambda_i'''} 
\le \xi \le 
\frac{\alpha}{2} - \frac{\phi_i}{\Delta x \, \lambda_i'''} 
\end{equation}
which is similar to the one derived by \citet{antuono2012numerical}, in the sense that the limiting value depends on $\phi_i /\Delta x \, \lambda_i'''$ which is related to the maximum resolved wavenumber.
Finally, even though the temporal solution of the linearized $\delta$-SPH cannot be expressed as an analytical function, Eqs.~\ref{eq_deltaSPH_Laplace_Y} and \ref{eq_Qp_deltaSPH} are validated in \ref{appendix_validation}.

\section{Application of LSA to the Transport Velocity Formulation (TV-SPH) \label{sec_TVSPH}}

The Transport-Velocity Formulation of SPH (referred to as TV-SPH in the following) was developed by \citet{adami2013transport}. The basic idea is to (i) advect particles at a modified transport velocity $\wt{u}$ that reduces the particle disorder, and (ii) to compute the physical source terms from the original fluid velocity $u$. The lower particle disorder increases the accuracy of the operators compared to traditional methods. This approach is based on the assumption of incompressible flow even though it is applied to WCSPH.
The set of equations is, in 1D:
\begin{subequations}
\label{eq_TVF_01}
\begin{align}[left = \empheqlbrace\,]
\rho_a = & \ m_a \sum\limits_b {W_{ab}}  \\
\frac{\wt{d} u_a} {d t} = & \  \frac{1}{m_a} \sum\limits_{b} (V_a^2 + V_b^2) \left[ -\wt{p}_{ab} W'_{ab} + \frac{1}{2} (T_a + T_b) W'_{ab} + \wt{\mu}_{ab}  \frac{u_{ab}}{x_{ab}} \, W'_{ab} \right] \\
p_a = & \  c^2 \, (\rho_a - \ol{\rho}) \\
\frac{d \wt{u}_a} {d t} = & \  \frac{\wt{d} u_a} {d t}  - \frac{p_{back}}{m_a} \sum\limits_{b} (V_a^2 + V_b^2) W'_{ab}\\
\ddv {x_a}{t} = & \ \wt{u}_a
\end{align}
\end{subequations}
where $\wt{d}/dt$ is the material derivative of a particle moving with the modified transport velocity and $T = \rho u (\wt{u} - u)$ is the convection of the particle momentum with the relative velocity $(\wt{u} - u)$. The dynamic viscosity and pressure are expressed with an inter-particle approach as $\wt{\mu}_{ab}=2\mu_a \, \mu_b / (\mu_a + \mu_b)$ and $\wt{p}_{ab} = (\rho_b \, p_a + \rho_a \, p_a) / (\rho_a+\rho_b)$, respectively. It is important to notice that even though $\wt{d}/dt$ and $d/dt$ are two material derivatives with different advection velocities for the Eulerian point of view, they are equivalent to $\partial / \partial t$ in the Lagrangian framework \citep{belytschko2000unified}. In Eqs.~\ref{eq_TVF_01}, the two independent variables are $u_a$ and $x_a$.
\\\\
It is assumed that the viscosity is homogenous, \ie $\wt{\mu}_{ab}=\mu$. The intermediate quantities are linearized to the first-order as:
\begin{subequations}
\label{eq_TVF_02}
\begin{align}[left = \empheqlbrace\,]
V_a^2 \approx  & \ \ol{V}^2 \left( 1 - 2\frac{\delta \rho_a}{\ol{\rho}} \right) \\
\wt{p}_{ab} \approx & \ \frac{\ol{\rho}_b \, \delta p_a + \ol{\rho}_a \, \delta p_b }{\ol{\rho}_a+ \ol{\rho}_b}\\
T_a \approx & \  \ol{\rho} \, \ol{u} \, \ol{w}     + \ol{\rho} \, \ol{u} \delta w_a + \ol{\rho} \, \ol{w} \delta u_a + \ol{u} \, \ol{w} \delta \rho_a
\end{align}
\end{subequations}
where the relative velocity is decomposed as $ w = \wt{u} - u =\ol{w} + \delta w$. Due to the quadratic dependency of $T_a$ on ${u}$, the mean velocity is considered positive only ($\ol{u} \ge 0$) to apply the linearization. The non-dimensionalization and the projection into the mode basis leads to:
\begin{subequations}
\label{eq_TVF03}
\begin{align}[left = \empheqlbrace\,]
\frac{1}{\Delta x} \, \ddv {\upsilon_i} t = & \ [ \ol{u} \, \ol{w} \lambda_i'  - (1 + \ol{u} \, \ol{w}) \Delta x \lambda_i'^2] \, y_i + [ (\ol{w} - \ol{u}) \lambda_i' + 2 \nu \lambda_i'''] \, \upsilon_i + \ol{u} \lambda_i' \ddv {y_i} {{t}} \\ 
\dsecdv {y_i} t = & \  \ddv {\upsilon_i} t +   2 \, p_{bg} \, \Delta x (\Delta x \lambda_i'^2 -  \lambda_i'' )  y_i 
\end{align}
\end{subequations}
where $\upsilon_i$ is the general term of the projection of $\uu{U}$ in the mode basis. Expressing the coordinate $y_i$ in the Laplace domain yields:
\begin{subequations}
\label{eq_TVF04}
\begin{align}
Y_i = & \  P_1(p) \, y_{i,0} + P_2(p) \, y'_{i,0} + P_3(p) \, \upsilon_{i,0}
\quad \text{with}  \\
P_1(p) = & \ p \, [p - (\ol{w} \lambda_i' + 2 \nu \lambda_i''') \Delta x] / Q(p) \\ 
P_2(p) = & \ [p -  ( [\ol{w} - \ol{u}] \lambda_i' + 2 \nu \lambda_i''') \Delta x ]  / Q(p) \\ 
P_3(p) = & \ ( [\ol{w} - \ol{u}] \lambda_i' + 2 \nu \lambda_i''') \Delta x  / Q(p)
\end{align}
\end{subequations}
and 
\begin{align}
\label{eq_TVSPH_Q_p}
Q(p) = & \  p^3 + \Delta x (\ol{w} \lambda_i' - 2 \nu \lambda_i''') \, p^2  \nonumber  \\
 & \ -  \Delta x [2 \, \pi_{bg} \, \varphi_i  +\Delta x \lambda_i'^2]   \, p  \nonumber  \\ 
& \ - 2  \Delta x^2 \, p_{bg}  \varphi_i  ( [\ol{w} - \ol{u}] \lambda_i' - 2 \nu \lambda_i''')
\end{align}
where $\pi_{bg} = p_{bg} - \ol{u} \ol{w}/2$ and $\varphi_i = \lambda_i''- \Delta x \lambda_i'^2$.
As in $\delta$-SPH, the constant part of $Q(p)$ ensures that $p$=0 is not a root. Therefore, $p_{back} > 0$ is a sufficient condition to ensure damping of any perturbations in the case of viscous flow ($\nu > 0$).
The Laplace variable of the physical velocity $\upsilon$ is labeled $\Upsilon$ and yields:
\begin{subequations}
\label{eq_TVF05}
\begin{align}
\Upsilon_i = & \  P_1(p) \, y_{i,0} + P_2(p) \, y'_{i,0} + P_3(p) \, \upsilon_{i,0}
\quad \text{with}  \\
P_1(p) = & \ [p^2 (1 - \ol{u} \Delta x \lambda_i') +  p \, \Delta x (\Delta x \lambda_i'^2  - \ol{u} \ol{w}\varphi_i)  - 2 p_{bg} \Delta x  \varphi_i  ]/ Q(p) \\ 
P_2(p) = & \ [\ol{u} \lambda_i' \, p  + \ol{u} \ol{w}\varphi_i - \lambda_i'^2 ] \Delta x  / Q(p) \\ 
P_3(p) = & \ [ p^2 + \ol{u} \Delta x  \lambda_i' \, p  - \Delta x (2 \pi_{bg} \varphi_i - \Delta x \lambda_i'^2 )  ]  / Q(p)
\end{align}
\end{subequations}
with $Q(p)$ as defined in Eq.~\ref{eq_TVSPH_Q_p}. Equations~\ref{eq_TVF04}, \ref{eq_TVSPH_Q_p} and \ref{eq_TVF05} are validated in \ref{appendix_validation}.
\\\\
\begin{figure}[!htb]
	\centering
	\includegraphics[width=0.49\textwidth,keepaspectratio]{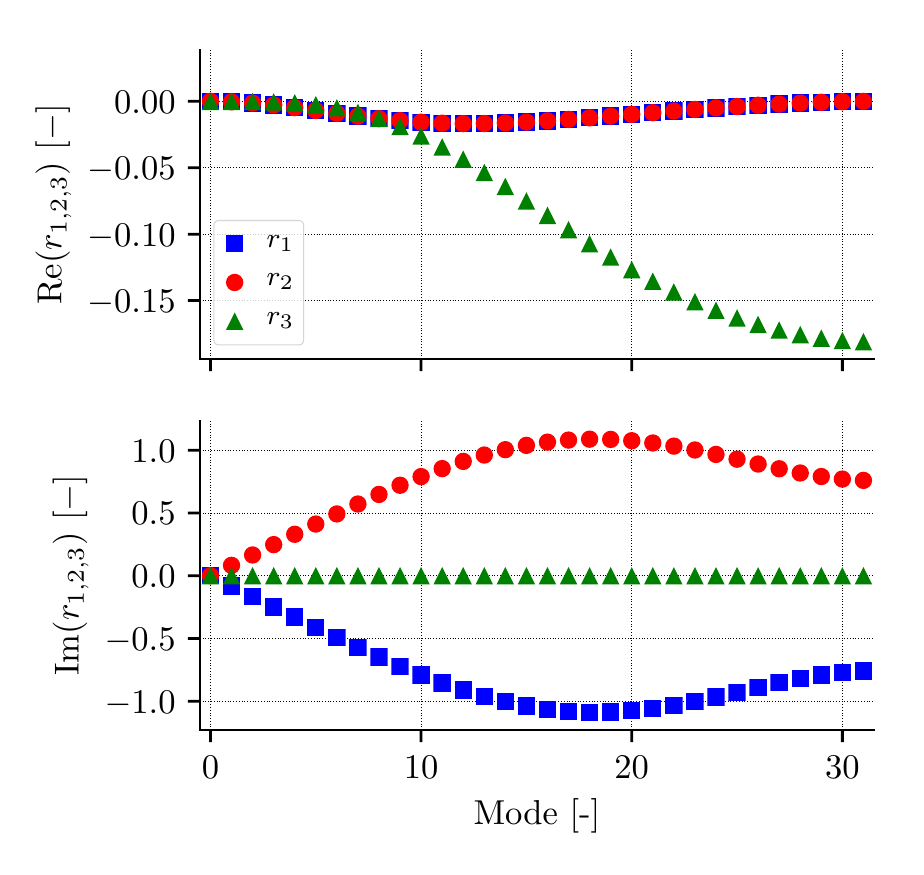}
	\includegraphics[width=0.49\textwidth,keepaspectratio]{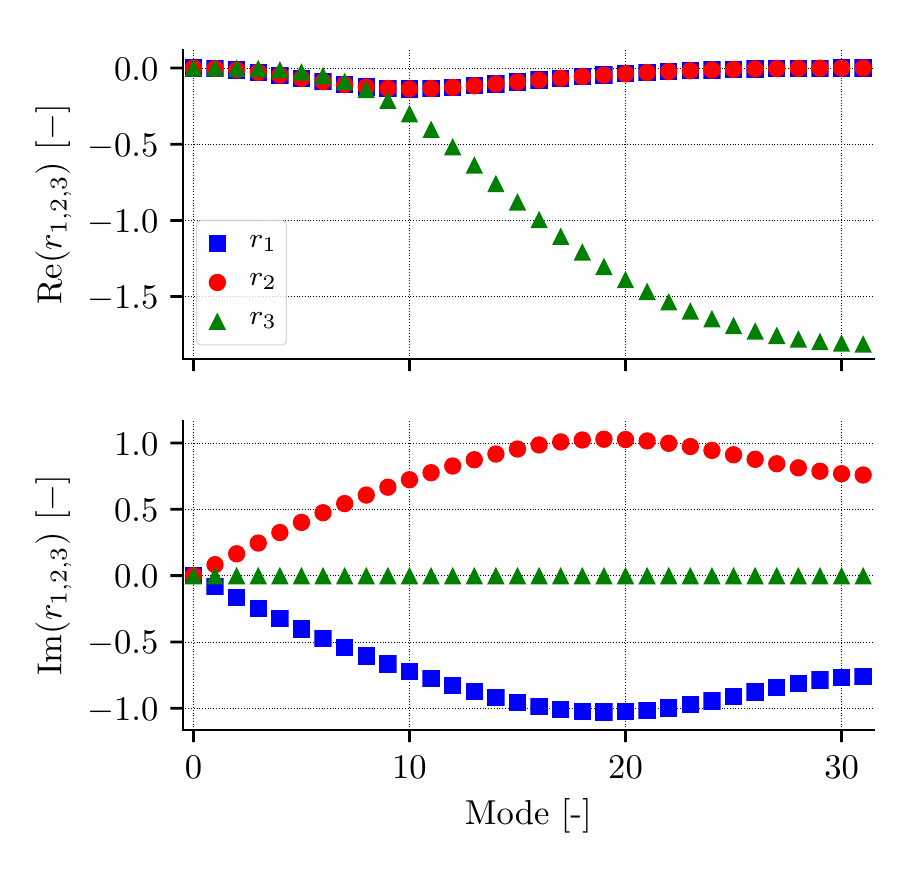}
	\caption{Real (top) and imaginary (bottom) parts of the roots $r_1$, $r_2$, $r_3$ of $Q$ for $\nu$ = 0.1 (left) and 1 (right). Other parameters are $p_{bg}$=1, $\tilde{\Delta} x$=0.143, N=63, ($\ol{u},\ol{w}$)=(0,0) and the quintic kernel.}
	\label{fig_TVSPH_modes_pbg1}
\end{figure}
Figure~\ref{fig_TVSPH_modes_pbg1} displays the real and imaginary part of the roots of $Q(p)$, for $\nu$=0.1 (left) and $\nu$=1 (right). Other parameters are $p_{bg}$=1, $\tilde{\Delta} x$=0.143, N=63, ($\ol{u},\ol{w}$)=(0,0), and the quintic kernel was used. A continuous solution ($r_3$) is observed for both values of the viscosity. The damping ratio is significantly larger than the oscillating solutions, for a  mode number $\gtrsim$~10. For a high mode numbers, the oscillating solutions show a damping ratio very close to zero, suggesting that high frequency disturbances might be weakly damped. Further investigations, especially concerning the influence of the velocities $\ol{u}$ and $\ol{w}$ are provided when discussing the dispersion curves in the next section. 

\section{Dispersion curves \label{sec_dispersion}}

In this section, the dispersion curves of the WCSPH methods are investigated as outlined in the previous sections. First, the viscosity is set to zero in order to study the non-damped sound propagation and second, the influence of the viscosity is studied. %

In order to quantify the deviation of discrete WCSPH to a continuum, the dispersion curves will be compared to the ones obtained from the linearized Navier-Stokes equations.
In this approach, the linearization of mass and momentum conservation leads to:
\begin{equation}
\psecdv {\delta u} t - c^2 \, \psecdv {\delta u} x = \nu \frac{\partial^3 \, \delta u}{\partial^2 x \, \partial t}
\label{eq_sound_dispersion_continuum} 
\end{equation}
Non-dimensionalizing Eq.~\ref{eq_sound_dispersion_continuum} with the variables $h$ and $c$, and assuming a solution of the form $\delta u(x,t) = \delta_0 \exp[\hat{i} (\omega t  - k x)]$ leads to dispersion relation:
\begin{equation}
\label{eq_continuum_dispersion}
\omega^2 - \hat{i} \nu \omega k^2 - k^2 = 0
\end{equation}
These solutions will compared to the SPH solutions in the following. The stability analysis is made in a temporal sense. The location is arbitrarily fixed at $x$=0 and the temporal evolution is investigated. Note that in this case, the stability is obtained for Im($\omega$)$>$0.

\subsection{Propagation in an inviscid medium}

\subsubsection{Classical WCSPH: sum-SPH}

\begin{figure}[!htb]
	\centering
	\includegraphics[width=\columnwidth,keepaspectratio]{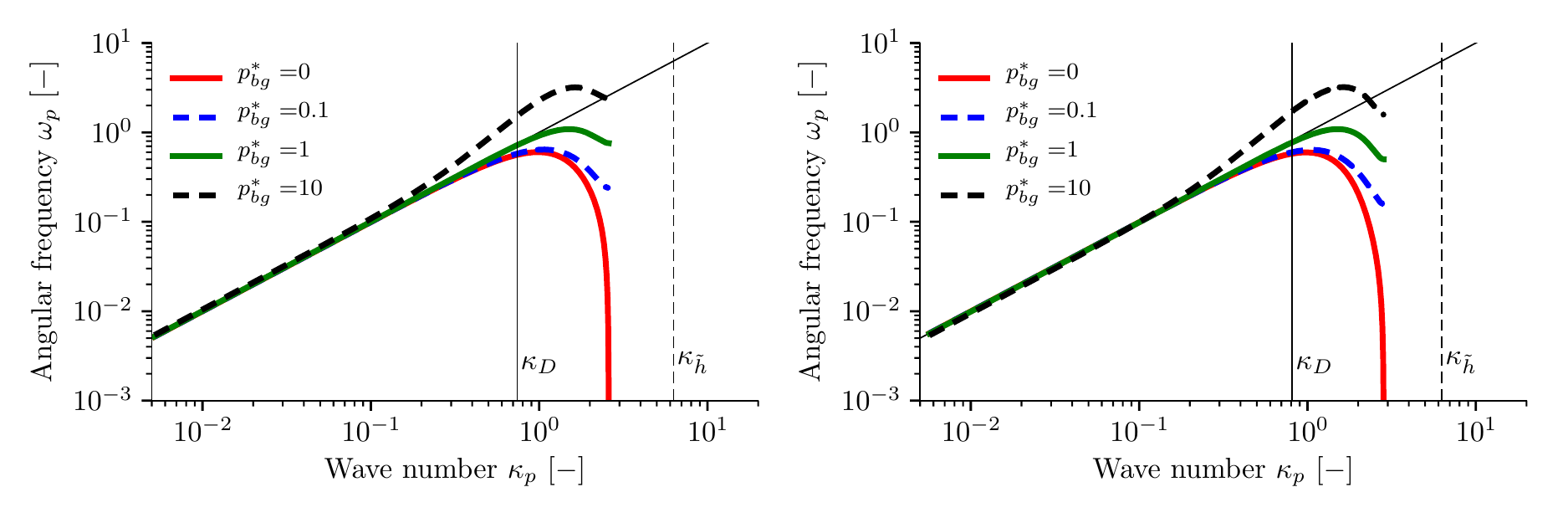}
	\caption{Dispersion curve for $N$=1000 and $\tilde{\Delta} x$=0.143. Left: quintic kernel. Right: Wendland kernel.}
	\label{fig_euler_dispersion01}
\end{figure}

\begin{figure}[!htb]
	\centering
	\includegraphics[width=\columnwidth,keepaspectratio]{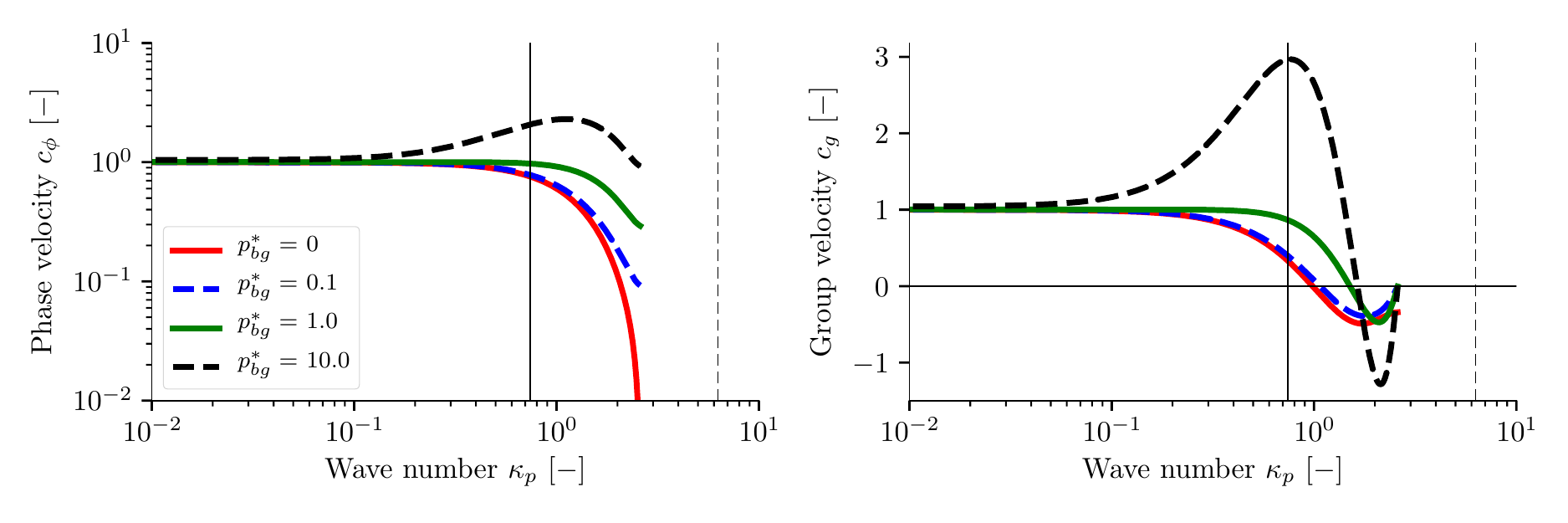}
	\caption{Phase (left) and group (right) velocity  for $N$=1000, $\tilde{\Delta} x$=0.143 and the quintic kernel.}
	\label{fig_euler_phase_group_velo01}
\end{figure}

\begin{figure}[h]
	\centering
	\includegraphics[width=\columnwidth,keepaspectratio]{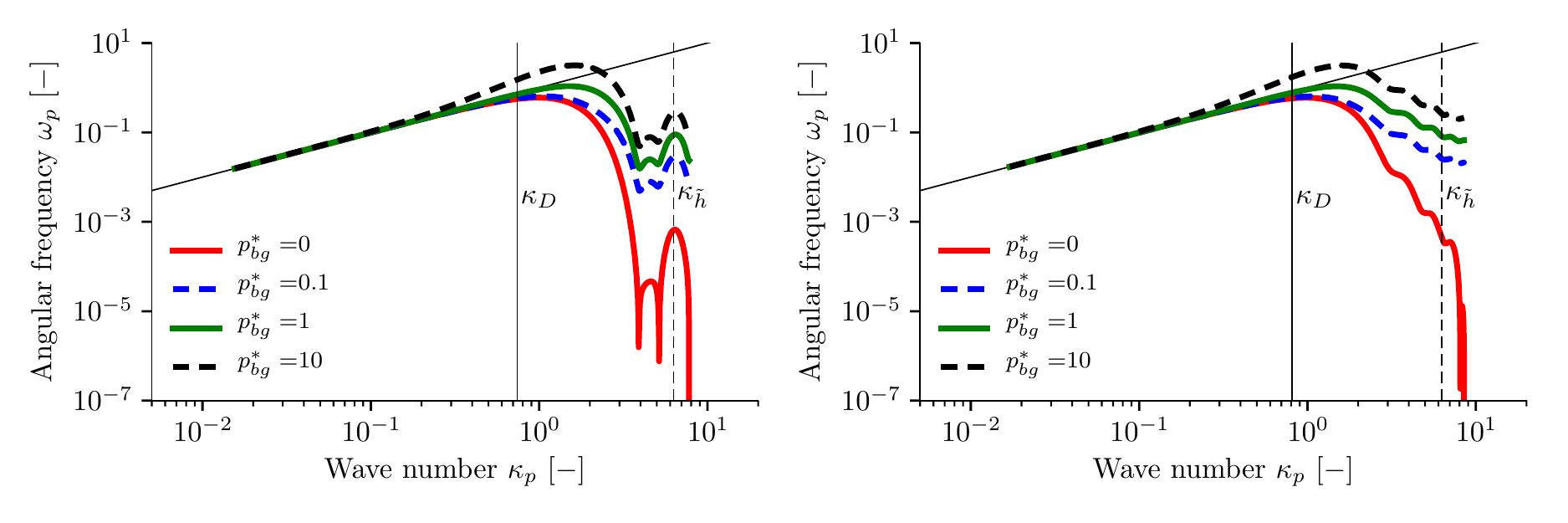}
	\caption{Dispersion curve for $N$=1000 and $\tilde{\Delta} x$=0.048. Left: quintic kernel. Right: Wendland kernel.}
	\label{fig_euler_dispersion02}
\end{figure}

\begin{figure}[h]
	\centering
	\includegraphics[width=\columnwidth,keepaspectratio]{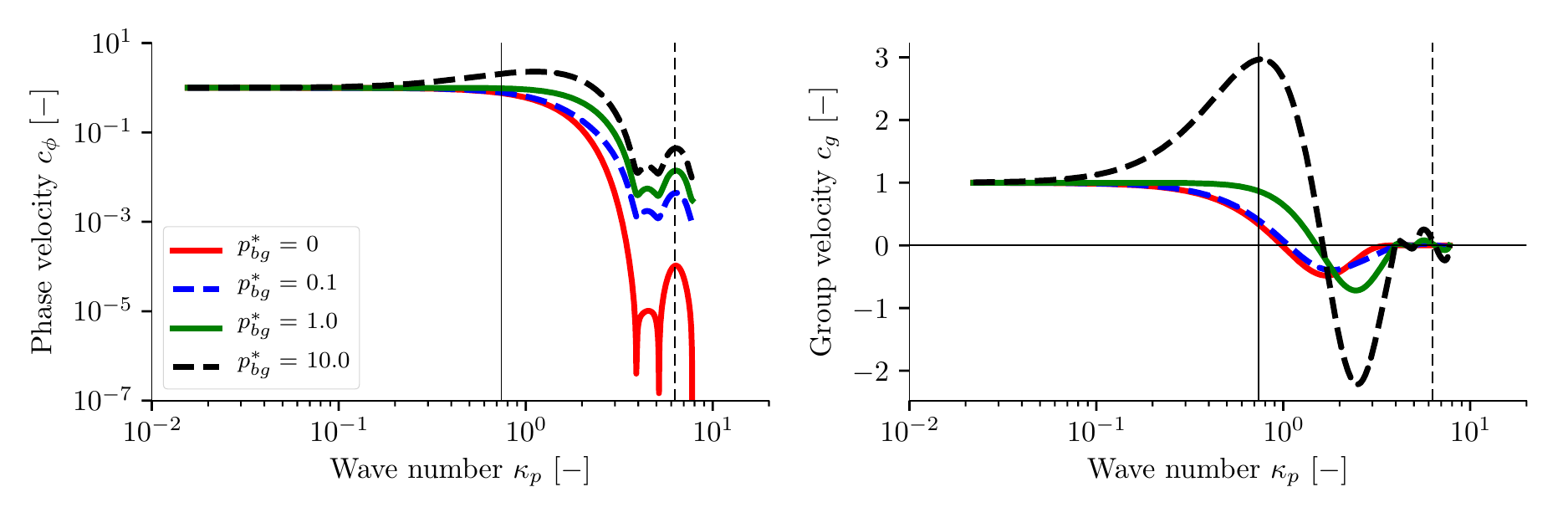}
	\caption{Phase (left) and group (right) velocity  for $N$=1000, $\tilde{\Delta} x$=0.048 and the quintic kernel.}
	\label{fig_euler_phase_group_velo02}
\end{figure}

The dispersion curves are obtained by plotting the angular frequency $\omega_p$ defined in Eq.~\ref{eq_euler_perturb_matrix6} versus the non-dimensional wavenumber $\kappa_p=2\pi \tilde{h}/\lambda_p$.
When the viscosity is zero, the angular frequency is given by $\omega_p = \phi_p$. 
Figure~\ref{fig_euler_dispersion01} displays the dispersion curve for the quintic kernel (left) and the Wendland kernel (right), and different $p_{bg}$. The domain parameters are $N$=1000 and $M$=3. The solid vertical line is related to the diameter of the sphere of influence $D=(2M+1)\Delta x$ and is labeled $\kappa_D$ while the dashed vertical line corresponds to the kernel standard deviation $\tilde{h}$ and is located at $\kappa_{\tilde{h}}$=$2\pi$. The highest resolved wavenumber is $\pi / \Delta x$.
The line of equation $y=x$ is the dispersion curve obtained from Eq.~\ref{eq_continuum_dispersion} and characterizes a non-dispersive medium.\\
First, it is observed that both kernels have a similar behavior. Hence, the superiority of one kernel compared to the other is not obvious here.
\majrev{Second, due to a low number of neighbors, $M=3$, the wavenumber $\kappa_{\tilde{h}}$ related to the kernel standard deviation cannot be resolved by the particle lattice.}\\
The ideal non-dispersive behavior is partly retrieved by the SPH scheme for low wavenumbers up to $\approx\kappa_D$. The upper bound of the ideal behavior depends on the background pressure, and the best case is observed for $p_{bg}$=1, independently of the kernel. For $\kappa>\kappa_D$, the angular frequency is not proportional to the wavenumber anymore. Finally, the spurious mode is observed at large wavenumber ($\kappa \approx \pi/\Delta x$) for $p_{bg}$=0.
\\\\
From the dispersion relation, it is possible to extract the phase velocity $c_{\phi}=\omega / \kappa$ and the group velocity  $c_{g}=\partial \omega / \partial \kappa$.
They are both shown for the quintic kernel in Fig.~\ref{fig_euler_phase_group_velo01}. On the left, the phase velocity is equal to 1 up to $\approx\kappa_D$, where it depends on $p_{bg}$, the best case being $p_{bg}$=1 as previously mentioned.
Contrary to the phase velocity $c_{\phi}$, the group velocity $c_g$ (Fig.~\ref{fig_euler_phase_group_velo01}, right) becomes negative at high wavenumbers. It is observed that a larger background pressure increases the wavenumber at which $c_{g}$ becomes negative.
Note that for a wavenumber leading to $c_g$<0, the previous results (\eg Fig.~\ref{fig_euler_dispersion01}) show no instability. Therefore, a negative group velocity does not imply an instability.
On the other hand, setting $p_{bg}>1$ increases the group velocity for $\kappa \lesssim \kappa_D$, meaning that the pressure information propagates at a speed faster than $c$.
As for the phase velocity, the best compromise is found for $p_{bg}$=1.
\\\\
The dispersion curve for $\tilde{\Delta} x$=0.048 ($M$=10) is shown in Fig.~\ref{fig_euler_dispersion02} for the quintic kernel (left) and the Wendland kernel (right). The same characteristics with regards to the background pressure are observed. However, in this case due to the larger number of neighbors, the particle lattice can resolve wavelengths smaller than $\tilde{h}$.
The spurious modes when $p_{bg}$=0 occurs several time for the quintic kernel, while it occurs only at $\kappa_{max}$ for the Wendland kernel. This suggests a better stability of the Wendland kernel with a large number of neighbors when $p_{bg}$=0.
The corresponding phase and group velocities are depicted in Fig.~\ref{fig_euler_phase_group_velo02} for the quintic kernel. The same findings as for $\tilde{\Delta} x$=0.143 are valid. In addition, for $\kappa>\kappa_{\tilde{h}}$, the group velocity is almost zero.

\subsubsection{Other methods without viscosity}

\begin{figure}[!htb]
	\centering
	\includegraphics[width=0.49\textwidth,keepaspectratio]{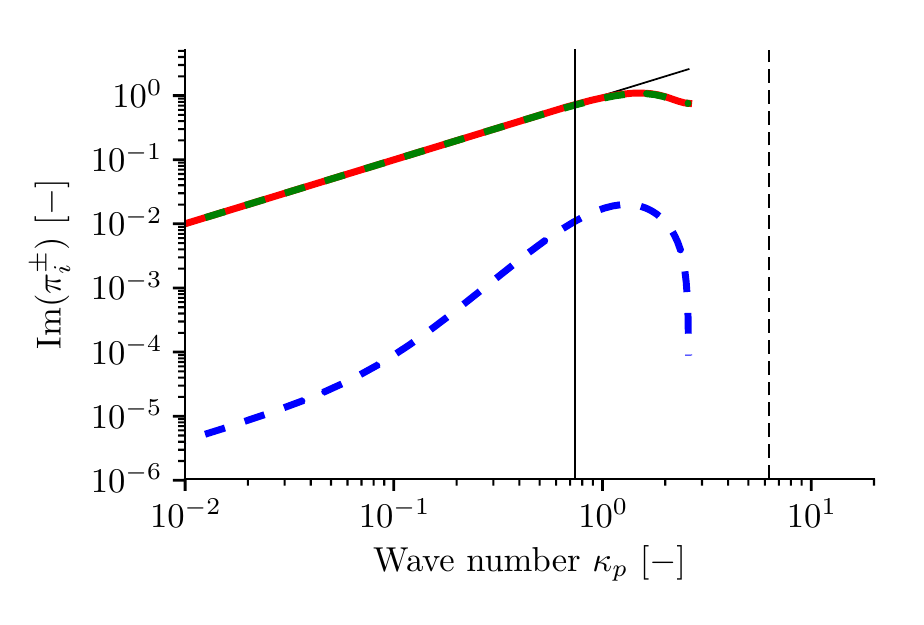}
	\caption{Absolute value of the imaginary part of the roots of $Q$ for $p_{bg}$=1, $\tilde{\Delta} x$=0.143, the quintic kernel and ($\ol{u},\ol{w}$)=(0.05,0)}
	\label{fig_TVSPH_upos_w0}
\end{figure}

When the viscosity is set to zero in div-SPH and $\delta$-SPH, the roots of $Q(p)$, labeled as $\pi_i^{\pm}$, are zero and $\pm \phi_i$, and the dispersion curve of these methods is the same as the one obtained from sum-SPH, with an additional continuous component.
Concerning TV-SPH, it also depends on the mean velocity $\ol{u}$ and the mean velocity difference $\ol{w}$. When both of them are zero, the roots of $Q(p)$ are also 0 and $\pm \phi_i$, leading to the regular dispersion curve plus a continuous component. However, when $\ol{w}$=0 and $\ol{u}>0$, the polynomial $Q$ is:
\begin{equation}
\label{eq_TVSPH_upos_w0_polyQ}
Q(p) = p^3 + \phi_i^2 \, p + 2 \, \Delta x^2 \, p_{bg} \varphi_i \ol{u} \lambda_i' 
\end{equation}
The roots of Eq.~\ref{eq_TVSPH_upos_w0_polyQ} were numerically computed for ($\ol{u},\ol{w}$)=(0.05,0), $p_{bg}$=1, $N$=1000 and $M$=3.
All roots have a real part equal to zero so no damping is found. The imaginary part of the roots of Eq.~\ref{eq_TVSPH_upos_w0_polyQ} are depicted in Fig.~\ref{fig_TVSPH_upos_w0}. It is obvious that, in addition to the regular oscillations following the line $y=x$, there are additional oscillations of lower frequency. This means that when both the physical and transport velocity are close (\ie $\ol{w} \approx 0$), a spatial excitation results in a superposition of several frequencies, which can be regarded as a non-linear behavior.
\\\\
When $\ol{w}\ne0$ and $\ol{u}>0$, the polynomial $Q$ is:
\begin{equation}
\label{eq_TVSPH_upos_wpos_polyQ}
Q(p) = p^3 + \Delta x \ol{w} \lambda_i' \, p^2  
-\Delta x [2 \, \pi_{bg} \, \varphi_i  +\Delta x \lambda_i'^2]   \, p  
-2  \Delta x^2 \, p_{bg}  \varphi_i  ( [\ol{w} - \ol{u}] \lambda_i')
\end{equation}
The roots of Eq.~\ref{eq_TVSPH_upos_wpos_polyQ} were numerically computed for ($\ol{u},\ol{w}$)=(0.05,-0.05), the other parameters being identical to the previous case. Like in the previous case, all roots have a real part equal to zero, and a additional mode of lower frequency is also found.
The particular case $\ol{u}=\ol{w}\ne0$ leads to a polynomial $Q$:
\begin{equation}
\label{eq_TVSPH_u_equal_w_polyQ}
Q(p) = p \left( p^2 + \Delta x \ol{w} \lambda_i' \, p + \phi_i^2  \right)
\end{equation}
Although this polynomial is different from the one obtained for $\ol{u}=\ol{w}=0$, the roots are almost the same and the real part is always zero. %

\subsection{Effects of viscosity}

\subsubsection{sum-SPH}

\begin{figure}[!htb]
	\centering
	\includegraphics[width=0.49\textwidth,keepaspectratio]{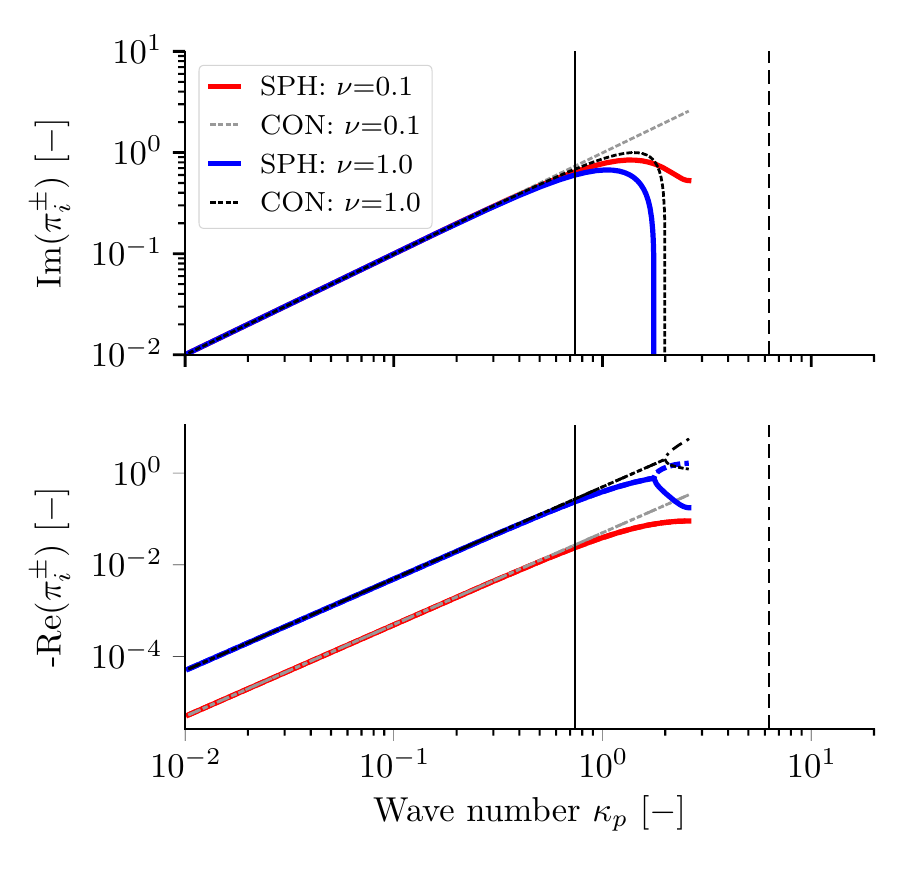}
	\includegraphics[width=0.49\textwidth,keepaspectratio]{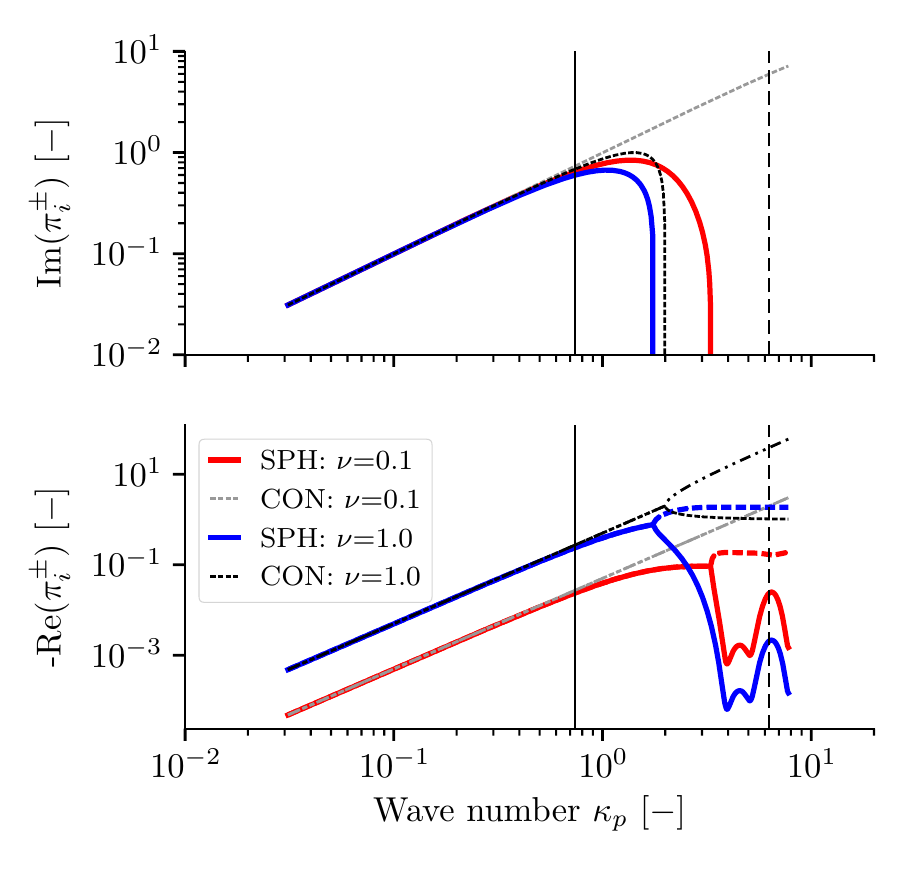}
	\caption{Imaginary and real part of angular frequency $\pi_i^{\pm}$ versus the wavenumber $\kappa$ for the quintic kernel and $p_{bg}$=0.5, superimposed with the continuum solution. For SPH, solid and dashed lines represent $\pi_i^{+}$ and $\pi_i^{-}$, respectively, whereas they are represented by dashed and dot-dashed lines for the continuum solution. Left: $\tilde{\Delta} x$=0.143. Right: $\tilde{\Delta} x$=0.048.}
	\label{fig_classical_dispersion_visco_01}
\end{figure}

Due to the damping ratio in the differential equation of motion, not only the imaginary part of $\pi_i$ but also its real part is presented and compared to the continuum solution.
The solutions of Eq.~\ref{eq_euler_perturb_matrix6} are presented in Fig.~\ref{fig_classical_dispersion_visco_01} for the quintic kernel, $N$=500, $p_{bg}$=0.5, $\tilde{\Delta} x$=0.143 (left) and $\tilde{\Delta} x$=0.048 (right), and compared to the continuum solution, labeled CON in figures.
For $\tilde{\Delta} x$=0.143, $p_{bg}$=0.5 leads to a good prediction in terms of damping ratio\improve{, but the best agreement for  angular frequency is found for $p_{bg}$=1}. For $\tilde{\Delta} x$=0.048, a large viscosity ($\nu$=1) shows a good agreement with the continuum solutions, whereas with $\nu$=0.1, the sum-SPH method predicts damping without oscillations starting at $\kappa \approx \kappa_{\tilde{h}}$ although the continuum solution shows damped oscillations.  The risk of instability is also observed for $\tilde{\Delta} x$=0.048 at $\kappa \approx \kappa_{\tilde{h}}$ where the spurious mode ($\omega_i$=0) shows a very low damping ratio.
Finally, it is to be mentioned that the prefactor of the Laplacian operator in Eq.~\ref{eq_WCSPH01b} was set to 2 \majrev{and not as proposed by \citet{macia2011theoretical}. According to their derivation, the proper pre-factor is $2(n_d + 2)$  where $n_d$ is the number of dimension, which would be equal to 6 here.} In this case, the damping ratio of sum-SPH would be overestimated by a factor of 3.

\begin{figure}[!htb]
	\centering
	\includegraphics[width=0.49\textwidth,keepaspectratio]{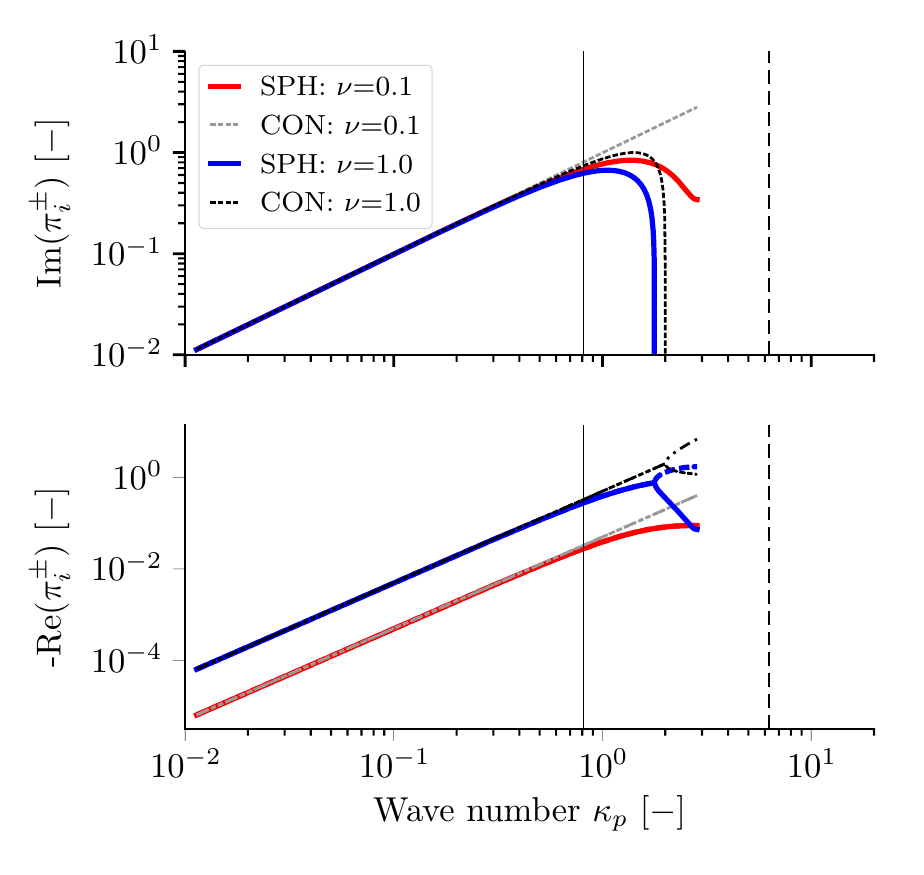}
	\includegraphics[width=0.49\textwidth,keepaspectratio]{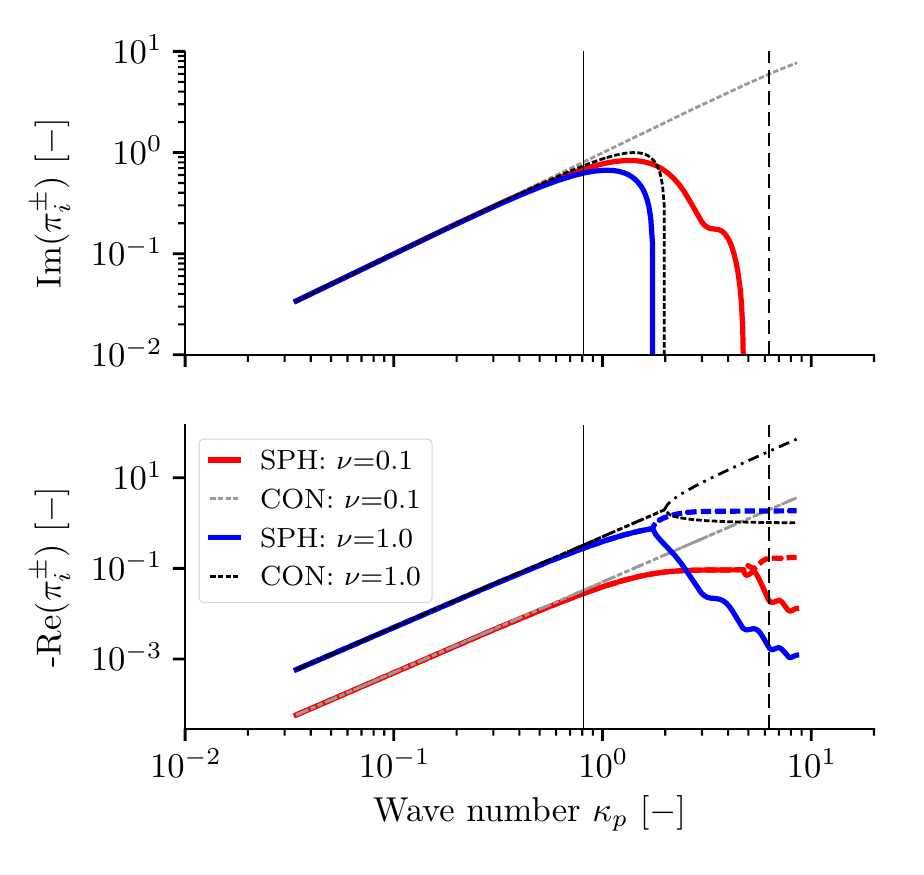}
	\caption{Imaginary and real part of angular frequency $\pi_i^{\pm}$ versus the wavenumber $\kappa$ for the Wendland kernel and $p_{bg}=0.5$, superimposed with the continuum solution. $\tilde{\Delta} x$=0.143 (left) and $\tilde{\Delta} x$=0.048 (right). For SPH, solid and dashed lines represent $\pi_i^{+}$ and $\pi_i^{-}$, respectively, whereas they are represented by dashed and dot-dashed lines for the continuum solution.}
	\label{fig_classical_dispersion_visco_02}
\end{figure}

The same investigation with the Wendland kernel is shown in Fig.~\ref{fig_classical_dispersion_visco_02}. When $\tilde{\Delta} x$=0.143, the Wendland kernel is very similar to the quintic kernel. When $\tilde{\Delta} x$=0.048, at high wavenumber ($\kappa \approx \kappa_{\tilde{h}}$), the Wendland kernel shows a monotonic decrease and a larger damping ratio than for the quintic kernel. For both kernels, the best value of $p_{bg}$ \improve{depends on the quantity to match. For a better prediction of the sound speed $p_{bg}$ should be one, whereas the best agreement for the damping is found for $p_{bg}\approx 0.5$.}

\subsubsection{$\delta$-SPH}

\begin{figure}[!htb]
	\centering
	\includegraphics[width=0.49\textwidth,keepaspectratio]{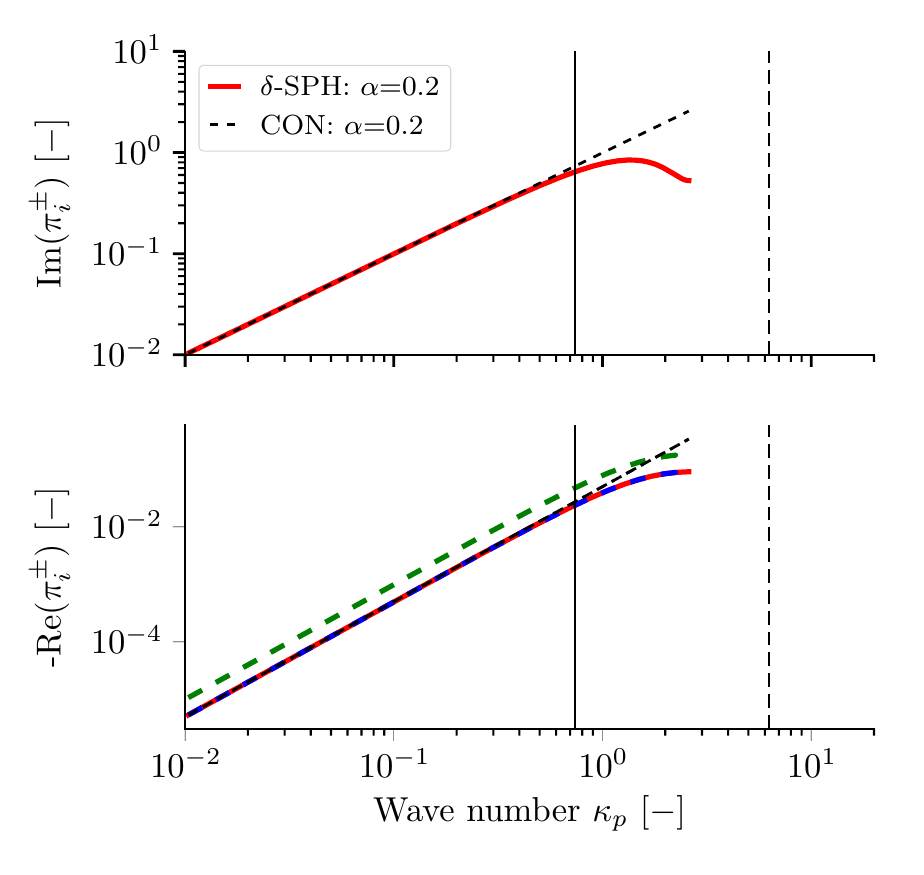}
	\includegraphics[width=0.49\textwidth,keepaspectratio]{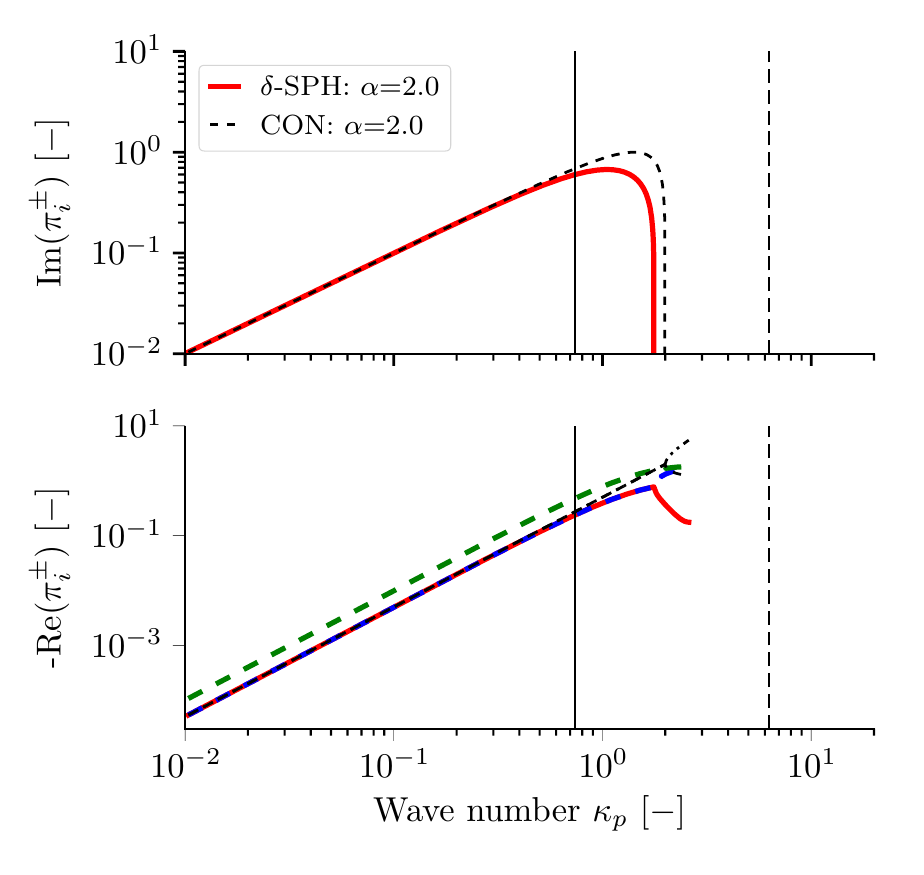}
	\caption{Imaginary and real part of angular frequency $\pi_i^{\pm}$ versus the wavenumber $\kappa$ for the quintic kernel, $p_{bg}=0.5$ and $\tilde{\Delta} x$=0.143, superimposed with the continuum solution. Left: $\alpha$=0.2 .Right: $\alpha$=2. For SPH, solid and dashed lines represent $\pi_i^{+}$ and $\pi_i^{-}$, respectively, whereas they are represented by dashed and dot-dashed lines for the continuum solution. The continuous mode of $\delta$-SPH is represented by a green dashed line.}
	\label{fig_deltaSPH_dispersion_visco_01}
\end{figure}

The dispersion curve of $\delta$-SPH method is presented in this subsection. As the poles of the transfer function cannot be analytically expressed in a compact form, they are computed numerically. Figure~\ref{fig_deltaSPH_dispersion_visco_01} shows the imaginary and real part of the solution for the quintic kernel, $p_{bg}=0.5$, $\Delta x$=0.86 and $\xi=\alpha/4$. Different values of the viscosity $\alpha$=0.2 (left) and $\alpha$=2 (right) were investigated.
The imaginary part of the damped continuous mode identified in Section~\ref{sec_deltaSPH} cannot be shown in a log-log plot because it is equal to zero. Its real part, \ie its damping ratio, is approximately two times larger than the damping ratio of the oscillating modes. \improve{Hence, the unphysical continuous mode is damped faster than physical oscillations.}
\improve{As in sum-SPH, the best agreement with the continuum for the sound speed is $p_{bg}=1$ whereas it is $p_{bg}\approx0.5$ for the damping coefficient.}
Investigations with the Wendland kernel showed very little difference to the quintic kernel. With $\Delta x$=0.29, the findings of sum-SPH apply also to $\delta$-SPH.

\subsubsection{TV-SPH}

\begin{figure}[!htb]
	\centering
	\includegraphics[width=0.49\textwidth,keepaspectratio]{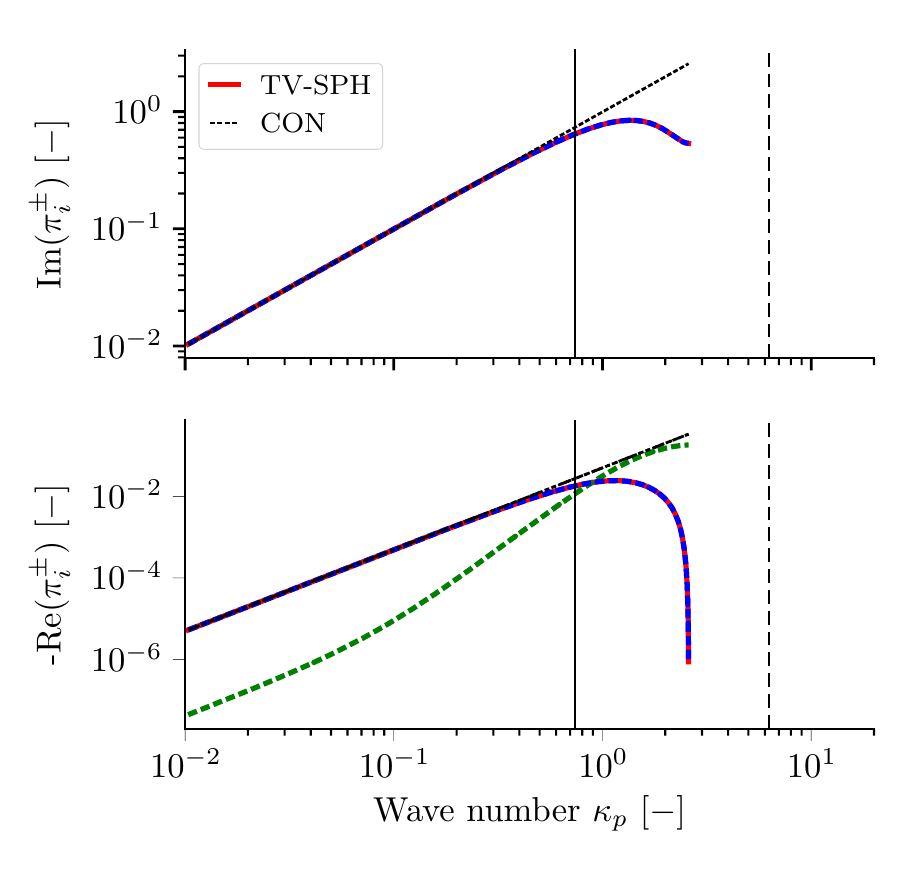}
	\includegraphics[width=0.49\textwidth,keepaspectratio]{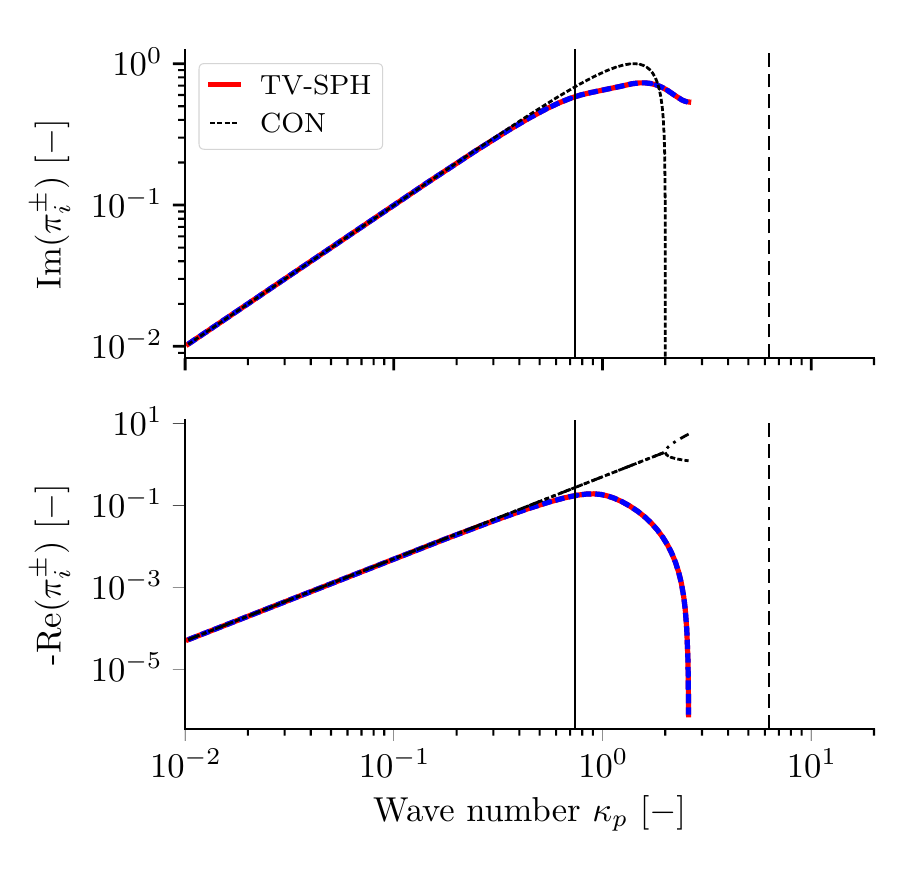}
	\caption{Imaginary and real part of angular frequency $\pi_i^{\pm}$ versus the wavenumber $\kappa$ for the quintic kernel, $p_{bg}=0.5$ and $\tilde{\Delta} x$=0.143, superimposed with the continuum solution. Particular parameters are ($\ol{u}$,$\ol{w}$)=(0,0). Left: $\nu$=0.1 .Right: $\nu$=1. For SPH, solid and dashed lines represent $\pi_i^{+}$ and $\pi_i^{-}$, respectively, while the continuous mode is represented by a green dashed line. For the continuum solution, they are represented by dashed and dot-dashed lines.}
	\label{fig_TVSPH_dispersion_visco_00}
\end{figure}

The poles of the transfer function (Eq.~\ref{eq_TVSPH_Q_p}) are computed numerically. They are depicted in Fig.~\ref{fig_TVSPH_dispersion_visco_00} for the quintic kernel, $p_{bg}$=0.5, $\Delta x$=0.86 and ($\ol{u}$,$\ol{w}$)=(0,0). The artificial continuous mode is still observed (not depicted in the log-log plot of the imaginary part) but it is damped. The damping ratio is significantly lower than that of the continuum solution up to $\kappa_D$, independently of the viscosity.
For $\nu$=1 (Fig.~\ref{fig_TVSPH_dispersion_visco_00} right), large wavenumbers show weakly damped oscillations whereas for the continuum solution a damped aperiodic regime is predicted. This phenomenon is always observed for other values of ($\ol{u}$,$\ol{w}$).
It was also observed but not depicted here, that a lower background pressure decreases the damping ratio of the non-physical mode and increases the damping ratio of the physical oscillations at high wavenumber, and \textit{vice versa}. This suggests that there exist an optimal background pressure for the prediction of acoustic waves with the TV-SPH method.

\begin{figure}[!htb]
	\centering
	\includegraphics[width=0.49\textwidth,keepaspectratio]{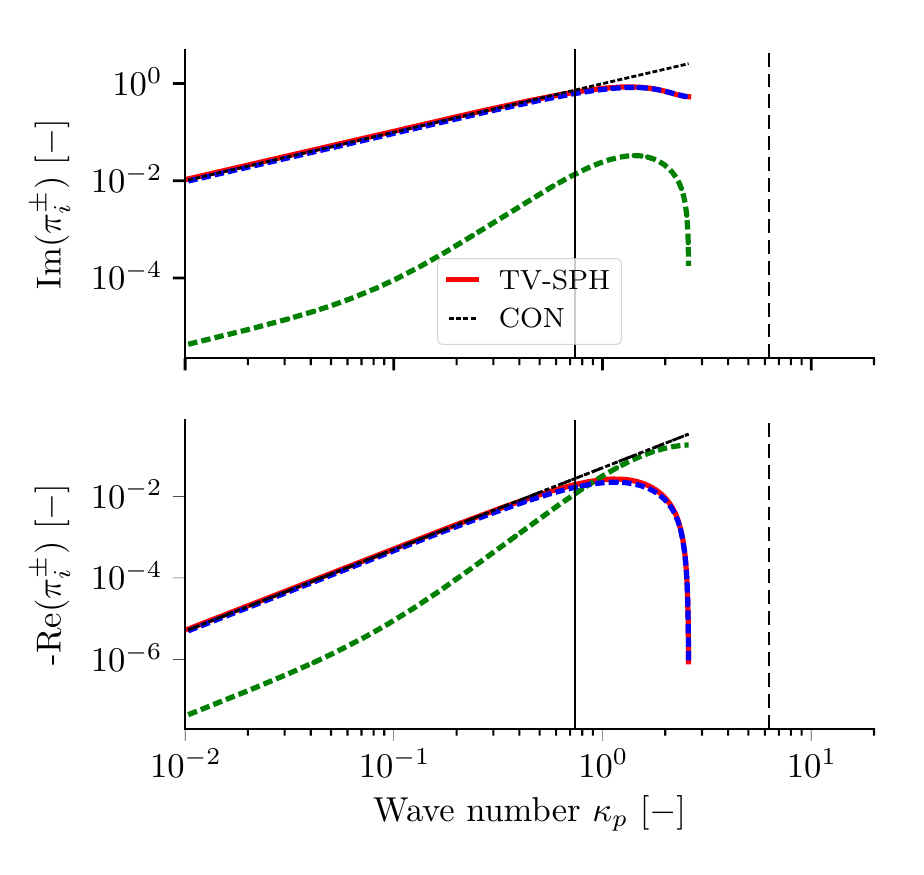}
	\includegraphics[width=0.49\textwidth,keepaspectratio]{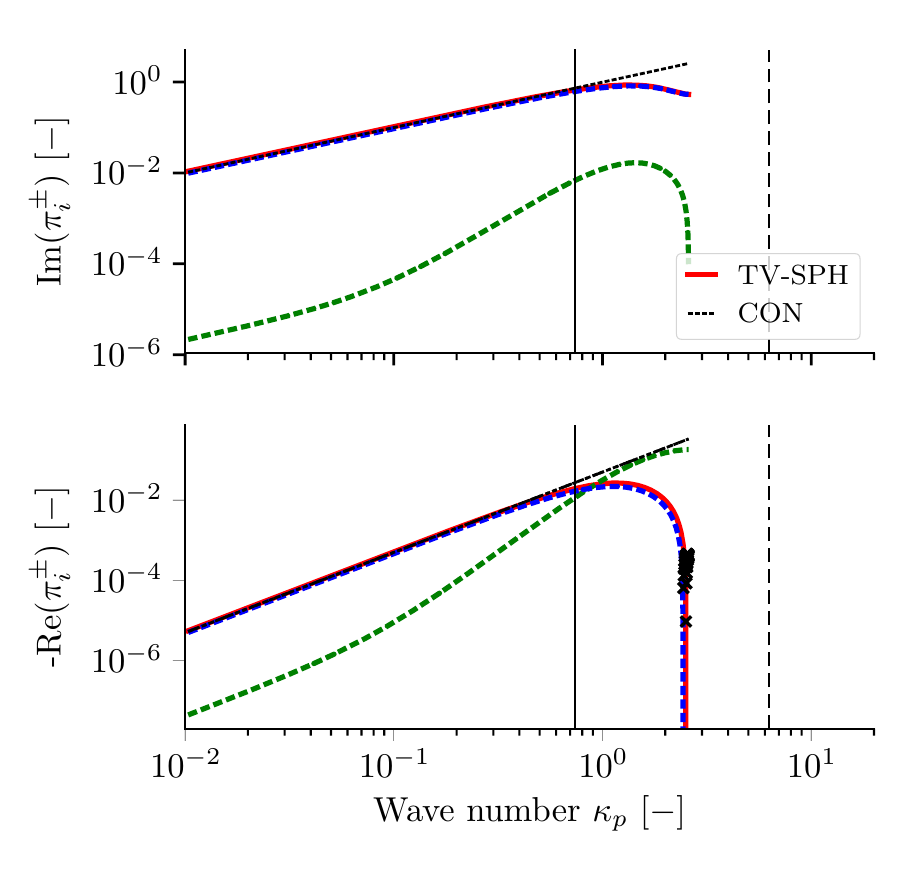}
	\caption{Imaginary and real part of angular frequency $\pi_i^{\pm}$ versus the wavenumber $\kappa$ for the quintic kernel, $p_{bg}=0.5$, $\tilde{\Delta} x$=0.143 and $\nu$=0.1, superimposed with the continuum solution. Left: ($\ol{u}$,$\ol{w}$)=(0,0.1). Right: ($\ol{u}$,$\ol{w}$)=(0.05,0.1). For SPH, solid and dashed lines represent $\pi_i^{+}$ and $\pi_i^{-}$, respectively, while the continuous mode is represented by a green dashed line. For the continuum solution, they are represented by dashed and dot-dashed lines.}
	\label{fig_TVSPH_dispersion_visco_01}
\end{figure}

Solutions for ($\ol{u}$,$\ol{w}$)=(0,0.1) and $\nu$=0.1 are depicted in Fig.~\ref{fig_TVSPH_dispersion_visco_01} (left). 
In this case, the non-physical continuous mode is turned into a low-frequency mode and shows the same damping characteristics as in Fig.~\ref{fig_TVSPH_dispersion_visco_00}.
The result of a positive fluid velocity ($\ol{u}$,$\ol{w}$)=(0.05,0.1) is 
shown in Fig.~\ref{fig_TVSPH_dispersion_visco_01} (right) and highlights the presence of instabilities at large wavenumber. They are depicted by cross symbols in Fig.~\ref{fig_TVSPH_dispersion_visco_01} and represent $+\Re(\pi_i)$ instead of $-\Re(\pi_i)$. This instability is illustrated in \ref{appendix_validation}.

\section{Illustration of the spurious mode with a large number of neighbors \label{sec_illus_spurious}}

\begin{figure}[h]
	\centering
	\includegraphics[width=\textwidth,keepaspectratio]{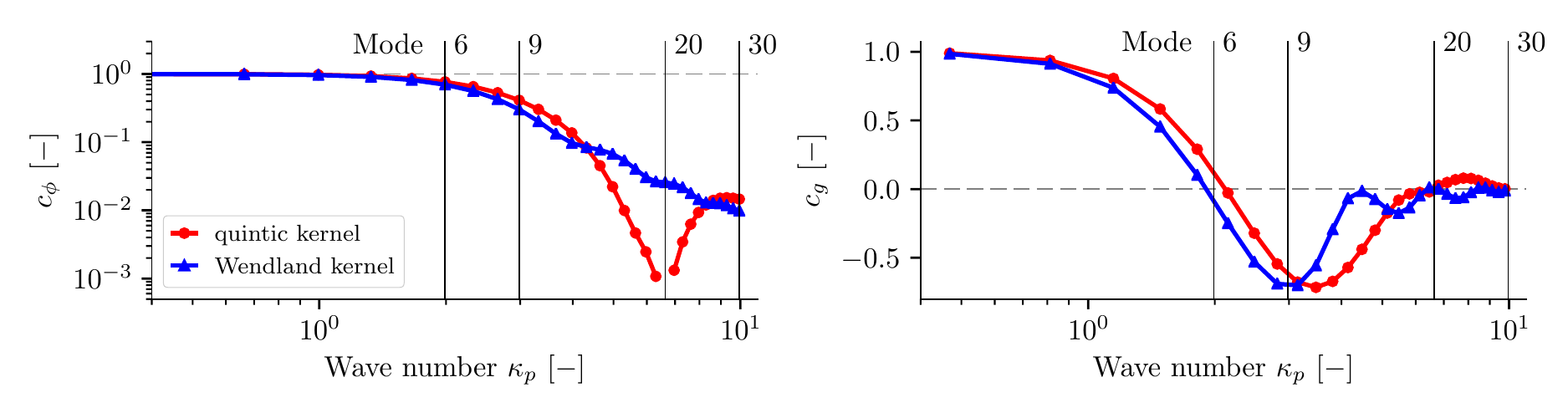}
	\caption{Dispersion curve for ($N$,$M$)=(60,9). Left: Phase velocity. Right: Group velocity. The vertical lines correspond to mode 6, 9, 20 and 30}
	\label{fig_dispersion_curve_N60M9}
\end{figure}

This section illustrates the difference between the quintic and the Wendland kernel if a large number of neighbor particles are present (M=9 $\Leftrightarrow \Delta x$=0.32) with $p_{bg}$=1 and $\nu$=0.
Special emphasis is put on the spurious mode as observed in Fig.~\ref{fig_phi_i2_map_M_vs_modes}.
The number of particle $N$ is 60. The dispersion curves $c_\phi$ and $c_g$ are plotted for the quintic and the Wendland kernel in Fig.~\ref{fig_dispersion_curve_N60M9} with several modes that represent different states.
\majrev{Note that in this figure, the normalizing length scale is $D$ in order to display the modes at the same position independently from the kernel.}
Mode 6 corresponds to a phase velocity close to the continuum solution and a group velocity $\approx$0, mode 9 shows a negative group velocity of magnitude $\approx$1.
Mode 20 and 30 have a phase velocity $\ll$ 1. In particular, mode 20 is the spurious mode for the quintic kernel. Mode 30 shows a group velocity $\approx$0.\\

\begin{figure}[!h]
	\centering
	\includegraphics[width=0.49\textwidth,keepaspectratio]{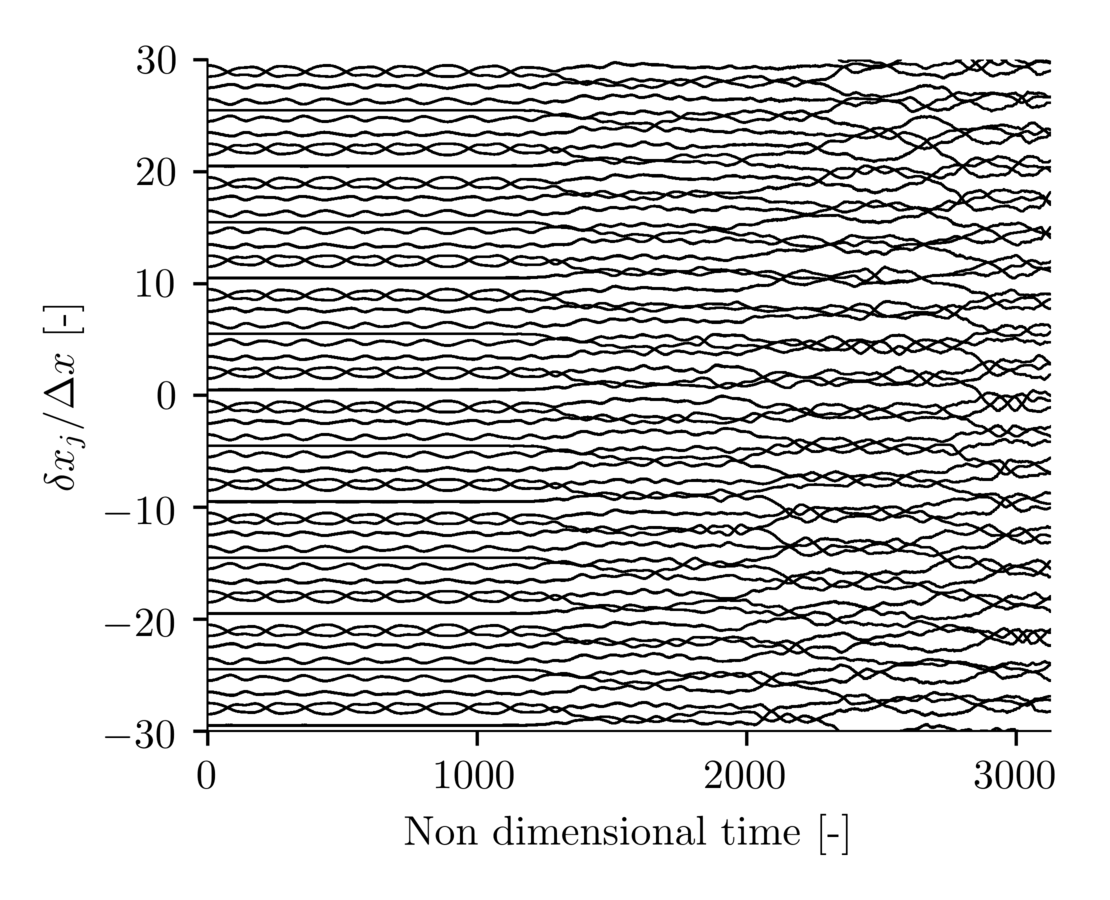}
	\includegraphics[width=0.49\textwidth,keepaspectratio]{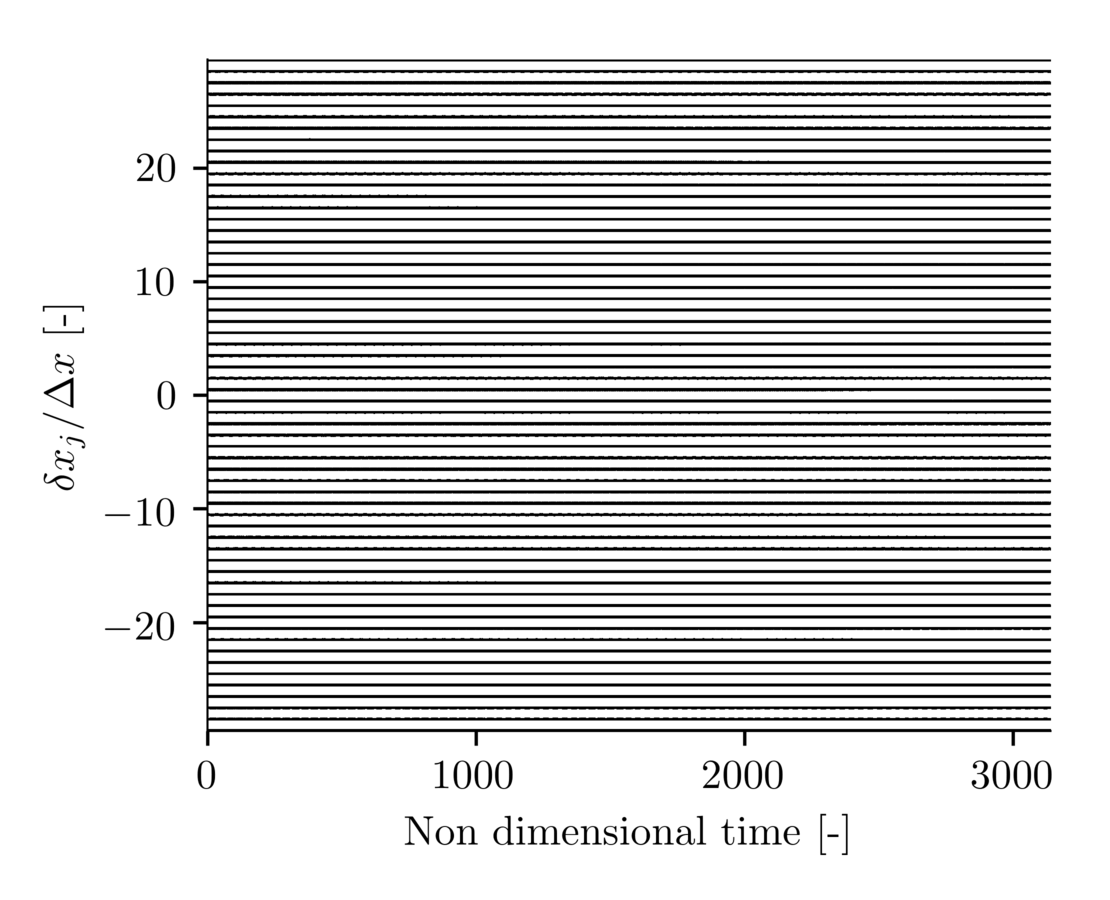}
	\caption{Mode 9: Graph of the particles position for $N$=60. Left: quintic kernel. Right: Wendland Kernel.}
	\label{fig_validate_spurious_mode9}
\end{figure}

\begin{figure}[!h]
	\centering
	\includegraphics[width=0.49\textwidth,keepaspectratio]{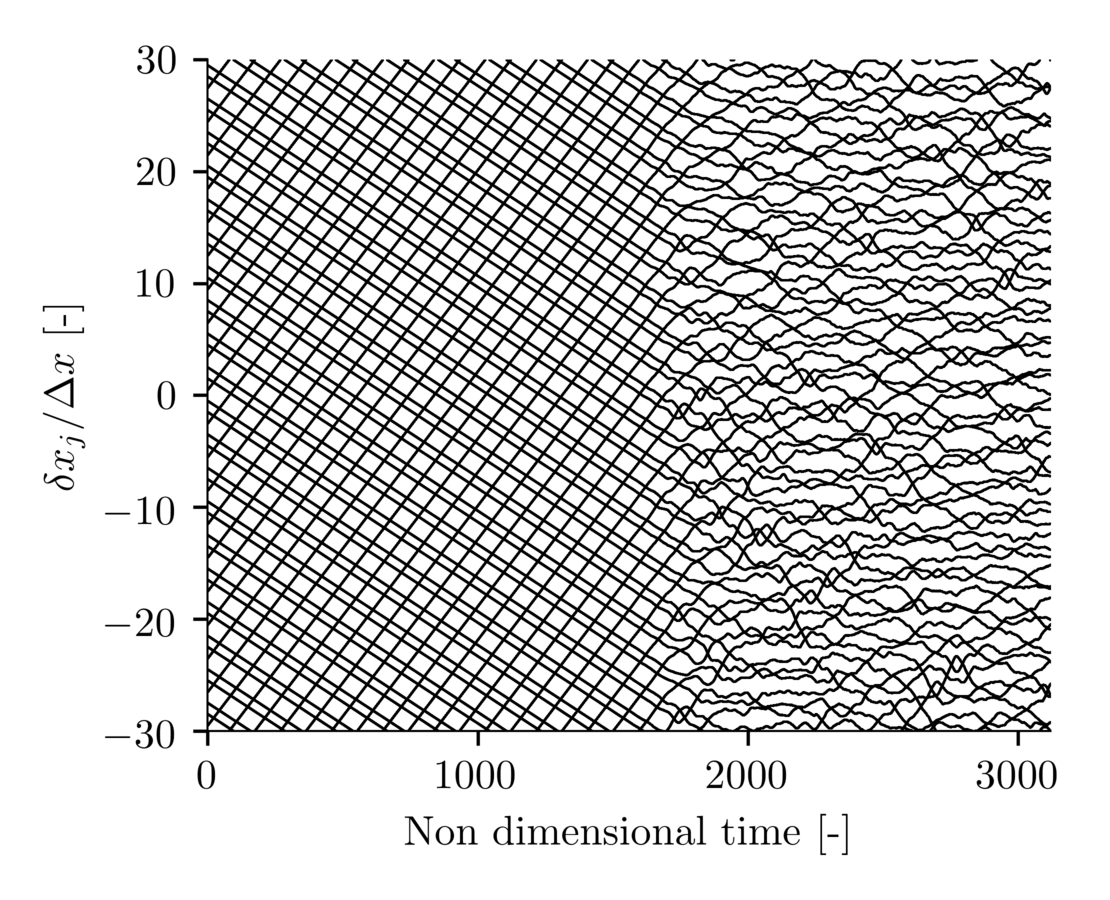}
	\includegraphics[width=0.49\textwidth,keepaspectratio]{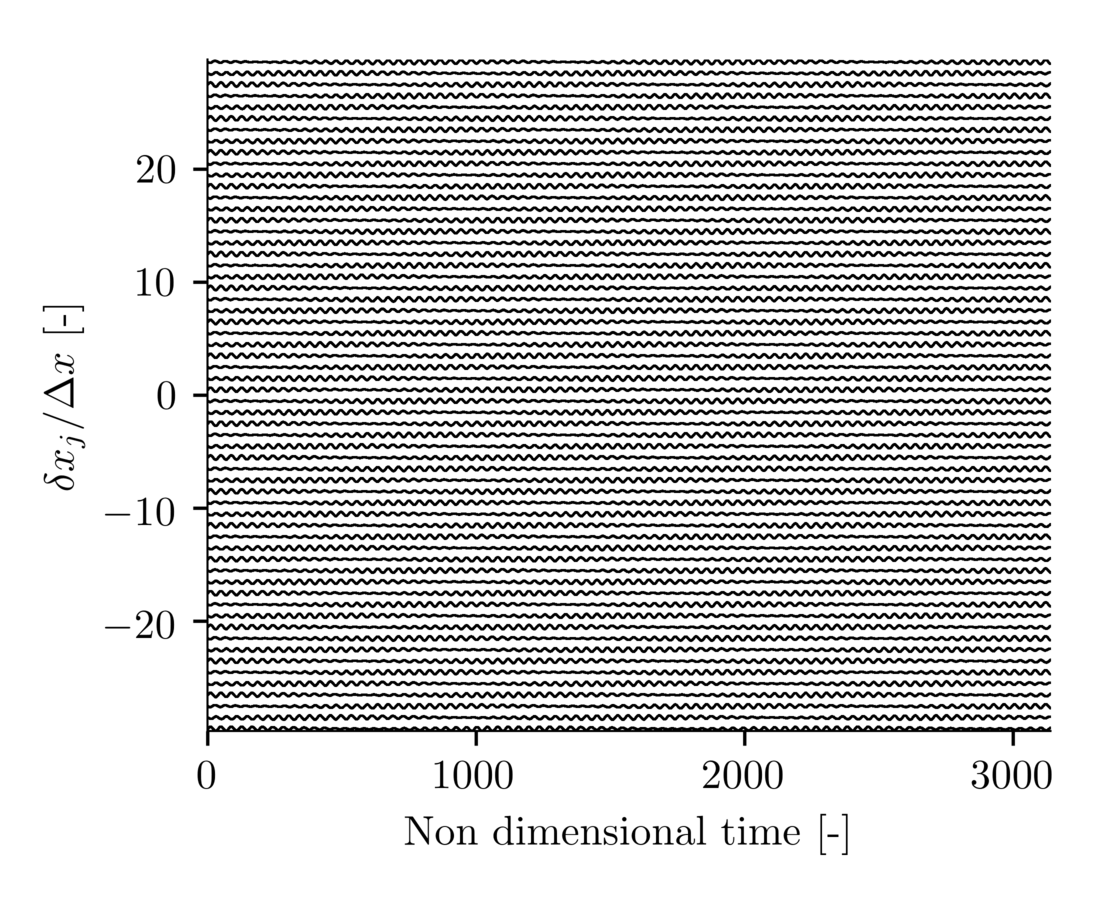}
	\caption{Mode 20: Graph of the particles position for $N$=60. Left: quintic kernel. Right: Wendland Kernel.}
	\label{fig_validate_spurious_mode20}
\end{figure}

\begin{figure}[!h]
	\centering
	\includegraphics[width=0.49\textwidth,keepaspectratio]{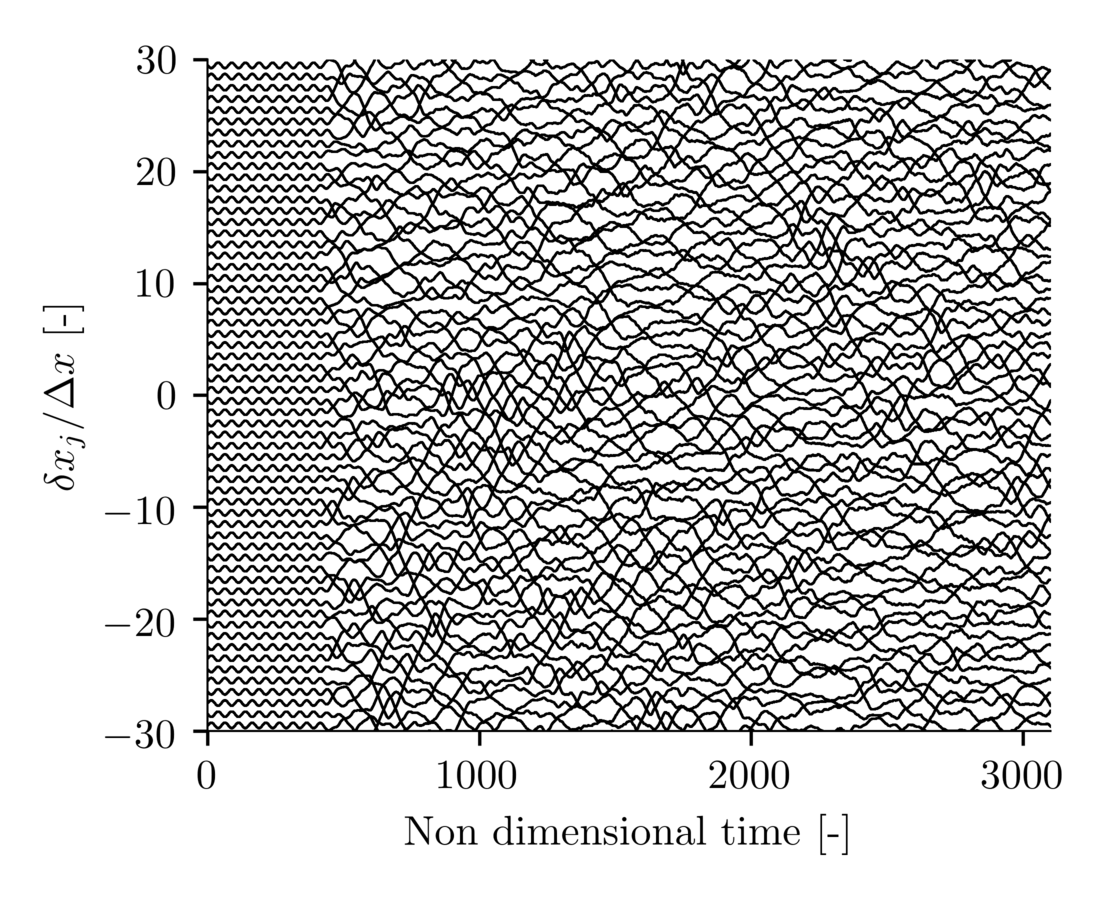}
	\includegraphics[width=0.49\textwidth,keepaspectratio]{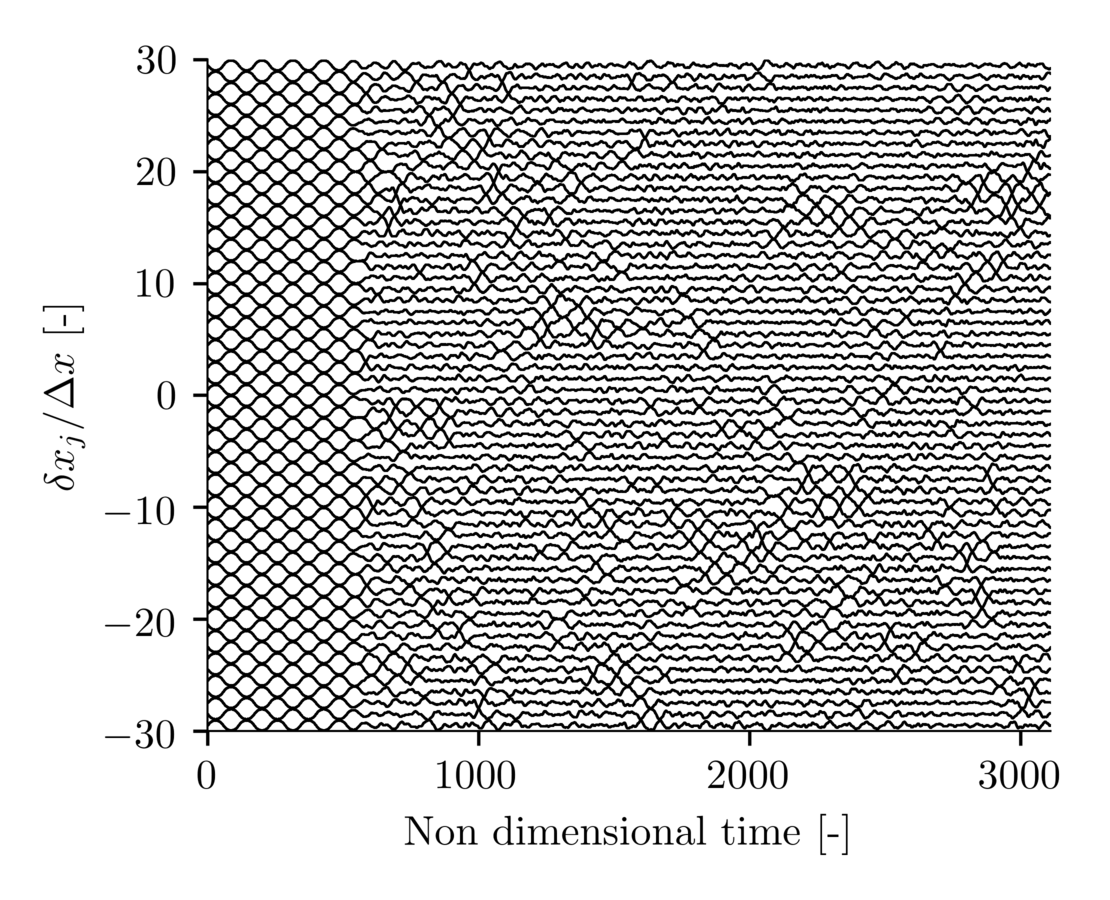}
	\caption{Mode 30: Graph of the particles position for $N$=60. Left: quintic kernel. Right: Wendland Kernel.}
	\label{fig_validate_spurious_mode30}
\end{figure}

In order to compare the LSA results with a real simulation, 1D sum-SPH simulations were run. The same numbers ($N$,$M$) were used and $p_{bg}$=1 and $\nu$=0 were set. \majrev{The rest of the numerical parameter are the same as those presented in \ref{appendix_validation}.}
An initial excitation velocity of 1\% of the speed of sound was applied, and the simulations were run for a long non-dimensional time. The focus is put on the long time instability with a large deviation from the equilibrium state. Hence, the temporal solution of the linearized perturbation may not apply.\\
In Fig.~\ref{fig_validate_spurious_mode9} to \ref{fig_validate_spurious_mode30} the normalized position of each particle is plotted versus non-dimensional time, for the quintic kernel (left figures) and the Wendland kernel (right figures). \\
With the mode 6, both kernels show a stable behavior even though the group velocity is almost zero. Therefore, the trajectories are not shown.
At mode 9 (Fig.~\ref{fig_validate_spurious_mode9}), the quintic kernel shows long time instability whereas the Wendland kernel is stable.
At mode 20 (Fig.~\ref{fig_validate_spurious_mode20}), the spurious mode of the quintic kernel is visible with all particles moving at their initial velocity, as predicted by Eq.~\ref{eq_spurious_inviscid}, until $t \approx$~1600.
The trajectories show that particles are crossing each other, in such an orchestrated way so that neither the pressure gradient nor the background pressure can counteract this motion. 
This is striking because even though the temporal solutions of the linearized equations (Eq.~\ref{eq_spurious_inviscid}) may not apply due to the large deviation from the initial position, the behavior highlighted in Fig.~\ref{fig_validate_spurious_mode20} still corresponds to the linearized solution. As observed in Fig.~\ref{fig_phi_i2_map_M_vs_modes}, the Wendland kernel is not subject to the spurious mode, and provides a stable solution.
For mode 30 (Fig.~\ref{fig_validate_spurious_mode30}), both kernels shows regular oscillations until $t \approx$~500. For $t \gtrsim$~500, the quintic kernel shows an unstable behavior with particle trajectories crossing each other whereas the Wendland kernel shows stable oscillations. These plots underline the more stable behavior of the Wendland kernel in case of large initial deviation from equilibrium. It also shows one more time that a negative group velocity does necessary lead to instability.

\section{Conclusion}

In this study, a \improve{new method} for studying the linear stability of the SPH method was presented and applied to different WCSPH methods on a 1D periodic domain. \improve{This method does not make any assumptions on the type of the perturbations and allows to resolve the transient temporal evolution of the perturbations}.
The equations of perturbations are written as a matrix ordinary differential equation in time where the coefficients are circulant matrices. The different SPH operators such a the gradient and the Laplacian can be represented by circulant matrices. Due to the properties of circulant matrices, the diagonalization of the system is equivalent to applying a spatial discrete Fourier transform. %
It is shown that the stability of the system depends on the discrete Fourier transform of the first and second derivative of the kernel, and the damping is determined by the discrete Fourier transform of $W'(x)/x$.

Four different WCSPH methods, namely sum-SPH, div-SPH, $\delta$-SPH and TV-SPH, were studied \majrev{with the help of circulant matrices.
All the derivations were validated with numerical simulations. It was also observed that some results obtained with linear approximations can hold even with large non-linear perturbations.}
When the background pressure is positive and of the order of magnitude of $\rho c^2$, no tensile nor pairing instability was found. 
\majrev{However, it was found for the first time with sum-SPH and div-SPH that when the background pressure exceeds a critical value, an instability arises. This critical background pressure was not investigated with $\delta$-SPH and TV-SPH.\\
The resolution of the transient state showed that with div-SPH, a perturbation of position and/or density would remain unaffected, which might explain the necessity to renormalize the density with this method. Therefore, it is advised to avoid the use of div-SPH and use $\delta$-SPH instead.}
Also, the TV-SPH method showed a slowly growing instability under particular conditions.\\ %
The dispersion curve of the four investigated methods was plotted for inviscid and viscous flows. It was observed that the sum-SPH method provides the best agreement to the continuum solution. 
\improve{The $\delta$-SPH method also provides a good prediction, because, due to its additional diffusion terms, the artificial continuous mode inherited from div-SPH is damped faster than physical oscillations.}
\improve{For these two methods, it was also found that the background pressure must be set to $\rho c^2$ to predict the correct sound speed over the largest range of wavelength. However, when the viscous effects are taken into account, the best damping ratio was found for a background pressure $\approx \rho c^2/2$.}
With a large number of neighbors, the Wendland kernel showed a more stable behavior featuring less spurious modes than the quintic kernel.
\\It was also demonstrated for the first time that the background pressure in combination with the uncorrected gradient operator counteracts any perturbations of equilibrium. In other words, particles will rearrange in a manner to counteract the particle disorder.
Also, the present results is different from those of \citet{fulk1994numerical} and \citet{swegle1995smoothed} who found that the stability depends on the second derivative of the kernel only, whereas our study showed that the stability depends on the first and second derivatives of the kernel.
\\The fact that no tensile nor pairing instabilities were found with sum-SPH and div-SPH can be explained by the fact that the base mean flow is incompressible.
The next steps of this study are 
(i) \majrev{to investigate the influence of the integration scheme, as done by \citet{violeau2014maximum}, and}
(ii) to apply the circulant matrix decomposition on a 2D configuration, in order to study the effect of a transverse instability.

\section{Acknowledgement \label{sec_mercitoutlemonde}}
The Authors like to thank the Helmholtz Association of German Research Centres (HGF) for funding (Grant No. 34.14.02).

\section{Nomenclature \label{sec_nomenclature}}

{\small

\setlength\tabcolsep{10pt}
\tablefirsthead{\textbf{Roman letter} & \textbf{Unit} & \textbf{Description} \\}
\begin{supertabular}{l l l}
	 $a$ & [$-$] & Particle of interest \\
	 $b$ & [$-$] & Neighbor particle \\
	$c$ & [\SI[per-mode=symbol]{}{\meter\per\second}]  & Speed of sound\\
	 $f$ & [$-$] & Temporal function of the initial position \\
	 $g$ & [$-$] & Temporal function of the initial velocity \\
	 $h$ & [$-$] & Temporal function of the initial density \\
	$h$ & [\SI{}{\meter}] & Smoothing length \\
	$\tilde{h}$ & [\SI{}{\meter}] & \majrev{Standard deviation of the kernel}\\
	$i,j$ & [$-$] & Indices \\
	$\hat{i}$ & [$-$] & Imaginary number \\
	$k$ & [\SI[per-mode=symbol]{1}{\per\meter}]  & Wave vector \\	
	$m$ & [\SI{}{\kilogram}] & Mass\\
	$p$ & [\SI{}{\pascal}] & Pressure \\
	$p$ & [$-$] & Laplace variable \\
	$r$ & [$-$] & Roots of a polynomial \\
	$u$ & [\SI[per-mode=symbol]{}{\meter\per\second}]  & Velocity \\
	$w$ & [\SI[per-mode=symbol]{}{\meter\per\second}]  & Relative velocity in TV-SPH\\
	$x$ & [\SI{}{\meter}] & Position \\
	$y$ & [$-$] & Position perturbation in the modal space \\
	$y_i$ & [$-$] & General term of the position perturbation vector \\
	$\uu{\uu{A}}'$ & [\SI[per-mode=symbol]{1}{\per\square\meter}] & Circulant matrix related compressibility \\
	$\uu{\uu{A}}''$ & [\SI[per-mode=symbol]{1}{\per\cubic\meter}] & Circulant matrix related to background pressure\\
	$\uu{\uu{A}}'''$ & [\SI[per-mode=symbol]{1}{\per\cubic\meter}] & Circulant matrix related to viscosity \\
	 $M$ & [$-$] & Number of neighbors on one side \\
	 $N$ & [$-$] & Total number of particles \\
	$\uu{\uu{P}}$ & [$-$] & Change-of-basis matrix \\
	$\uu{P}$ &  [\SI{}{\pascal}] & Vector of pressure perturbation\\
	$P(p)$ &  [$-$] &  Numerator of a rational function \\
	$Q(p)$ &  [$-$] &  Denominator of a rational function \\
	$\uu{R}$ &  [\SI[per-mode=symbol]{}{\kilogram\per\cubic\meter}] & Vector of density perturbation\\
	$V$ &  [\SI{}{\cubic\meter}] & Volume, length in 1D \\
	$W$ & [\SI[per-mode=symbol]{1}{\per\meter}] & Kernel \\
	$\uu{X}$ &  [\SI{}{\meter}] & Vector of position perturbation\\
	$\uu{Y}$ &  [$-$] & Vector of position perturbation in the modal space \\
	$Y_i$ &  [$-$] & Position perturbation in the Laplace domain \\
	$Z_i$ &  [$-$] & Density perturbation in the Laplace domain \\
\end{supertabular}

\tablefirsthead{\textbf{Greek letter} & \textbf{Unit} & \textbf{Description} \\}
\begin{supertabular}{l l l}
	$\gamma$ & [$-$] & Polytropic ratio \\
	$\delta$ & [$-$] & Perturbation \\
	$\kappa$ & [$-$] & Wave number \\
	$\lambda_p$ & [$-$] & Wave length \\
	$\lambda'$ &  [\SI[per-mode=symbol]{1}{\per\square\meter}] & Eigenvalue of $\uu{\uu{A}}'$ \\
	$\lambda''$ &  [\SI[per-mode=symbol]{1}{\per\cubic\meter}] & Eigenvalue of $\uu{\uu{A}}''$ \\
	$\lambda'''$ &  [\SI[per-mode=symbol]{1}{\per\cubic\meter}] & Eigenvalue of $\uu{\uu{A}}'''$ \\
	$\nu$ &  [\SI[per-mode=symbol]{}{\square\meter\per\second}] & Kinematic viscosity \\
	$\rho$ &  [\SI[per-mode=symbol]{}{\kilogram\per\cubic\meter}] & Density \\
	$\phi_i$ &  [$-$] & Undamped angular frequency \\
	$\psi_{1,i}$ &  [$-$] & First term of $\phi_i^2$ \\
	 $\psi_{2,i}$ & [$-$] & Second term of $\phi_i^2$ \\
	$\omega$ &  [$-$] & Angular frequency \\
	$\Delta_i$ &  [$-$] & Discriminant \\
	$\Delta t$ &  [\SI[per-mode=symbol]{}{\meter}] & Time step \\
	$\Delta x$ &  [\SI[per-mode=symbol]{}{\meter}] & Particle spacing\\
	$\uu{\uu{\Delta}}'$ &  [$-$] & Diagonal matrix equivalent to $\uu{\uu{A}}'$  \\
	$\uu{\uu{\Delta}}''$ &  [$-$] & Diagonal matrix equivalent to $\uu{\uu{A}}''$  \\
	$\uu{\uu{\Delta}}'''$ &  [$-$] & Diagonal matrix equivalent to $\uu{\uu{A}}'''$  \\
	$\Upsilon_i$ &  [$-$] & Velocity perturbation in the Laplace domain \\
\end{supertabular}

}

\section{References \label{sec_biblio}}
\bibliography{SPH_LSA_1D.bbl}

\appendix

\section{Expressing perturbation equation as a matrix system \label{appendix_circulant_matrix}}

This section illustrates how the perturbation equations can be cast into a matrix system for a simplified case. Let us consider Eq.~\ref{eq_sumSPH_momentum} where only the term proportional to $W''$ is present:
\begin{equation}
\label{eq_pback_single_effect_3}
\dsecdv {\delta x_a} t = - 2 \, \frac{\ol{p} \,  \ol{V}}{\ol{\rho}} \, \sum\limits_{b}
 (\delta x_{ab}) \, W''_{\overline{ab}}
\end{equation}
Focusing on particle $0$ at the left of the domain (see Fig.~\ref{fig_1D_domain_01}) and considering a sphere of influence containing 5 particles (\ie M=2), the equation of motion is explicitly expressed as:
\begin{align}
\frac{\ol{\rho}}  {2 \, \ol{p} \, \ol{V}} \, \dsecdv {\delta x_0} t 
= + & \ \delta x_{N-2} \  W''(2 \Delta x)  \nonumber \nonumber \\
+ & \ \delta x_{N-1} \  W''(\Delta x) \nonumber \\
-  & \  \delta x_0 \  [ W''(2\Delta x) + W''(\Delta x) + W''(-\Delta x) + W''(-2\Delta x) ] \nonumber \\
+ & \ \delta x_{1} \  W''(-\Delta x) \nonumber \\
+ & \ \delta x_{2} \  W''(-2\Delta x)
\end{align}
Gathering all particle perturbations into the vector $\uu{X} = (\delta x_0, \delta x_1, \delta x_2, ... , \delta x_{n-1})^T$, Eq.~\ref{eq_pback_single_effect_3} can be expressed as a matrix system:
\begin{equation}
\dsecdv {} t  ({\uu{X}}) = 2 \, \frac{\ol{p} \, \ol{V}}{\ol{\rho}} \,  \uu{\uu{A}}'' \,  \uu{X}
\label{eq_motion_simple_example}
\end{equation}
where the matrix $\uu{\uu{A}}''$ is given by:
{\footnotesize
\begingroup
	\renewcommand*{\arraystretch}{1.5}
	\begin{equation*}
	\uu{\uu{A}}'' = \left[ \begin{array}{c c c c c c c c c c}
	c_0 & W''(-\Delta x) &W''(-2 \, \Delta x) & 0 &... & 0 & W''(2 \, \Delta x) & W''(\Delta x) \\
	W''(\Delta x) & c_0 & W''(-\Delta x) &W''(-2 \, \Delta x) & 0 &... & 0 & W''(2 \, \Delta x)  \\
	W''(2 \, \Delta x) &W''(\Delta x) & c_0 & W''(-\Delta x) &W''(-2 \, \Delta x) & 0 &... & 0  \\
	... & ...& ...& ...&... &... &... & ...\\
	W''(-2 \, \Delta x) & 0 &... & 0 & W''(2 \, \Delta x) & W''(\Delta x) & c_0  & W''(-\Delta x)\\
	W''(-\Delta x) &W''(-2 \, \Delta x)  & 0 &... & 0 & W''(2 \, \Delta x) & W''(\Delta x) & c_0 \\
	\end{array} \right ] 
	\end{equation*}
\endgroup
}
with:
\begin{equation}
c_0 = - [ W''(-2\Delta x) + W''(-\Delta x) + W''(\Delta x) + W''(2\Delta x) ]
\label{eq_c0}
\end{equation}
$\uu{\uu{A}}''$ is a circulant matrix which is entirely defined by one line or one column. For instance, the vector of the first row:
\begin{equation}
C = (c_0, W''(-\Delta x), W''(-2 \, \Delta x), 0 , ... , 0 , W''(2 \, \Delta x) , W''(\Delta x))
\end{equation}
contains all the matrix coefficients. Circulant matrices enjoy the property to be diagonalizable in the same basis, composed of the n-th root of unity \citep{SWB-011280778}.
Therefore, $\uu{\uu{A}}''$ can written as:
\begin{equation}
\uu{\uu{A}}''  = \uu{\uu{P}}^{-1}  \, \uu{\uu{\Delta}}''  \, \uu{\uu{P}}
\label{eq_diago_Asecond}
\end{equation}
with $\uu{\uu{P}}$ is the change-of-basis matrix of general term:
\begin{equation}
p_{ij} = \frac{1}{\sqrt{N}} \, \exp \left(  \hat{i} \, \frac{2 \pi}{N} i j \right)
\end{equation}
where $\hat{i}$ is the imaginary number.
The diagonal matrix 
$\uu{\uu{\Delta}}'' = \text{diag}(\lambda_0'', \lambda_1'', \lambda_2'', ..., \lambda_{N-1}'')$
in Eq.~\ref{eq_diago_Asecond} is expressed in the general case by:
\begin{equation}
\lambda_j'' = \sum\limits_{k=0}^{N-1} c_k'' \,  \exp \left(  \hat{i} \, \frac{2 \pi}{N} k j \right)
\end{equation}

\section{Validation of temporal solutions \label{appendix_validation}}

\improve{This section provides a validation of the derivation of the perturbations equation for each of the four methods. Therefore the temporal solutions provided for each method are compared to the results of a simple numerical scheme.}
Two types of perturbations are investigated. First, an inviscid flow ($\nu$=$\alpha$=0), where only the particle at the center of the domain is perturbed.
Second, a viscous flow ($\nu$=0.1, equivalently $\alpha$=0.2 for $\delta$-SPH), where all particles are following an initial sinusoidal perturbation. In this case, two different modes are investigated.\\
In both cases, the initial perturbed value as well as the intensity of the perturbation was varied depending on the method (Table~\ref{tab_test_case}). This was done in order to illustrate different initial configurations, but every SPH method was extensively tested with all types of initial perturbations. 
The domain is a periodic line composed of 63 particles (N=63) and each particle has 6 neighbors (M=3). The time integration of the numerical scheme is a first-order sequential explicit scheme \citep{violeau2014maximum}, and the time step $\Delta t$ is selected according to:
\begin{equation}
\Delta t =  C_{\Delta t} \, \min ( \Delta t_{\nu},\Delta t_{C \! F \! L} )
\quad \text{where} \quad
\Delta t_{\nu} = \frac{1}{8} \, \frac{\Delta x^2} {\nu}
\quad \text{and} \quad
\Delta t_{C \! F \! L} = \frac{1}{4} \, \frac{\Delta x} {c + u_{max}}
\label{eq_CFL_condition_validation}
\end{equation}
The constant $C_{\Delta t}$ was introduced because the different SPH methods
show different accuracies depending on the time integration.
Its values are recalled in Table~\ref{tab_test_case}.
\improve{Please note that these differences of accuracy 
depicts the sensitivity of 
each method to the numerical integration scheme
used to produce the reference temporal solution.
The analytical temporal solution provided by our derivation do not require any time integration.}
\majrev{According to the analysis by \citet{violeau2014maximum} on the stability of the integration scheme, the combination of the sequential scheme with the constants in Eqs.~\ref{eq_CFL_condition_validation} and the constants in Table~\ref{tab_test_case} ensures that the scheme is stable.}
Other parameters are $p_{bg}$=1, $c$=1 and $\gamma$=7.\\
Specific parameters for $\delta$-SPH are $\alpha=2\nu$, $\xi=\alpha/4$, $\chi=\xi$ and $\Gamma=0.151$. The latter was estimated as proposed by \citet{antuono2010free} by taking into account the background pressure in the equation of state. For TV-SPH, both physical and transport velocities are set to zero unless it is mentioned.
\begin{table}[htb]
	\centering
	\caption{Perturbed quantities and integration time constant used in test cases 1 and 2.}
	\begin{tabular}{c|c|c|c|c}
		Parameter          & sum-SPH & div-SPH &  $\delta$-SPH &  TV-SPH \\
		\hline
		Perturbed value & $u$ &  $x$ & $\rho$ & $x$ \\
		$\delta_0$ & 0.1 $c$ & 0.1 $\Delta x$ & 0.1  $\ol{\rho}$ & 0.1 $\Delta x$ \\
		$C_{\Delta t}$ & 1 & 0.5 & 0.1 & 1  \\      
	\end{tabular}
	\label{tab_test_case}
\end{table}
\\\\
\noindent For sum-SPH and div-SPH, the analytical solution is directly compared to the results of the simulation. Therefore, Eqs.~ \ref{eq_WCSPH_solution}, \ref{eq_sumSPH_solution_of_perturbations} and \ref{eq_WCSPH_F_and_G} are compared to the numerical solution of the system~\ref{eq_WCSPH01} for sum-SPH, and Eqs.~\ref{eq_divSPH_temporal_funcs}, \ref{eq_divSPH_sum_temporal} and \ref{eq_divSPH_to_geom_H} are compared to
\improve{the numerical solution of}
the system~\ref{eq_divu01} for div-SPH.
Concerning $\delta$-SPH and TV-SPH, no analytical solutions were derived. Therefore, the transfer functions in the Laplace domain were decomposed into simple complex rational functions using partial fraction decomposition for each temporal mode $i$:
\begin{equation}
P_i(p) = \frac{z_1}{p - z_2} + \frac{z_3}{p - z_4} + \frac{z_5}{p - z_6}
\end{equation}
where $z_j$ are complex numbers.
Then, each simple rational function is transformed into a temporal function using the identity:
\begin{equation}
\mathscr{L}^{-1} \left[\frac{\alpha}{p - \beta} \right]  = \alpha \, \exp(\beta t)
\quad \text{for} \quad
p > \Re(\beta)
\quad \text{and} \quad
(\alpha, \beta) \in \mathbb{C}^2
\end{equation}
The temporal functions are then transformed into the geometrical space using the general change-of-basis formula for circulant matrix:
\begin{equation}
\label{eq_general_projection_to_geom}
F_{ij}(t) = \frac{1}{N} \left[\sum\limits_{k=0}^{N}  f_k(t) \ \exp \left( \hat{i} \, \frac{2 \pi}{N} k (i-j)  \right) \right]
\end{equation}
This last operation is equivalent to expressing the matrices $\uu{\uu{A}}'$, $\uu{\uu{A}}''$ and $\uu{\uu{A}}'''$ into their original basis. One can observe that Eqs.~\ref{eq_WCSPH_F_and_G} and \ref{eq_divSPH_to_geom_H} are particular cases of Eq.~\ref{eq_general_projection_to_geom}. The functions $F_{ij}(t)$ are finally compared to the numerical solutions of the original system, Eqs.~\ref{eq_delta01} for $\delta$-SPH and Eqs.~\ref{eq_TVF_01} for TV-SPH.

\subsection{Case 1: inviscid flow with a single perturbed particle}

\begin{figure}[!htb]
	\centering
	\includegraphics[width=0.49\columnwidth,keepaspectratio]{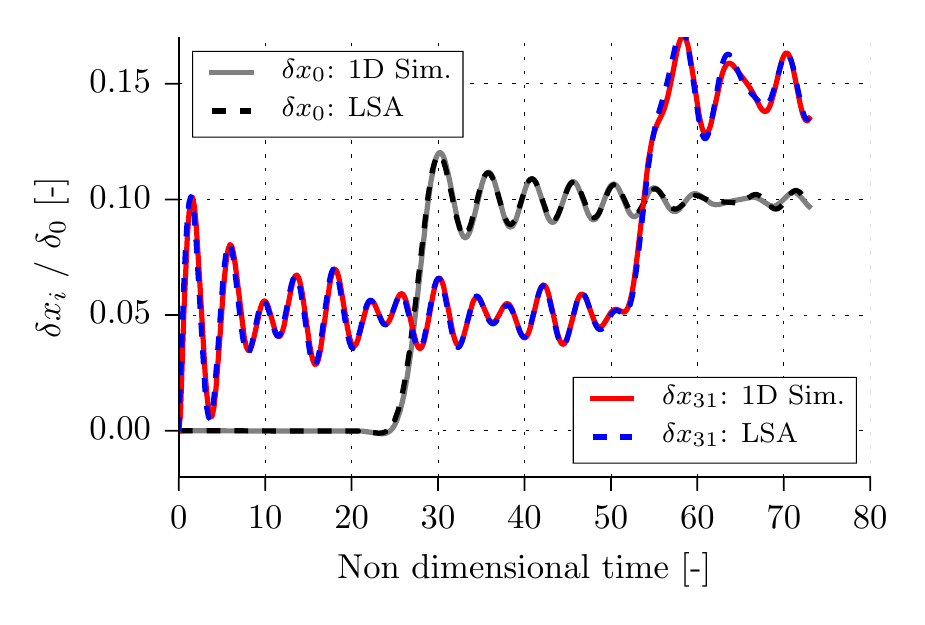}
	\includegraphics[width=0.49\columnwidth,keepaspectratio]{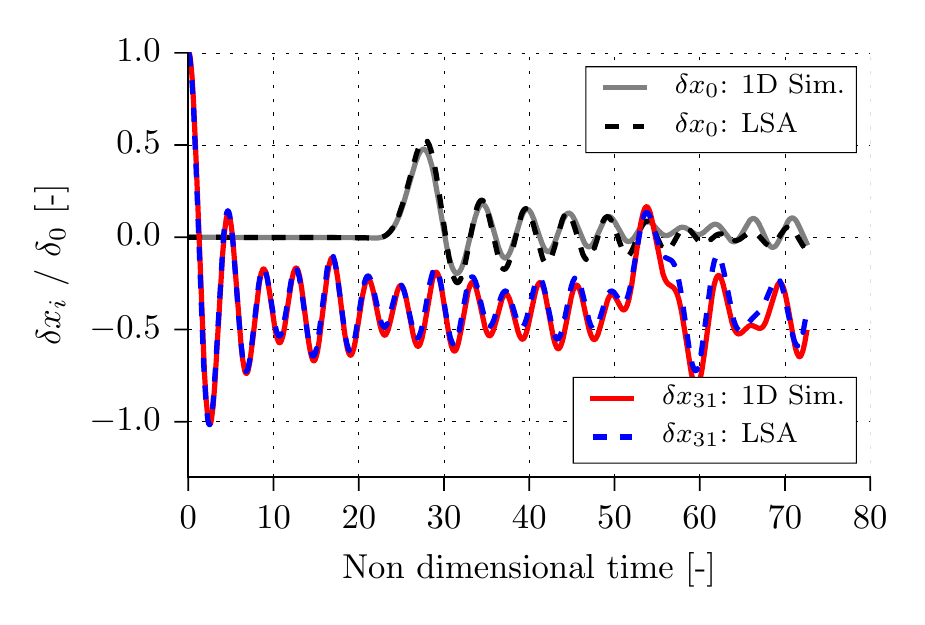}
	\caption{Comparison of LSA solution with a 1D SPH simulation. Left: sum-SPH. Right: div-SPH.}
	\label{fig_validate_case1_01}
\end{figure}

\begin{figure}[!htb]
	\centering
	\includegraphics[width=0.49\columnwidth,keepaspectratio]{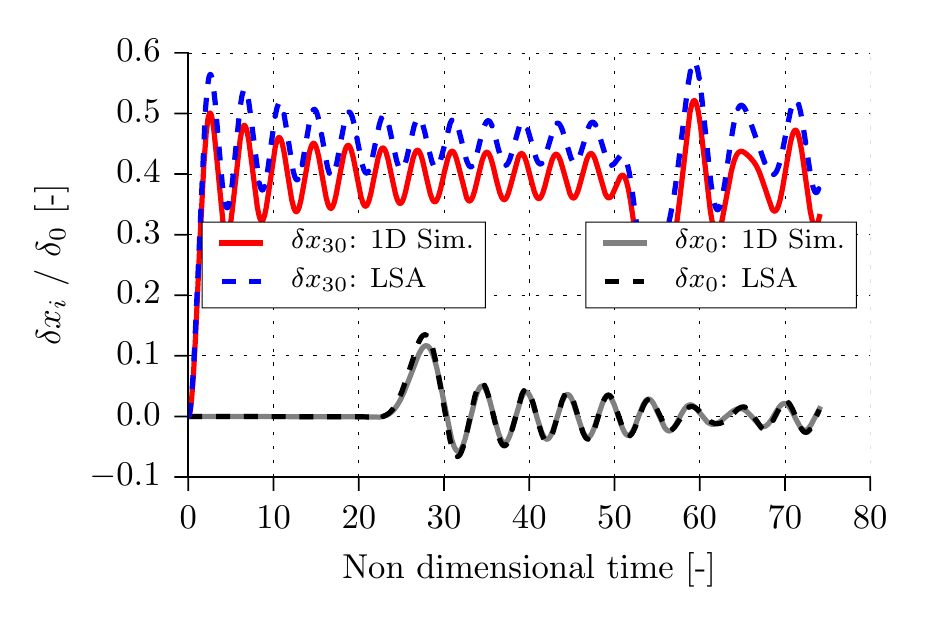}
	\includegraphics[width=0.49\columnwidth,keepaspectratio]{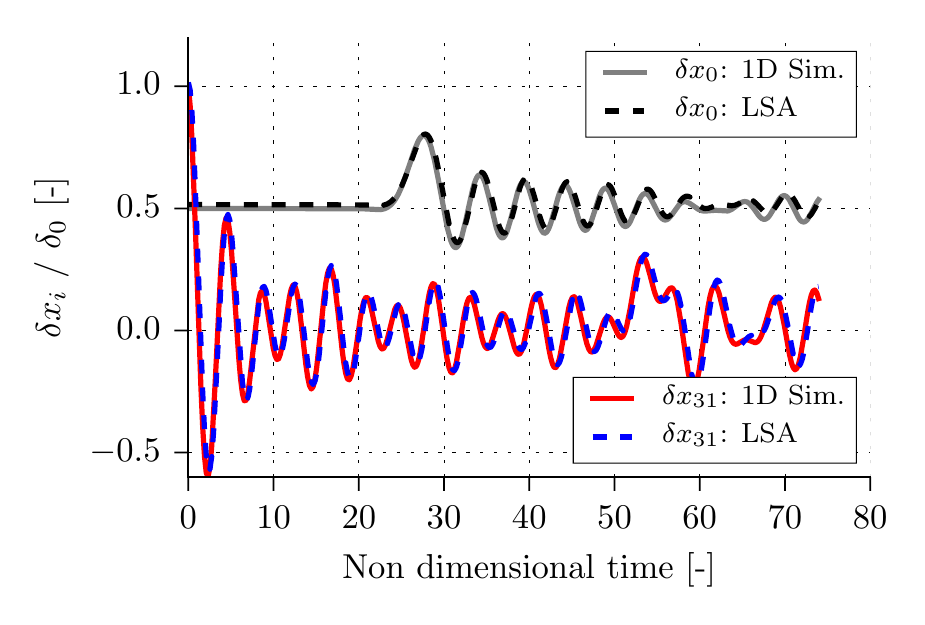}
	\caption{Comparison of LSA solution with a 1D SPH simulation. Left: $\delta$-SPH. Right: TV-SPH, ($\ol{u}$,$\ol{w}$)=(0,0), $\delta x_0$ was shifted of +0.5 for the sake of clarity.}
	\label{fig_validate_case1_02}
\end{figure}

In this subsection, the position of the initially perturbed particle ($\delta x_{31}$) and the particle which is located furthest away ($\delta x_{0}$) are shown. Figure~\ref{fig_validate_case1_01} (left) shows the comparison with sum-SPH, where the initial velocity was perturbed by 0.1c.
The comparison with div-SPH is shown in Fig.~\ref{fig_validate_case1_01} (right). It is observed that the perturbed particle ($\delta x_{31}$) does not oscillate around the equilibrium position, but around a position shifted by $\approx$ -0.5, as predicted by the constant term in Eq.~\ref{eq_divSPH_temporal_modal_position}.
\improve{The temporal solution shows a superposition of several modes, which are retrieved by our derivations.}
For both methods, the agreement is excellent,
even for a large perturbation of 0.1c and 0.1$\Delta x$. The time delay of the information propagation, visible on particle 0, is well taken into account.
\\The comparison with $\delta$-SPH and TV-SPH are shown in Fig.~\ref{fig_validate_case1_02}. For $\delta$-SPH, the perturbed position $\delta x_{31}$ of particle 31 was constant over the time so the position $\delta x_{30}$ of the next particle is presented. The slight discrepancy for $\delta x_{30}$ is attributed to the sensitivity of the $\delta$-SPH method to the time integration. For instance, an initial perturbation of 0.01 $\ol{\rho}$ gives an excellent agreement.
\majrev{This sensitivity to the integration scheme highlights the need for a subsequent study on the influence of the integration scheme on this particular method, as proposed by \citet{violeau2014maximum}.}
In the comparison with TV-SPH (Fig.~\ref{fig_validate_case1_02} right), $\delta x_0$ was artificially shifted by +0.5 for the sake of clarity. It also shows a good agreement between LSA and SPH simulation.

\subsection{Case 2: viscous flow with a spatial sinusoidal perturbation}

\begin{figure}[h]
	\centering
	\includegraphics[width=0.49\textwidth,keepaspectratio]{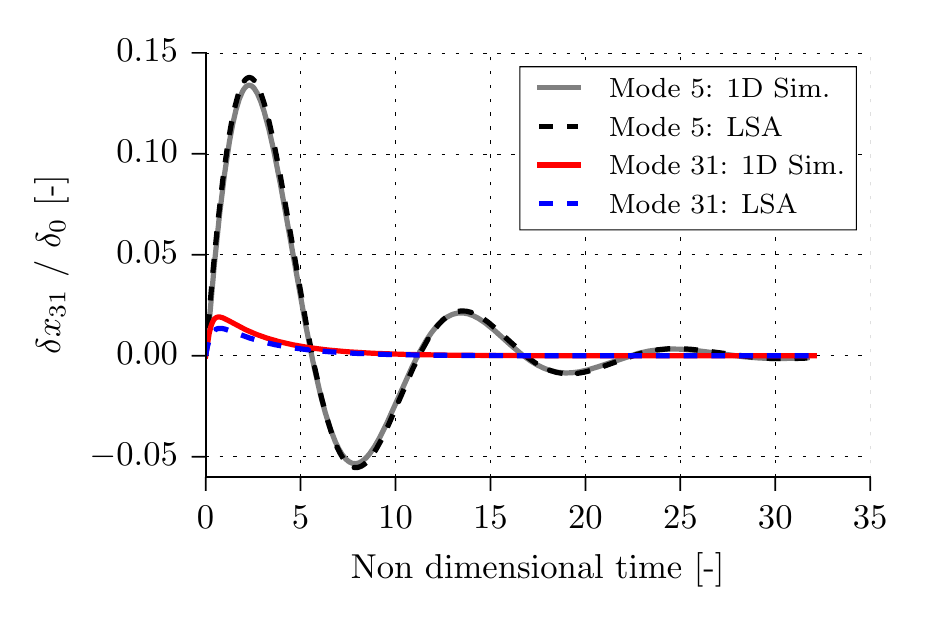}
	\includegraphics[width=0.49\textwidth,keepaspectratio]{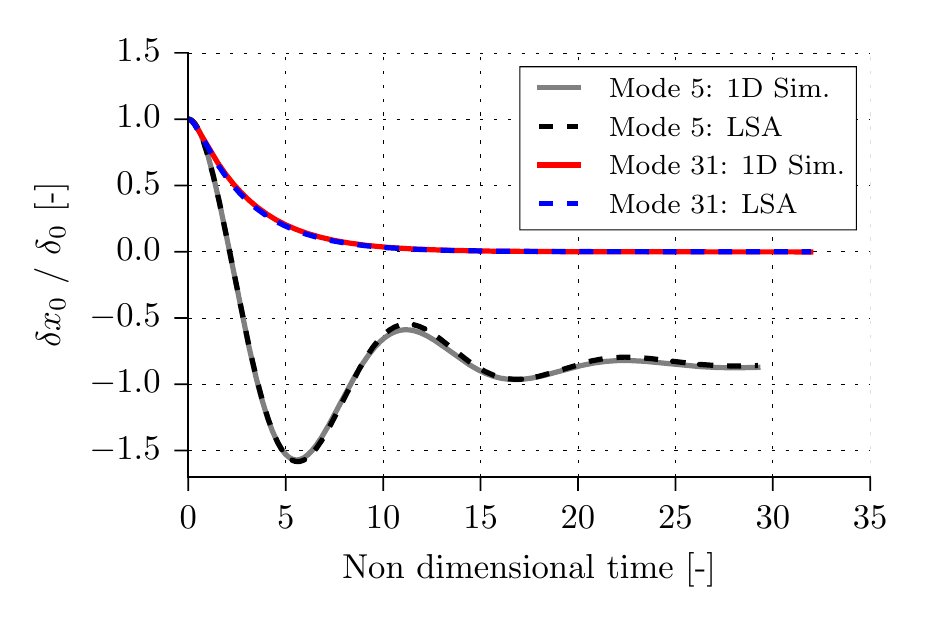}
	\caption{Left: sum-SPH. Right: div-SPH}
	\label{fig_validate_case2_01}
\end{figure}
\begin{figure}[h]
	\centering
	\includegraphics[width=0.49\textwidth,keepaspectratio]{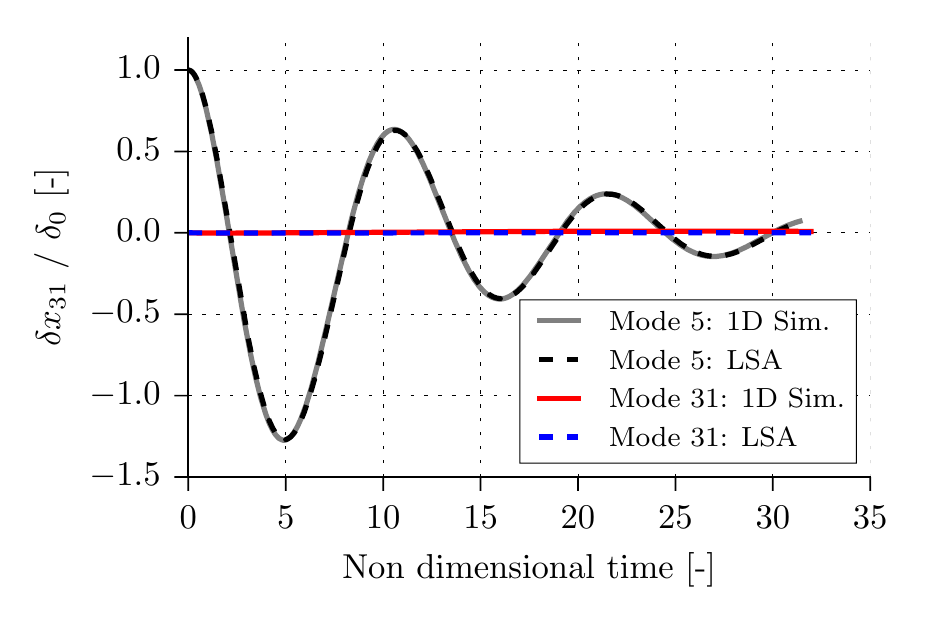}
	\includegraphics[width=0.49\textwidth,keepaspectratio]{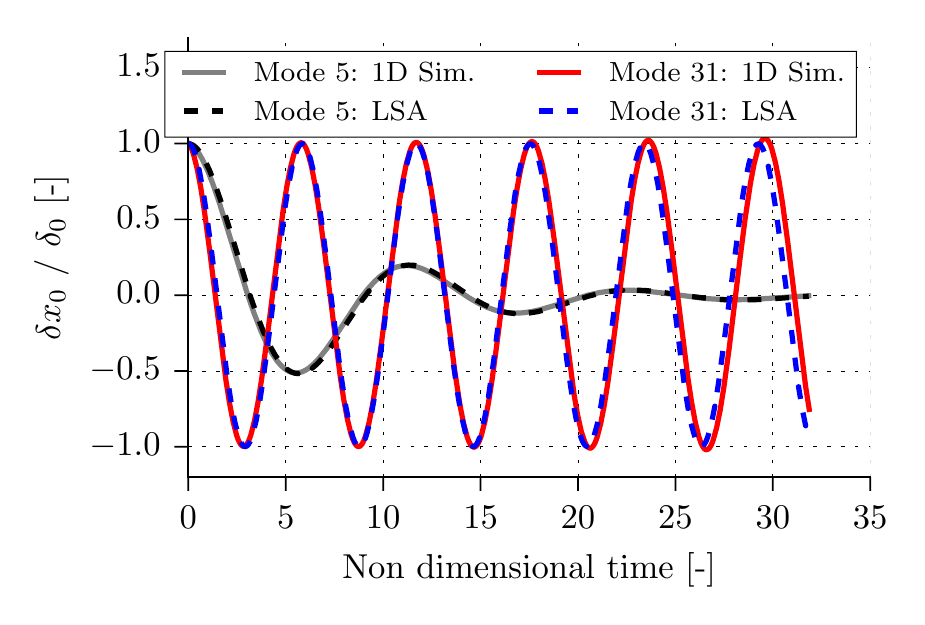}
	\caption{Left: $\delta$-SPH. Right: TV-SPH}
	\label{fig_validate_case2_02}
\end{figure}
For this test case, all particles are spatially perturbed according to a cosine function. Two different modes are investigated. With a wavelength of almost two time the diameter of the sphere of influence, mode 5 represents \textit{resolved} oscillations. Mode 31 is investigated as it represents high frequency spatial oscillations which are prone to instability. Since in mode 31, every consecutive particle is shifted in the opposite 'direction', the perturbation amplitude is set to $\delta_0/2$ in this case compared to $\delta_0$ for mode 5.\\
The comparison between LSA and SPH simulation is presented in Figs.~\ref{fig_validate_case2_01} and \ref{fig_validate_case2_02} where the agreement is excellent for all methods. For TV-SPH (Fig.~\ref{fig_validate_case2_02} right), the mode 31 is not damped, as it was highlighted by the dispersion curve discussion in Section~\ref{sec_dispersion}. 
\begin{figure}[!h]
	\centering
	\includegraphics[width=0.49\textwidth,keepaspectratio]{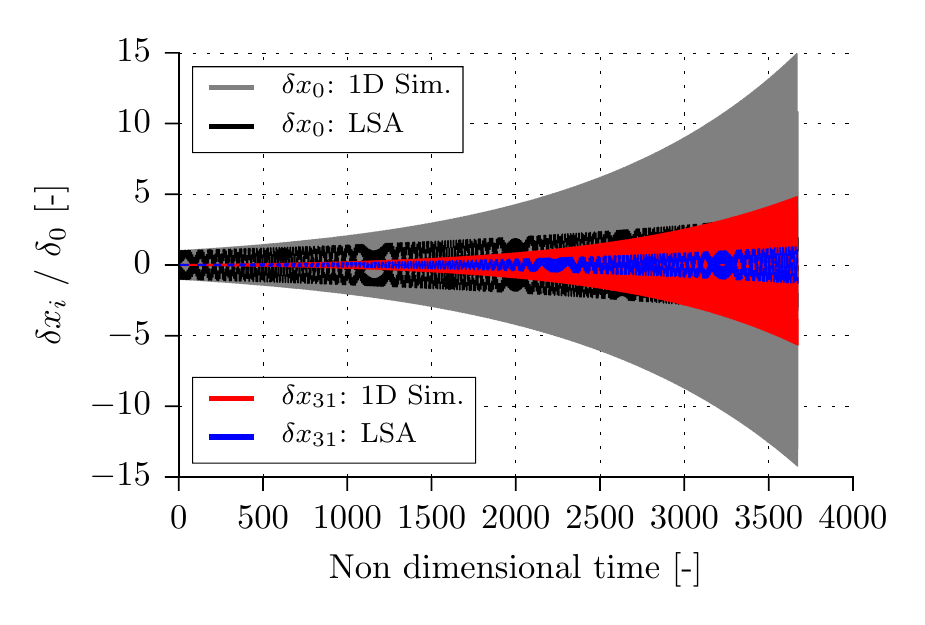}
	\caption{Slow divergence of TV-SPH in case of constant velocities: ($\ol{u}$,$\ol{w}$)=(0.05,0.1)}
	\label{fig_validate_case2_03}
\end{figure}
\\Finally, in Fig.~\ref{fig_validate_case2_03} mode 5 and 31 are shown for TV-SPH with ($\ol{u}$,$\ol{w}$)=(0.05,0.1) during a long time simulation. The instabilities exhibited in Fig.~\ref{fig_TVSPH_dispersion_visco_01} are visible but with a larger growth rate in the 1D simulation compared to LSA. This deviation of growth rate is probably due to the first-order time integration scheme of the numerical method.
\\\\
\improve{As a conclusion of this validation part, the temporal solutions provided by our derivations perfectly match the solutions provided by the numerical resolution of the original systems of equations.
In the case of a single perturbed particle, which corresponds to a continuous spectrum of excitations, the temporal solutions shows a superposition of all modes. This means that all modes predicted by our derivation are well retrieved in the numerical simulation. In addition, the caveats highlighted by the dispersion curves are also well illustrated by the numerical simulations. Indeed, the div-SPH method fails to damp a perturbation of position (Figs.~\ref{fig_validate_case1_01} right and \ref{fig_validate_case2_01} right), while the TV-SPH method do not damp the mode of higher frequency (Figs.~\ref{fig_validate_case2_02} right).}

\majrev{
\section{Validation of the instability at large background pressure \label{appendix_pback_lim}}

In this subsection the instability at large background pressure presented in Section~\ref{sssec_critical_pback} is verified. The SPH method is sum-SPH, and the other numerical parameters are the same as in the previous subsection.
The setup consists of a line of 500 particles perturbed at lower modes (1 $\to$ 9) where the instability arises. Two additional modes are tested, mode 50 and 125. The Wendland kernel is used with a $M$ parameter of 3.
The particles are perturbed by a sinusoidal velocity whose amplitude is equal to 1\% of the sound speed, and the simulation is run for ten thousand time step. For each investigated mode, the background pressure is set 50, 70, 90, 100, 110, 130 and 150\% of the predicted critical pressure.
For modes 9, 50 and 125, the critical pressure was set after preliminary tests.
The stability of the system is monitored with the value of the time step. Due to the CFL condition, the time step sharply decreases in case of a larger velocity triggered by an instability.
\begin{figure}[h]
	\centering
	\includegraphics[width=0.49\textwidth,keepaspectratio]{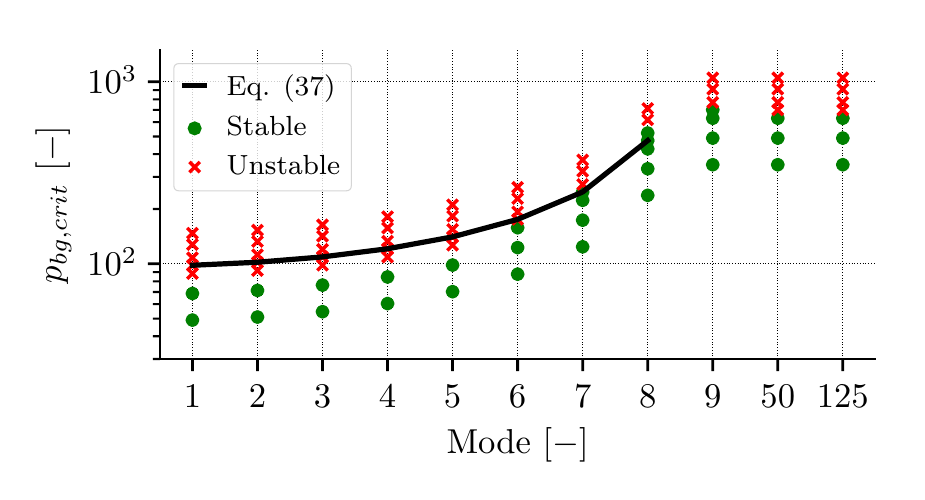}
	\caption{Comparison between the analytical value of Eq.~\ref{eq_phii2_negative} and the results of the numerical simulation.}
	\label{fig_pback_lim_valid}
\end{figure}

The results are shown in Fig.~\ref{fig_pback_lim_valid}. The agreement is good. Up to mode 5, the instability arises for $p_b$ between 70 and 90\% of the theoretical value, and between 90 and 110\% for modes 6, 7, and 8. It is found that the numerical simulations are unstable for larger modes whereas the theory does not predict any instability.
Its possible origin could be due to the deterioration of the spatial perturbation. Due to many  oscillations, the accuracy of the particle positions would decrease because of the truncation error of float numbers. This would lead to a spatial perturbation composed of different modes, including the unstable ones.

}

\end{document}